\documentclass[aps,pre,twocolumn,superscriptaddress,longbibliography,nofootinbib]{revtex4-1}  
\usepackage{graphicx} 
\usepackage{float} 
\usepackage{dcolumn} 
\usepackage{bm,color}
\usepackage{amsmath,amssymb,dsfont,amstext,amsfonts}
\usepackage{extarrows}
\usepackage[colorlinks=true,linkcolor=blue,urlcolor=blue,citecolor=blue]{hyperref}
\usepackage{xcolor}
\usepackage{wasysym}
\usepackage{mathtools}
\usepackage{bbold}
\usepackage{multirow}
\usepackage{longtable}

\usepackage{tikz-cd}
\usepackage[normalem]{ulem} 
\usepackage{color}
\usepackage{mathrsfs}

\def\bra#1{\mathinner{\langle{#1}|}}
\def\ket#1{\mathinner{|{#1}\rangle}}

\def\sgn{\mathrm{sgn}}

\def\mod{\:\mathrm{mod}\:}
\def\imi{\mathrm{i}}
\renewcommand{\paragraph}[1]{{\par\it #1.---}\ignorespaces}
\def\bs#1{\boldsymbol{#1}}

\usepackage[normalem]{ulem} 
\usepackage{soul}                   
\definecolor{TB}{rgb}{1,0.5,0}

\begin{document}
\title{Geometric approach to fragile topology beyond symmetry indicators}
\author{Adrien Bouhon}
\affiliation{Nordic Institute for Theoretical Physics (NORDITA), Stockholm, Sweden}
\affiliation{Department of Physics and Astronomy, Uppsala University, Box 516, SE-751 21 Uppsala, Sweden}
\author{Tom\'{a}\v{s} Bzdu\v{s}ek}
\address{Condensed Matter Theory Group, Paul Scherrer Institute, CH-5232 Villigen PSI, Switzerland}
\address{Department of Physics, University of Z\"{u}rich, Winterthurerstrasse 190, 8057 Z\"{u}rich, Switzerland}
\author{Robert-Jan Slager}
\affiliation{TCM Group, Cavendish Laboratory, University of Cambridge, J. J. Thomson Avenue, Cambridge CB3 0HE, United Kingdom}
\affiliation{Department of Physics, Harvard University, Cambridge MA 02138, USA}

\date{\today}

\begin{abstract}
We present a framework to systematically address topological phases when finer partitionings of bands are taken into account, rather than only considering the two subspaces spanned by valence and conduction bands. Focusing on $C_2\mathcal{T}$-symmetric systems that have gained recent attention, for example in the context of layered van-der-Waals graphene heterostructures, we relate these insights to homotopy groups of Grassmannians and flag varieties, which in turn correspond to cohomology classes and Wilson-flow approaches. We furthermore make use of a geometric construction, the so-called Pl\"ucker embedding, to induce windings in the band structure necessary to facilitate non-trivial topology. Specifically, this directly relates to the parametrization of the Grassmannian, which describes partitioning of an arbitrary band structure and is embedded in a better manageable exterior product space. From a physical perspective, our construction encapsulates and elucidates the concepts of fragile topological phases beyond symmetry indicators as well as non-Abelian reciprocal braiding of band nodes that arises when the multiple gaps are taken into account. The adopted geometric viewpoint most importantly culminates in a direct and easily implementable method to construct model Hamiltonians to study such phases, constituting a versatile theoretical tool.
\end{abstract}

\maketitle
\section{Introduction} 
Whereas the conceptional discovery of topological insulators \cite{Rmp1,Rmp2} is nearing a fifteen-year anniversary, the research into their properties and material realizations remains increasingly active. 
The consideration of spatial symmetries and of gapless systems has by now resulted in a rich variety of topological phases and characterizetions \cite{Schnyder08,Kitaev,InvTIBernevig, Clas1, InvTIVish, Clas2, probes_2D,Chenprb2012,Morimoto_2013, Shiozaki14,SchnyderClass,Mode2,Bouhon_CurrentInversion,Codefects1, Wi1,Wi3,Nodal_chains, Modebbcst,Clas3,ShiozakiSatoGomiK,Wi2,NodalLines1,Clas4, Clas5, BzduSigristRobust,BbcWeyl,Bouhon_HHL,UnifiedBBc,HolAlex_Bloch_Oscillations,Bzdusek_conversion,ShiozakiSatoGomiK,Codefects2,semimetals}. Recently, consistency equations for representations in momentum space were used to describe the possible topological band configurations~\cite{Clas3,Wi2}, which has provided several schemes to compare these configurations against those that have an atomic limit~\cite{Clas4,Clas5,HolAlex_Bloch_Oscillations}. 
More specifically, band representations that cannot be written as an integer sum of band structures corresponding to atomic orbitals are diagnosed as topological. 

There is a possibility that a non-trivial band representation amounts to a \emph{difference} of two trivial (i.e.~atomic) configurations, inducing the so-called \emph{fragile topology}~\cite{Ft1}. Following this discovery, there has been an intense activity in the characterization of fragile topology when it is indicated by the irreducible representations of crystalline symmetries \cite{bouhon2018wilson,AlexHol_nogo,Bradlyn_fragile,song2019fragile,Hwang_inversion_fragile,alex2019crystallographic,Elcoro_SmithDecomp}. Further advances in unveiling the physical properties of such symmetry-eigenvalue-indicated fragile topology have been achieved with the prediction and observation of \textit{twisted} bulk-boundary correspondence \cite{Song794,Peri797}.

Conventionally, both the stable and the fragile topology of band structures are characterized under the condition of a single spectral gap. This can be thought of as partitioning the bands into two subspaces, i.e.~an ``occupied'' subspace spanned by  states with energies below the energy gap, and the complementary ``unoccupied'' subspace spanned by states with energies above the energy gap. However this is in fact the coarsest partitioning of bands that can enable nontrivial topology.

In this work, we consider a finer characterization of band topology, which is obtained by assuming multiple spectral gaps. Such a refined partitioning of energy bands has been recently applied to certain $C_2\mathcal{T}$-symmetric and $\mathcal{PT}$-symmetric systems ($C_2$ is $\pi$-rotation, $\mathcal{T}$ is time reversal, and $\mathcal{P}$ is space inversion) when symmetry indicators are not necessarily available. Indeed, information from the irreducible representations~\cite{Clas3,Wi2,Clas4} and elementary band representations~\cite{Clas5}, may not be sufficient to diagnose the fragile criterion, rather similar to how they cannot detect Chern number, or the Kane-Mele $\mathbb{Z}_2$ invariant~\cite{KaneMele_Z2} and the $\mathbb{Z}_2$ nested Berry phases~\cite{BJY_nielsen,Wieder_axion,Kooi_nested}, in certain scenarios. In this context, the consideration of multiple spectral gaps recently provided new insights into the fragile band topology characterized by Wilson loop winding (Euler class)~\cite{bouhon2018wilson,BJY_linking,BJY_nielsen}, and has led to the prediction of
a new kind of reciprocal braiding of band nodes inside the momentum space~\cite{Wu1273,BJY_nielsen,Tiwari:2019, bouhon2019nonabelian}. 

Remarkably, the topological insights obtained from such refined multi-gap partitioning of bands and their interplay with $C_2\mathcal{T}$-symmetry touch upon several experimentally viable systems, as they constitute the key elements in the modelling of twisted layer graphene systems~\cite{Potwisted,Cao2018} and of non-Abelian braiding of Dirac points therein~\cite{Kang_Dirac_node_braiding,BJY_nielsen} and of Weyl points in ZrTe \cite{bouhon2019nonabelian}. Very recently Euler class has also been reported to produce robust signatures in quenched optical lattices~\cite{Unal_quenched_Euler}. Fragile topology has furthermore been shown to play a role in the new field of higher-order topology and axion insulators~\cite{Wieder_HOTI,Wieder_axion} where we foresee a prospective utilization of our geometric approach.

The main achievement of the present study is a systematic geometric construction of fragile topological phases beyond symmetry indicators in $C_2\mathcal{T}$-symmetric systems. Specifically, we consider the so-called Pl\"ucker embedding which enables us to parametrize real oriented Grassmannians that classify the Bloch Hamiltonians and band structures in question. As a next step, we can then address the topology by considering the homotopy classes of these objects. Such homotopy evaluations allows us to construct representative Hamiltonians for each topological phase~\cite{abouhon_EulerClassTightBinding}, while also intimately relating to Wilson flow arguments that provide in many circumstances a readily implementable viewpoint to discern band topologies \cite{bouhon2018wilson,WindingKMZ2, InvTIBernevig, PointGroupsTI, Wi1, Wi3,  Alex_BerryPhase, Wi2, AlexAdiabaticSOC, BBS_nodal_lines, HolAlex_Bloch_Oscillations}. 

\begin{figure}[t]
\centering
\begin{tabular}{c}
	\includegraphics[width=0.98\linewidth]{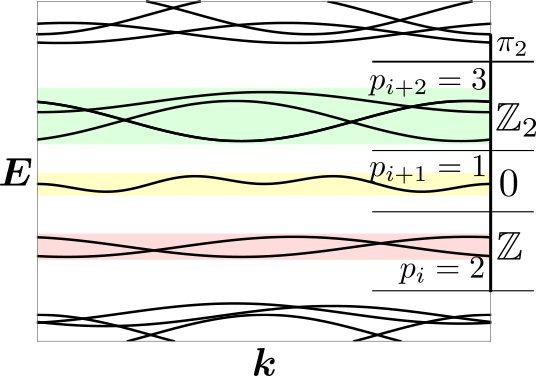}
\end{tabular}
\caption{\label{fragile} Band partitioning with multiple gap conditions. Each block of energy bands (colored strips) is separated from all other bands by energy gaps (white regions) both from above and from below. The stable topology of the $i$-th subspace with a number $p_i$ of bands is classified by cohomology classes which we show correspond to elements of the second homotopy group of a Grassmannian, $\pi_2[\mathrm{Gr}_{p_i,N}]$. When the bands are orientable, i.e.~when the subspace does not carry $\pi$-Berry phase (see text), one-band subspaces are trivial, two-band subspaces are classified by the Euler class in $\mathbb{Z}$ (reduced to $\mathbb{N}$ after dropping the orientation), and three(or more)-band subspaces are classified by the second Stiefel-Whitney class in $\mathbb{Z}_2$, see Sec.~\ref{sec:homotopy-groups}. 
}
\end{figure}

The manuscript is organized as follows. We begin in Sec.~\ref{sec:band-structures} by specifying the symmetry settings and the assumptions on tight-binding models of band structures. In this context we also introduce the notion of a total Bloch bundle. In Sec.~\ref{sec:gaps-class} we define several notions related to vector bundles and frame bundles, including the appropriate classifying spaces (the \emph{Grassmannians}) which provide the natural language to completely characterize the studied band topology. In Sec.~\ref{sec:orientability} we discern the notions of \emph{orientedness} vs.~\emph{orinetability}, which later translate to a difference between \emph{based} and \emph{free} homotopy classes. We also comment on several related but distinct mathematical notions, attempting to resolve possible sources of misconception. We continue in Sec.~\ref{sec:homotopy-groups} by discussing the homotopy groups of the classifying spaces of vector subbundles, and we relate the identified topological invariants to the Euler and the Stiefel-Whitney characteristic classes. In Sec.~\ref{sec:partitioning} we generalize the mathematical description to the presence of multiple band gaps (cf.~Fig.~\ref{fragile}) and relate the obtained topological invariants again to the characteristic classes. This generalized ``multi-gapped'' context allows us to define fragile topology via repartitioning of energy bands. We argue that an observable signature of both the Euler and the second Stiefel-Whitney class of a band subspace is given by a minimum number of stable nodal points formed within the band subspace.

After introducing this set of key mathematical notions, we use the developed machinery to generate physical models corresponding to various fragile topological phases. First, in Sec.~\ref{geo} we discuss our strategy in a general abstract setting. We show that a representative of any topological class can be obtained as a pullback of the tautological total gapped bundle on the classifying space, where explicit Hamiltonians are parametrized through the Pl\"{u}cker embedding. We then turn our attention to specific few-band examples. Specifically, in Sec.~\ref{3_bands} we focus on the case of three bands that are partitioned into a two-band and single-band subspace. We similarly perform this analysis for the four-band case in Sec.~\ref{4_bands}, where the extra band gives rise to various different partitionings in terms of single-band and two-band blocks. In both instances we use our general insights to address the classification aspects as well as their topological stability, resp.~fragility, that are of direct physical interest. In Sec.~\ref{sec:generalizations} we set the basis of the study of systems with more bands and gaps, as well as of higher dimensional fragile topological phases, hence underpinning the generality of the framework. Finally, in Sec.~\ref{sec:conclusions} we turn to the conclusions and discussions, where we outline several directions of extension. 

We exported the tight-binding models produced by the described mathematical machinery as \texttt{Mathematica} notebooks, which we made publicly available online~\cite{abouhon_EulerClassTightBinding}. These models were also used to produce the numerical results presented in Secs.~\ref{3_bands} and~\ref{4_bands}, as well as to study the signatures of the fragile topology in quenched optical lattices by Ref.~\cite{Unal_quenched_Euler}.

\section{Real band structures}\label{sec:band-structures}

We model crystalline systems through a Hermitian Bloch Hamiltonian $\mathcal{H} = \sum_{\mu\nu,\boldsymbol{k}\in B} \vert \phi_{\mu},\boldsymbol{k} \rangle H_{\mu\nu}(\boldsymbol{k}) \langle \phi_{\nu},\boldsymbol{k} \vert$, where the Bloch state $\vert \phi_{\mu},\boldsymbol{k} \rangle = \sum_{\boldsymbol{R}} e^{\imi \boldsymbol{k}\cdot (\boldsymbol{R}+\boldsymbol{r}_{\mu})} \vert w_{\mu} , \boldsymbol{R}+\boldsymbol{r}_{\mu} \rangle $ is the Fourier transform of the Wannier state $\vert w_\mu, \boldsymbol{R}+\boldsymbol{r}_{\mu}\rangle$ that represents the physical orbital $\mu$ at site $\boldsymbol{R}+\boldsymbol{r}_{\mu}$ (possibly with a spin), where $\boldsymbol{R}$ is a Bravais lattice vector, and $\boldsymbol{r}_{\mu}$ is the (sub-lattice) position within the $\boldsymbol{R}$-th unit cell. The Bloch wave vector $\boldsymbol{k}$ is a point of the Brillouin zone $B$, that is a 2-torus ($B=\mathbb{T}^2$) for two-dimensional crystals.\footnote{In later sections, we sometimes replace the base space $B$ by a $2$-sphere $\mathbb{S}^2$.} In this work we assume that the Bloch states $\vert \phi_{\mu},\boldsymbol{k} \rangle$ are fully trivial, i.e.~they carry no Berry phase and their Wannier representations $\langle \boldsymbol{r} \vert w_{\mu}, \boldsymbol{R}+\boldsymbol{r}_{\mu}\rangle = w_{\mu}(\boldsymbol{R}+\boldsymbol{r}_{\mu}-\boldsymbol{r})$ are exponentially localized.  
This implies that the real-space hopping amplitudes $H_{\mu\nu}(\bs{R}-\bs{R}')$ have an exponential decay in $\vert \bs{R}-\bs{R}'\vert$, such that the Fourier transform $H_{\mu\nu}(\boldsymbol{k})$ is smooth in $\boldsymbol{k}$. In practice the hopping amplitudes in the tight-binding models of materials are cut-off beyond a finite support. We remark that in this convention the states $\ket{\phi_\mu,\bs{k}}$ and the Bloch Hamiltonian $H(\bs{k})$ are not necessarily periodic in reciprocal lattice vectors.

It is known that $C_2\mathcal{T}$-symmetry~with $(C_2\mathcal{T})^2=+1$ implies the existence of a basis in which $H(\boldsymbol{k})$ is real~\cite{bouhon2019nonabelian}, irrespective of the spinfulness. 
In the subsequent text we assume this 
choice of basis, i.e.~$H(\boldsymbol{k})$ is an $N\times N$ real and symmetric matrix where $N\geq 2$ is the number of degrees of freedom per unit cell. 
This property implies that all eigenstates of $H(\boldsymbol{k})$ can be gauged to be real vectors \cite{BJY_linking}, allowing us to drop the difference between bra-states and ket-states, $\bra{u_n(\bs{k})}^\top = \ket{u_n(\bs{k})} \equiv u_n(\bs{k})$. 

From the eigenvalue problem $H(\boldsymbol{k}) u_{n}(\boldsymbol{k}) = E_{n}(\boldsymbol{k}) u_{n}(\boldsymbol{k})$, we get the spectral decomposition $H(\boldsymbol{k}) = R(\boldsymbol{k})  \mathcal{D}(\boldsymbol{k})  R(\boldsymbol{k})^T$, with eigenvalues $\mathcal{D}(\boldsymbol{k})=\mathrm{diag}\left[ E_1(\boldsymbol{k}),\dots,  E_N(\boldsymbol{k}) \right]$, and the diagonalizing matrix $R = \left(u_1\cdots u_N\right)$ formed by the real column eigenvectors, i.e.~$u_n \in \mathbb{R}^N$. 
The 
eigenvectors define a rank-$N$ orthonormal frame, $R\!\in \!\mathsf{O}(N)$, and serve as a basis of the real vector space 
$V_{\boldsymbol{k}} = \mathrm{Span} \{u_1,\dots ,u_N\}_{\boldsymbol{k}} \cong \mathbb{R}^N$ at each point $\boldsymbol{k}\!\in\! B$. 
The collection of fibers $V_{\boldsymbol{k}}$ at each point $\boldsymbol{k}$ of the base space $B$ allows us to construct a real vector bundle \cite{Hatcher_2}. 

More precisely, we define the \textit{Bloch bundle} \cite{Panati_Chern} as the union of the fibers, $\mathcal{E}_{N,N}=\bigcup_{\boldsymbol{k} \in B}  V_{\boldsymbol{k}}$, with the continuous projection onto the base space, i.e.~$\pi: \mathcal{E}_{N,N}  \rightarrow B $, and so that it is locally homeomorphic to a direct product space, i.e.~$\phi : \pi^{-1}(U) \rightarrow U \times \mathbb{R}^N$ for any contractible open subset $U\!\subset\! B$. By virtue of the later property we say that $\mathcal{E}_{N,N}$ is \textit{locally trivializable}. In the following we fix the ordering of the eigenvalues, $E_1 \!\leq\! \dots \!\leq\! E_N$, and we assume the same ordering for the eigenvectors in $R$.

\section{Gap condition and classifying spaces}\label{sec:gaps-class}

\subsection{Vector subbundles and total gapped bundle}\label{sec:bundles-defs}

In this work we assume that the ``total'' Bloch bundle $\mathcal{E}_{N,N}$ as defined in Sec.~\ref{sec:band-structures} is trivial, which corresponds to situations in which the Bloch Hamiltonian can be brought to real-symmetric form periodic in reciprocal lattice vectors.\footnote{Some reasons and a simple example of when the assumption on the triviliaty of the total Bloch bundle fails are discussed in Sec.~\ref{clarifications} below.} Nontrivial topology may then arise by considering subbundles defined through a spectral gap condition~\cite{Budich_DensityMatrix}. Under the condition of a single energy gap 
\begin{equation}
\label{eq_gap_cond}
\begin{split}
	E_1 \leq \dots &\leq E_{p} < E_{p+1} \leq \dots\leq E_{N}\, \\
	&\textrm{with $1 \leq p < N$},
\end{split}
\end{equation} 
i.e.~with a finite gap $\delta(\boldsymbol{k}) =  E_{p+1}(\boldsymbol{k}) - E_{p}(\boldsymbol{k}) > 0 $ for all $\boldsymbol{k}\in B$, the total frame $R = (R_I R_{II})$ splits into subframes $R_I = (u_1\cdots u_{p})$ and $R_{II} = (u_{p+1}\cdots u_{N})$. The collection of all $p$-component subframes of $\mathbb{R}^N$ is called \emph{the Stiefel manifold}, labelled $\mathrm{P}_p(\mathbb{R}^N)$ \cite{Milnor:1974}.
We now define the rank-$p$ ``occupied'' vector subbundle
\begin{equation}
\mathcal{B}_{I}(p) \!=\!\! \bigcup_{\boldsymbol{k}\in B} \! V_{I,\boldsymbol{k}} \;\;\textrm{with}\;\;V_{I,\boldsymbol{k}} \!=\! \mathrm{Span}\{u_1(\boldsymbol{k})\dots u_p(\boldsymbol{k})\},\label{eqn:occ-vec-bundle}
\end{equation} 
and the rank-$(N-p)$ ``unoccupied'' subbundle $\mathcal{B}_{II}(N-p)$ similarly via $V_{II,\boldsymbol{k}} = \mathrm{Span}\{u_{p+1}(\boldsymbol{k})\dots u_N(\boldsymbol{k})\}$. We will occasionally consider a restriction of the vector subbundle $\mathcal{B}_I(p)$ to a loop $l\subset B$ in the Brillouin zone, 
i.e.~$\{V_{I,\boldsymbol{k}} \,\vert\, \boldsymbol{k} \in l \} \equiv \mathcal{B}_I(p)\vert_l$. Furthermore, we sometimes call rank-1 subbundles as \textit{line bundles}.

While it is customary to consider only one vector subbundle at a time, a band structure with an energy gap
really consists of the \emph{ordered collection} of two subbundles $\mathcal{B}_{I}(p)$ and $\mathcal{B}_{II}(N-p)$, which we write as $\mathcal{E}_{p,N} = \mathcal{B}_{I}(p) \cup \mathcal{B}_{II}(N-p)$. We call this the \textit{total gapped bundle}. Importantly, $\mathcal{E}_{p,N} \neq \mathcal{B}_{I}(p) \oplus \mathcal{B}_{II}(N-p) \cong \mathcal{E}_{N,N}$. Indeed, the direct sum allows us to take arbitrary intra- and inter-subspace linear combinations of eigenvectors, i.e.~mixing the vectors of $\mathcal{B}_{I}(p)$ with those of $\mathcal{B}_{II}(N-p)$, see Sec.~\ref{clarifications}, while only intra-subspace linear combinations of eigenvectors are allowed in $\mathcal{E}_{p,N}$. In other words, the direct sum $\mathcal{E}_{N,N}$ ``forgets'' about the gap condition.

We finally consider the isomorphism (i.e.~equivalence) classes of the introduced bundles under continuous deformations that preserve the gap condition.
Assuming a fixed choice of base space $B$, we write $[\mathcal{B}_{I}(p)]$ for the 
isomorphism classes of rank-$p$ vector bundles 
that are subbundles of $\mathcal{E}_{N,N}$. 
We further write $[\mathcal{E}_{p,N}]$ for the 
isomorphism classes of total gapped bundles that split into the vector subbundles $\mathcal{B}_{I}(p)$ and $\mathcal{B}_{II}(N-p)$. 
Labeling the isomorphism classes with integers, 
we indicate the trivial class by $0$. We point out that by assumption the total Bloch bundle is a trivial rank-$N$ bundle, thus $[\mathcal{E}_{N,N}] = 0$.
It is important to note that for us, and contrary to what is usually done in the classification schemes based on $K$-theory (e.g. Ref.~\cite{Thiang_KTheory}), we keep $N$, i.e.~the rank of the underlying band structure, finite and fixed.

\subsection{Unoriented and oriented Grassmannians}\label{subsec:un_oriented_grass}

By flattening the spectrum, i.e.~$\mathrm{diag}[E_1,\dots,E_{p}]\rightarrow -\mathbb{1}$ and $\mathrm{diag}[E_{p+1},\dots,E_{N}]\rightarrow \mathbb{1}$, we get the flattened Hamiltonian $Q = R\cdot\left[ -\mathbb{1}_{p} \oplus \mathbb{1}_{N-p} \right]\cdot R^T $. The constructed $Q$ is invariant under any orthogonal gauge transformation $R\mapsto R \cdot [G_I \oplus G_{II}]$ with $G_{I} \in \mathsf{O}(p)$ and $G_{II} \in \mathsf{O}(N-p)$. The \emph{classifying space} of the flattened Hamiltonian is then obtained as the space of $R$ ``divided'' by the group of gauge symmetries, resulting in the quotient space 
\begin{equation}
\mathrm{Gr}_{p,N} = \mathsf{O}(N) / [\mathsf{O}(p)\times \mathsf{O}(N-p)] ,
\end{equation}
called the \emph{real Grassmannian}, having the property $\mathrm{Gr}_{p,N}=\mathrm{Gr}_{N-p,N}$ (the reason why it is called the classifying space will become clear in Sec.~\ref{sec:homotopy-groups} and \ref{geo}).

We note that any matrix $R\in \mathsf{O}(N)$ can be taken in $\mathsf{SO}(N)$ by a gauge transformation. It follows that the Grassmannian can be conveniently rewritten for $R\in\mathsf{SO}(N)$ as $\mathrm{Gr}_{p,N} = \mathsf{SO}(N) / \mathsf{S}[\mathsf{O}(p)\times \mathsf{O}(N-p)]$ by restricting the group of gauge transformations $R\mapsto R\cdot G$ to the subgroup with $\det G = +1$. More specifically, the point of the Grassmannian corresponding to the matrix $R\in\mathsf{SO}(N)$ is defined as the left coset
\begin{eqnarray}
\label{eq:leftcoset}
[R] &=& \left\{ R \cdot [G_{I} \oplus G_{II}], \;\;\textrm{such that} \right.\nonumber \\ 
&\phantom{=}& G_{I}\in \mathsf{O}(p) \;\;\textrm{and}\;\; G_{II}\in \mathsf{O}(N-p),\; \textrm{and}\nonumber \\
&\phantom{=}& \left. \det (G_I \oplus G_{II}) = +1 \right\}.
\end{eqnarray}
From now on we always assume that $R\in \mathsf{SO}(N)$.\footnote{Since $\pi_i[\mathsf{O}(N)] = \pi_i[\mathsf{SO}(N)]$ for all $i \geq 1$, there is no topological obstruction for injecting the frames $R$ from $\mathsf{O}(N)$ to $\mathsf{SO}(N)$ (assuming that the base space $B$ is connected).}

To any orthogonal matrix $G\in \mathsf{O}(N)$ we can associate an orientation through $\det G =\pm1$, and to any subframe $R_{I}$ we can associate an oriented exterior product $\omega_p = u_1 \wedge \cdots \wedge u_p$ that is invariant under $\mathsf{SO}(p)$ gauge transformations of the eigenvectors, resp. $\omega_{N-p} = u_{p+1} \wedge \cdots \wedge u_N$ for $R_{II}$. (These forms will be particularly useful when discussing the Pl\"{u}cker embedding in Sec.~\ref{sec:Plucker}.) By definition the coset $[R] = [(R_I R_{II})] \in \mathrm{Gr}_{p,N}$ is invariant under the orientation reversal of the subframes $R_I$ and $R_{II}$, i.e.~$(\omega_p,\omega_{N-p}) \rightarrow -(\omega_p,\omega_{N-p})$, hence $\mathrm{Gr}_{p,N}$ is called the real \textit{unoriented} Grassmannian. 

One can similarly consider the \textit{oriented} Grassmannian \begin{equation}
    \mathrm{Gr}^+_{p,N} = \mathsf{SO}(N)/[\mathsf{SO}(p) \times \mathsf{SO}(N-p) ],   
\end{equation}
where the gauge symmetries do not include orientation reversal of the subframes. More specifically, the point of the oriented Grassmannian corresponding to $R\in \mathsf{SO}(N)$ is defined as the left coset  
\begin{eqnarray}
\label{eq:leftcoset_oriented}
[R]^+ &=& \left\{ R \cdot [G_{I} \oplus G_{II}], \;\;\textrm{such that} \right.\nonumber \\ 
&\phantom{=}& \left.G_{I}\in \mathsf{SO}(p) \;\;\textrm{and}\;\; G_{II}\in \mathsf{SO}(N-p) \right\}.
\end{eqnarray}
Considerations within the oriented Grassmannian allow us to define the \emph{subframe-orientation reversal} operator
\begin{equation}
\label{eq:orientation_reversal_sr}
    \mathrm{sr}:R\mapsto R^{\mathrm{sr}} = R\cdot G_{\mathrm{sr}},
\end{equation}
e.g.~with $G_{\mathrm{sr}} = \left[(-1\oplus \mathbb{1}_{p-1}) \oplus (-1\oplus \mathbb{1}_{N-p-1}) \right]$ or any other transformation that reverses the orientation of the two subframes $R_I$ and $R_{II}$ at the same time. One should note that $[R]^+ \neq [R^\textrm{sr}]^+$. We further observe that by forgetting orientation every pair of points of opposite orientation in $\mathrm{Gr}^+_{p,N}$ is mapped to a single point in $\mathrm{Gr}_{p,N}$, i.e. there is a natural 2-to-1 injection $\bar{q}:\{[R]^+,[R^{\text{sr}}]^+\}\mapsto [R]$ from the oriented Grassmannian to the unoriented Grassmannian. This hints to the fact that $\mathrm{Gr}^+_{p,N}$ is the \textit{orientable double cover} of $\mathrm{Gr}_{p,N}$, with $\bar{q}$ called the \textit{covering map}, see Appendix \ref{ap:homotopy_grassmannian} where we review in more detail the geometric and topological properties of Grassmannians.

In the following, we often consider loops and spheres inside the Grassmannian as obtained through continuous maps respectively from the unit interval $\mathbb{I} = [0,1]$, and the unit square $\mathbb{I}^2=[0,1]\times[0,1]$. More precisely, we have the loop image $\ell : \mathbb{I} \rightarrow \mathrm{Gr}_{p,N} : s\mapsto \ell(s)$ with a base point $[R(\boldsymbol{k}_0)] = \ell(0) = \ell(1)$, and the sphere image $f : \mathbb{I}^2 \rightarrow \mathrm{Gr}_{p,N}: (s_1,s_2) \mapsto f(s_1,s_2) $ with a base point $[R(\boldsymbol{k}_0)] = f(\partial \mathbb{I}^2) $ ($\partial \mathbb{I}^2$ is the boundary of the unit square), and similarly for the oriented Grassmannian $\mathrm{Gr}^+_{p,N}$. The homotopy equivalence classes of loops $[\ell]$, and of spheres $[f]$, inside the Grassmannian constitute the elements of the first homotopy group $\pi_{1}[\textrm{Gr}^{(+)}_{p,N}]$, resp.~of the second homotopy group $\pi_{2}[\textrm{Gr}^{(+)}_{p,N}]$ (see Fig.~\ref{homotopy_square} of Appendix \ref{ap:homotopy_action}).

\subsection{The projective plane}\label{subsec:projective_plane}

\begin{figure*}[t!]
\centering
\begin{tabular}{ccc}
    \includegraphics[width=0.32\linewidth]{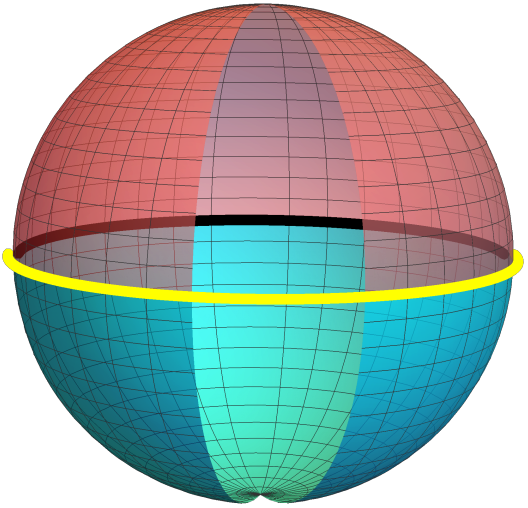} 
    &
	\includegraphics[width=0.33\linewidth]{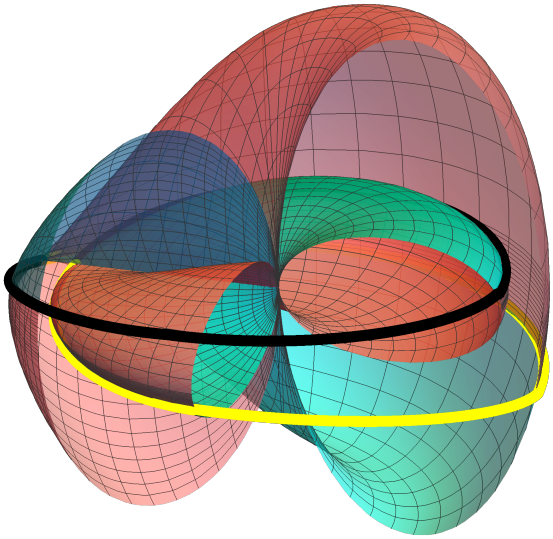}
	&
	\includegraphics[width=0.33\linewidth]{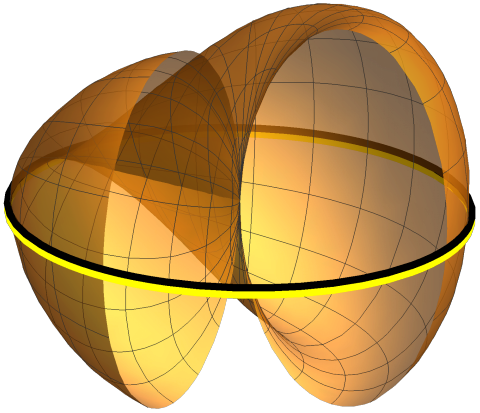} \\
	$\mathbb{S}^2$ & $\mathbb{S^2} \rightarrow \mathbb{R}\mathrm{P}^2$ &  $\mathbb{R}\mathrm{P}^2$ 
\end{tabular}
\caption{\label{double_cover_RP2}Orientable double cover $\mathbb{S}^2\rightarrow \mathbb{R}\mathrm{P}^2$ obtained through the folding of the sphere onto itself. It is dictated by the twisted folding of the equator, i.e.~we form the shape `8' and fold the black loop onto the yellow loop. This results in identifying every antipodal pair of points of the sphere to a single point of the projective plane $\mathbb{R}\mathrm{P}^2$.
} 
\end{figure*}

It is very instructive for the understanding of Grassmannians in general to look at the special case of the projective plane $\mathrm{Gr}_{2,3} = \mathbb{R}\mathrm{P}^2$ as it can be grasped pictorially. $\mathbb{R}\mathrm{P}^2$ is obtained from the sphere by identifying antipodal pairs of points, i.e.~$\mathbb{R}\mathrm{P}^2 = \mathbb{S}^2/\sim$ with $x \sim -x$. We show in Fig.~\ref{double_cover_RP2} the orientable double cover $\mathbb{S}^2 \rightarrow \mathbb{R}\mathrm{P}^2$ obtained by twisting the sphere in a way that its equator is folded in half onto itself. 

More concretely, we first form the shape `8' with the equator and then fold its two halves (black and yellow in Fig.~\ref{double_cover_RP2}) onto each other. Identifying every antipodal pair of points, we obtain the surface (known as ``cross-cap'') displayed in the right panel of Fig.~\ref{double_cover_RP2}. Comparing the middle panel with the right panel, we see that every open subset $U$ of $\mathbb{R}\mathrm{P}^2$ is covered by two disjoint open subsets of $\mathbb{S}^2$ (the \textit{sheets} of the covering over $U$). It is remarkable that locally there are two disjoint sheets over any region $U$, while globally the covering sheets belong to a single connected sphere. Furthermore, any sheet of the covering can be mapped to the other sheet under the action of inversion on the sphere ($x\rightarrow -x$). We then readily find that every path within the sphere that connects two antipodal points is mapped to a non-contractible loop, say $\ell$, of $\mathbb{R}\mathrm{P}^2$ (e.g.~the black or yellow loop in $\mathbb{R}\mathrm{P}^2$). If we then compose the black and the yellow loops in $\mathbb{R}\mathrm{P}^2$, i.e.~$\ell\cdot \ell$, it can be lifted to a loop on the sphere (namely the entire equator) which in turn can be contracted to a point (as any other loop on the sphere). By continuity of the covering map we then find that $\ell \cdot \ell$ can also be contracted to a point in $\mathbb{R}\mathrm{P}^2$. We thus conclude that $\pi_1[\mathbb{R}\mathrm{P}^2] = \mathbb{Z}_2$ with $[\ell]$ as the generator. 

We finally address the second homotopy group of $\mathbb{R}\mathrm{P}^2$. Let us define an orientation at each point of the sphere through a normal vector pointing outwards. Focusing on the points of the black-yellow equator in the middle panel of Fig.~\ref{double_cover_RP2}, we readily see that the normal vectors of the two sheets (before to be identified by the double covering) point in the same direction. Therefore, the twofold oriented wrapping of $\mathbb{R}\mathrm{P}^2$ by the sphere is additive (i.e.~the normal vectors pointing in the same direction) and it is not contractible. Let us count this wrapping as `2'. We can instead design the double covering by twisting or folding the equator in the opposite direction, in which case we count the wrapping as `-2'. By doubling the wrapping of Fig.~\ref{double_cover_RP2} we obtained a fourfold wrapping of $\mathbb{R}\mathrm{P}^2$ which we count as `4', and so on. We have thus intuitively found the second homotopy group of the projective plane to be $\pi_2[\mathbb{R}\mathrm{P}^2] = 2\mathbb{Z}$ (note the group isomorphism $\bar{q}_*:\pi_2[\mathbb{S}^2] \rightarrow \pi_2[\mathbb{R}\mathrm{P}^2] :  \beta^+ \mapsto \beta=2\beta^+$, defined in terms of the covering map through $\bar{q}_*\beta^+ = \bar{q}_*[f^+] = [\bar{q}(f^+)] = [f] = \beta $).

\section{Orientability of bands and bundles}\label{sec:orientability}

\subsection{Orientable versus oriented bundles}

We associate to the vector subbundle $\mathcal{B}_I(p)$ certain orthonormal frame bundle.
Using $\mathcal{O}[R_I(\bs{k})]\subset \mathrm{P}_p(\mathbb{R}^N)$ to indicate the orbit of subframe $R_I(\bs{k})$ under the right transitive action of $G_I \in \mathsf{O}(p)$,
we define the \emph{associated frame subbundle} 
\begin{equation}
F_I(p) = \bigcup_{\boldsymbol{k}\in B} \mathcal{O}[R_{I}(\boldsymbol{k})].\label{eqn:Bloch-frame-bundle}
\end{equation}
Each fiber of $F_I(p)$ is isomorphic to the structure group $\mathsf{O}(p)$, thus making $F_I(p)$ a principal $\mathsf{O}(p)$-bundle~\cite{Hatcher_2,Milnor:1974}. 
An analogous construction can be carried for the unoccupied sector, defining the associated frame subbundle $F_{II}(N-p)$. Each fiber of $F_I(p)$ can moreover be equipped with the $\mathsf{SO}(p)$-invariant exterior product $\omega_p = u_1 \wedge \dots \wedge u_p$, resp.~$\omega_{N-p} = u_{p+1} \wedge \dots \wedge u_N$ for $F_{II}(N-p)$.

The associated frame subbundle allows us to introduce the notion of \emph{orientability}. 
Given a local trivialization $\phi:\pi^{-1}(U) \rightarrow U\times (\mathbb{R}^p \oplus \mathbb{R}^{N-p})$ of a total gapped bundle $\mathcal{E}_{p,N}$, the pushforwards 
$\phi_*\omega_p = \mathfrak{o}_I\vert_U e_1\wedge \dots \wedge e_p$
and $\phi_*\omega_{N-p} = \mathfrak{o}_{II}\vert_U e_{p+1}\wedge \dots \wedge e_N$, where $(e_1, \dots,e_N)$ are orthogonal coordinate vectors on $\mathbb{R}^N$, allow us to define $\mathfrak{o}_{I,II}\vert_U = \pm1$ called the \emph{local orientations} of the vector/frame subbundles. Considering a good open cover $\{U_i\to B\}$ of the base space with local trivializations $\phi_i$, every non-empty pairwise overlap $U_i \cap U_j \neq \varnothing$ is characterized by $\mathbb{Z}_2$-valued functions $t^{ij}_I$ and $t^{ij}_{II}$. 
More precisely, starting with an arbitrary subframe $R_I^{ij}(\bs{k})$, one defines \emph{transition functions} $t^{ij}_I = (\mathfrak{o}_I\vert_{U_i})(\mathfrak{o}_I\vert_{U_j})=\pm 1$, and similarly for the unocccupied bands.

Change of a local trivialization $\phi_i$ or frames $R^{ij}_{I(II)}$ may lead to a reversal of $t^{ij}_{I(II)}$. 
We say that a vector subbundle $\mathcal{B}_{I(II)}$ is  \textit{unorientable} if for all trivializations $\phi_i$ (and for all choices of $R_{I(II)}^{ij}$) there are some transition functions $t^{ij}_{I(II)}\neq +1$.
We call the total gapped bundle $\mathcal{E}_{p,N}$ unorientable if either the occupied or unoccupied vector subbundle is unorientable. The classifying spaces of the corresponding gapped band structure is the unoriented Grassmannian $\mathrm{Gr}_{p,N}$. In contrast, when local trivializations can be found such that simultanously all transition functions are equal to $+1$ the vector subbundle is called \textit{orientable}. The total gapped bundle is called orientable if both the occupied and the unoccupied vector (frame) subbundles are orientable. 
In the case of a trivial total bundle  
$\mathcal{E}_{N,N}$, as it is assumed in this work, the subbundles $\mathcal{B}_I(p)$ and $\mathcal{B}_{II}(N-p)$ are either both orientable or both unorientable. 

Fixing the orientation of the subframe over the whole base space in a consistent manner, we obtain an \textit{oriented} vector subbundle, written $\mathcal{B}^+_{I}(p)$ ($\mathcal{B}^+_{II}(N-p)$). Taking the two oriented subbundles together, we form the oriented total gapped bundle $\mathcal{E}^+_{p,N} = \mathcal{B}^+_{I}(p) \cup \mathcal{B}^+_{II}(N-p)$ that has the oriented Grassmannian $\mathrm{Gr}^+_{p,N}$ as its classifying space.  

Importantly, the classifying space of an orientable gapped bundle is the \emph{unoriented} Grassmannian and not the oriented one. Indeed, the choice of an orientation for both subframes is a gauge freedom of gapped Hamiltonians, while it is not a gauge symmetry for the elements of the oriented Grassmannian [Sec. \ref{subsec:un_oriented_grass}]. Nevertheless~\cite{Hatcher_2} a total gapped bundle $\mathcal{E}_{p,N}$ is orientable iff its classifying map\footnote{This is the map that assigns to a point $\boldsymbol{k}\in B$ with Hamiltonian $H(\boldsymbol{k})$ the coset $ [R(\boldsymbol{k})] \in \mathrm{Gr}_{p,N}$, see Sec.~\ref{geo} for more detail on such maps.} $f : B\rightarrow \mathrm{Gr}_{p,N}$ can be lifted to a classifying map $f^+ : B\rightarrow \mathrm{Gr}^+_{p,N}$, i.e.~the map that assigns to each $\boldsymbol{k}$ the subframe-orientation-preserving coset $[R(\boldsymbol{k})]^+\in \mathrm{Gr}^+_{p,N}$.\footnote{We note that all bundles on $B=\mathbb{S}^2$ are orientable and thus can be lifted. } 

The lift induces the choice of an orientation of both subframes over the whole base space, and this can be made continuously (i.e.~consistently over the whole base space) by virtue of the assumed orientability of the total gapped bundle. More specifically, there is a gauge freedom in the choice of an orientation for the subframes at an initial point, say $\boldsymbol{k}_0$, where, for a given matrix $R(\boldsymbol{k}_0)$, we can lift the image $f(\boldsymbol{k}_0)=[R(\boldsymbol{k}_0)]$ in $\mathrm{Gr}_{p,N}$ either to $f^+_a(\boldsymbol{k}_0)=[R(\boldsymbol{k}_0)]^+$ or to $f^+_b(\boldsymbol{k}_0)=[R(\boldsymbol{k}_0)^\mathrm{sr}]^+$ in $\mathrm{Gr}^+_{p,N}$. Once this initial choice is made, the orientation over the rest of the base space is enforced by continuity, thus unfolding the whole lifted map $f^+$ (see in Appendix \ref{ap_0} the explicit example of a hedgehog structure emerging for the case of $\mathbb{R}\mathrm{P}^2$). We note that the fixing of an orientation is equivalent to the fixing of gauge as discussed in Ref.~\cite{bouhon2019nonabelian}.

We thus conclude that an orientable occupied (unoccupied) vector subbundle $\mathcal{B}_{I(II)}$, characterized by a classifying map $f : B \rightarrow \mathrm{Gr}_{p,N}$, can be equipped with a subframe orientation through the lifted map $f^+:B \rightarrow \mathrm{Gr}^+_{p,N} $, which in turn characterizes an oriented vector subbundle $\mathcal{B}^+_{I(II)}$ and, taken together, an oriented total gapped bundle $\mathcal{E}^+_{p,N}$. It is important to keep in mind though the arbitrariness when assigning a subframe orientation to an orientable vector subbundle. We indeed show in Sec.~\ref{sec:homotopy-groups} that the orientation must be dropped for the topological classification of band structures, as there exists in some cases adiabatic transformations between distinct oriented homotopy classes.

Our strategy to unfold the topological classification of band structures and to derive their representative tight-binding models, which is the content of the following sections, is to first represent the topological phases as oriented gapped bundles classified by $\mathrm{Gr}^+_{p,N}$. Then we address the effect of forgetting the orientation, i.e.~projecting the classifying space from $\mathrm{Gr}^+_{p,N}$ to $\mathrm{Gr}_{p,N}$, which is that two distinct oriented bundles can be continuously deformed into a single orientable bundle. This culminates with the explicit derivation of three-band and four-band tight-binding models in Sec.~\ref{3_bands} and \ref{4_bands} from which all homotopy classes can be represented. From now on we simplify the terminology, whereas a band sector characterized by an orientable (oriented) vector subbundle would be called \emph{orientable} (oriented) bands.

\subsection{Conceptual clarifications}\label{clarifications}

We importantly remark that total Bloch bundle $\mathcal{E}_{N,N}$ as defined in Sec.~\ref{sec:band-structures} is not necessarily trivial. A simple example of such a non-trivial case is provided by the two-band 2D Mielke model discussed in Ref.~\cite{Montambaux_2018} which exhibits total $\pi$-Berry phases in both directions of the Brillouin zone torus \cite{BJY_nielsen}, making its two-band total Bloch bundle non-orientable (see the definition of the first Stiefel-Whitney class in Sec.~\ref{sec:homotopy-groups}). This can be understood as an effect of the body-centered lattice structure of the model (this will be elaborated elsewhere). Nevertheless, a theorem in vector bundle theory asserts that any vector subbundle $\mathcal{B}_I(p)$ can be trivialized through the direct sum with an appropriate vector subbundle $\overline{\mathcal{B}_I}(N'-p)$, i.e.~$\mathcal{B}_I(p)\oplus \overline{\mathcal{B}_I}(N'-p)  \cong B\times \mathbb{R}^{N'}$. This especially also applies to a non-trivial total Bloch bundle $\mathcal{E}_{N,N}$, for which there exists a vector bundle $\overline{\mathcal{E}}(N'-N)$ such that $\mathcal{E}_{N,N} \oplus \overline{\mathcal{E}}(N'-N)\equiv \mathcal{E}'_{N',N'}\cong B \times \mathbb{R}^{N'}$. Then $\mathcal{E}'_{N',N'}$ can be interpreted as a total \emph{trivial} Bloch bundle, of which the original $\mathcal{E}_{N,N}$ and the trivializing $\overline{\mathcal{E}}(N'-N)$ are two complementary subbundles. 
For instance, for the two-band Mielke model \cite{Montambaux_2018}, the triviality of the total bundle is achieved for a four-band model obtained through the direct sum of two Mielke models.

We now comment on the relevance of the concept of vector bundle for band structures. We defined $F_{I}(p)$ in Eq.~(\ref{eqn:Bloch-frame-bundle}) by gluing together the orbits $\mathcal{O}[R_I(\bs{k})]$ of the $p$-frame spanned by the occupied eigenvectors under the action of the gauge group $\mathsf{O}(p)$. One could instead consider the finer notion of an \emph{eigenbundle}~\cite{gottlieb_2003}, which corresponds to gluing together the \emph{ordered collection of eigenvectors}, rather than their orbit.
While local trivializability belongs to the axioms of fiber 
bundles, the eigenbundle may not have this property.
This notably happens when the eigenvalues form a topologically stable crossing, i.e.~the nodal points discussed in Sec.~\ref{nodal_points}, in which case the eigenvectors cannot be expressed in a locally continuous gauge~\cite{bouhon2019nonabelian}. The discontinuities of the gauge for eigenstates is often modelled by introducing \emph{Dirac strings} that terminate at the nodes~\cite{BJY_nielsen}.
One thus finds that the eigenbundle is not locally trivializable at the band nodes, i.e.~it does not meet the axioms of a fiber bundle when the base space $B$ contains a band node.

In contrast, the frame subbundle $F_{I}(p)$ remains trivializable even at band nodes. More concretely a smooth section of $F_{I}(p)$ can be formed at a band node by forming linear combinations of the $p$ eigenvectors, i.e.~$[v_{n'}(\boldsymbol{k})]_l = \sum_{n=1,\dots,p}  [u_n(\boldsymbol{k})]_{l}  g_{nn'}(\boldsymbol{k})$, with $  g_{nn'}(\boldsymbol{k}) = [G_I]_{nn'}(\boldsymbol{k})$ and $G_I\in \mathsf{SO}(p)$ (here $[u_n]_{l}$ is the $l$-th component of the vector $u_n$). 
Clearly, the vectors $v_{n'}(\boldsymbol{k})$ need not be eigenvectors in general. Since a section of a $p$-frame bundle is essentially an ordered collection of $p$ pointwise orthonormal vector bundles, the vector subbundle $\mathcal{B}_I(p)$ is also locally trivializable. Therefore, in contrast to eigenbundles, the higher flexibility of the vector and frame subbundles permits the local trivialization, as has been scrupulously analyzed e.g.~in the Supplementary Material of Ref.~\cite{bouhon2019nonabelian}. 

It also follows that the occupied subbundle of a topological semimetal does not form a vector bundle over the whole Brillouin zone, while it does so over any closed manifold that avoids the semimetallic nodes.\footnote{These considerations are reflected in the fact that while the eigenstate of a band with nodes is never orientable (see Sec.~\ref{subsec:first}), the eigenbundle of a band subspace with nodes that is disconnected from the other bands (by a band gap) is orientable whenever it corresponds to an orientable vector subbundle (see Sec.~\ref{subsec:second}).}

\section{Homotopy classification and homotopy invariants}\label{sec:homotopy-groups}

\subsection{General description}
The topological classification of gapped band structures is given by the set of all allowed isomorphism classes $[\mathcal{E}_{p,N}]$ of 
total gapped bundles.  
The later is isomorphic to the set of free homotopy classes of continuous maps from the base space (the Brillouin zone $B= \mathbb{T}^2$) to the classifying space of gapped band structures. We denote the set of such homotopy classes as $[\mathbb{T}^2,\mathrm{Gr}_{p,N}]$. These can be expressed~\cite{Simon_homotopy,Kitaev,Kennedy_homotopy,Wojcik:2019} as 
\begin{equation}
\label{eq:homotopy_classes}
[\mathbb{T}^2,\mathrm{Gr}_{p,N}] = \bigcup_{\alpha_x,\alpha_y}[\mathbb{I}^2,\mathrm{Gr}_{p,N}]^{(\alpha_x,\alpha_y)}.
\end{equation}
Here, the weak invariants $\alpha_{x(y)}\in\pi_1[\textrm{Gr}_{p,N}]$ characterize the total gapped bundle $\mathcal{E}_{p,N}=\mathcal{B}_I(p)\cup \mathcal{B}_{II}(N-p)$ along the two non-contractible loops $l_x$ (resp.~$l_y$) of $\mathbb{T}^2$, as discussed in Sec.~\ref{subsec:first}. 
Further, $[\mathbb{I}^2,\cdot]^{(\alpha_x,\alpha_y)}$ is the set of free homotopy classes of maps from a square $\mathbb{I}^2$ (the inside of the BZ) to the space ``$\cdot$'' which are compatible with the weak invariants $(\alpha_x,\alpha_y)$ on the BZ boundary $\partial \mathbb{I}^2$. These homotopy classes are studied in detail in Sec.~\ref{subsec:second}. The decomposition in Eq.~(\ref{eq:homotopy_classes}) mimics the CW-complex decomposition of $\mathbb{T}^2$, namely the wedge sum of the two non-contractible loops $l_x \vee l_y$ together with a two-dimensional sheet $\mathbb{I}^2$ with its boundary glued along the loops \cite{Hatcher_1}.

When the total number of bands $N$ is large enough, the homotopy groups of the classifying space do not depend on $N$. This is called the \emph{stable limit}. In contrast, the homotopy groups for few-band models may depend on $N$, in which case we speak of \emph{unstable topology}.  
Note that in our definition of stability of topological invariants, and contrary to works based on $K$-theory, we keep the number $p$ of occupied bands fixed. In this section we discuss the stable results, while an in-depth analysis of the unstable topology of 3-band and 4-band systems is presented in Secs.~\ref{3_bands} and~\ref{4_bands}.

\subsection{Topology in one dimension}\label{subsec:first}
The stable limit for the first homotopy group is reached for $N \geq 3$, when $\pi_1[\mathrm{Gr}_{p,N}] = \mathbb{Z}_2$.
The element $\alpha_l$ in the first homotopy group for a  non-contractible \textit{base loop} $l\in B$ coincides with the first Stiefel-Whitney (SW) class $w_1[\mathcal{B}_I(p)\vert_l]\in H^1(\mathbb{S}^1,\mathbb{Z}_2)$~\cite{BJY_linking} (i.e.~the characteristic class of the bundle that is captured by the first cohomology group of $l\simeq \mathbb{S}^1$ with $\mathbb{Z}_2$ coefficients), 
which is known to capture the orientability of the vector subbundle $\mathcal{B}_{I}(p)$ restricted to $l$~\cite{Hatcher_2,Milnor:1974}. An example of a non-orientable bundle is a line bundle (i.e.~rank-1 eigenbundle) over a loop encircling a nodal point~\cite{BJY_linking,bouhon2018wilson}. Considering now the occupied vector subbundle $\mathcal{B}_I(p)$ inside the full Brillouin zone, one can independently study the first SW class on the two non-contractible paths of the torus, which define an element $(\alpha_x,\alpha_y) \equiv w_1[\mathcal{B}_I(p)]\in H^1(\mathbb{T}^2,\mathbb{Z}_2) = \mathbb{Z}_2\oplus \mathbb{Z}_2 $. Accordingly, the vector subbundle $\mathcal{B}_I(p)$ and the band subspace it represents are orientable iff $\alpha_x = \alpha_y  = 0 $. The first SW class can be computed through the Berry phase factor $\alpha_l = e^{\imi \gamma_I [l]} = \det \mathcal{W}_I[l] \in \{+1,-1\}$ along a loop $l$, where the $\mathsf{O}(p)$ Wilson loop $\mathcal{W}_I$ is obtained from the $p$ eigenvectors spanning $\mathcal{B}_{I}$. 

The first SW class resp.~the Berry phase can also be computed for the unoccupied vector subbundle. From the assumed triviality of $\mathcal{E}_{N,N}$ and from the Whitney sum formula for the cup product of cohomology classes, the first SW class satisfies the sum rule $ 0=w_1[\mathcal{B}_I(p) \oplus \mathcal{B}_{II}(N-p)] = (w_1[\mathcal{B}_I(p)] + w_1[\mathcal{B}_{II}(N-p)]) \mod 2 $~\cite{BJY_linking}, so that 
\begin{equation}
    w_1[\mathcal{B}_I(p)] = w_1[\mathcal{B}_{II}(N-p)] \;,
\end{equation}
and similarly for the Berry phase, i.e.~$(\gamma_I [l]  = \gamma_{II} [l]) \mod 2\pi$ for both non-contractible loops of the Brillouin zone torus. This relation clarifies our statement below Eq.~(\ref{eq:homotopy_classes}) that $\alpha_{x(y)}$ characterize the topology of the total gapped bundle (rather than of just the occupied or unoccupied vector subbundle) -- at least along the non-conctractible loops $l_{x(y)}$. A similar relation is found in Sec.~\ref{subsec:2D-cohomology-classes} also for the topological classification over the two-dimensional Brillouin zone square.

\subsection{Topology in two dimensions}\label{subsec:second}

Ascending now one dimension higher, Eq.~(\ref{eq:homotopy_classes}) suggests that the classification of vector subbundles $\mathcal{B}_I(p)$, and $\mathcal{B}_{II}(N-p)$ depends on the weak indices $(\alpha_x,\alpha_y)$. 
In the nonorientable case, the free homotopy set on the Brillouin zone square is 
\begin{equation}
[\mathbb{I}^2\!,\mathrm{Gr}_{p,N}]^{(1,0)} \!=\! [\mathbb{I}^2\!,\mathrm{Gr}_{p,N}]^{(0,1)} \!=\! [\mathbb{I}^2\!,\mathrm{Gr}_{p,N}]^{(1,1)} \!=\! \mathbb{Z}_2
\end{equation}
where the $\mathbb{Z}_2$ invariant corresponds to the second SW class~\cite{BJY_nielsen}, $w_2[\mathcal{B}_I(p)]\in H^2(\mathbb{T}^2,\mathbb{Z}_2)=\mathbb{Z}_2$, which we discuss in more detail in Sec.~\ref{subsec:2D-cohomology-classes}.

In the following, we focus on the more interesting \emph{orientable} case, i.e.~when $\alpha_x=\alpha_y=0$, such that the Berry phases are zero along both non-contractible loops of the Brillouin zone. Then the topological classification is given by the free homotopy set $[\mathbb{I}^2,\mathrm{Gr}_{p,N}]^{(0,0)}$. The triviality of the weak invariants implies that the mapping to the classifying space can be deformed into a constant on the boundary $\partial \mathbb{I}^2$ of the Brillouin zone square. This allows us to identify the boundary as a single point, resulting in $\mathbb{I}^2/\partial \mathbb{I}^2 \simeq \mathbb{S}^2$, i.e.~a sphere. Therefore,
\begin{equation}
    [\mathbb{I}^2,\cdot]^{(0,0)}=[\mathbb{S}^2,\cdot]\;
\end{equation}
which differs from the second homotopy group $\pi_2[.]$ only by the absence of a base point. 

It is worth reminding that the base point of the homotopy group $\pi_2[\mathrm{Gr}_{p,N}]$ is meant to be constant over all homotopy classes. This together with an implicit orientation of the sphere image $f(\mathbb{I}^2)$ for $[f]\in \pi_2[\cdot]$ (see Appendix \ref{ap:homotopy_action}) equips the composition of homotopies with a group structure. However, by removing the constraint on the base point, the \emph{free} homotopy set may loose the group structure. More precisely, the unoriented Grassmannian $\textrm{Gr}_{p,N}$ contains non-contractible loops $\ell$ (i.e.~with $[\ell]$ the generator of $\pi_1[\textrm{Gr}_{p,N\geq 3}] = \mathbb{Z}_2$), and evolving the base point along this loop induces an automorphism $\triangleright_\ell :\pi_2[\textrm{Gr}_{p,N}]\to \pi_2[\textrm{Gr}_{p,N}]$ on the based homotopy group~ \cite{Sun:2019,Wojcik:2019}. The latter is called the action of the element $[\ell]\in\pi_{1}[\mathrm{Gr}_{p,N}]$ on an element $\beta = [f]  \in\pi_{2}[\mathrm{Gr}_{p,N}]$ and is induced by a homotopy of the map $f$ that traces out $\ell$ when restricted to the base point of $f$, see Appendix~\ref{ap:homotopy_action} for a precise definition. The automorphism acts as $\triangleright_\ell:\beta\mapsto \beta^{-1}$ on elements $\beta\in \pi_2[\textrm{Gr}_{p,N}]$. Since $\beta = \beta^{-1}$ for $p\neq2$ though, it acts non-trivially only for rank-2 bundles. 

Knowing that second (and higher) homotopy groups are Abelian, one can always represent them as a direct sum of several $\mathbb{Z}$ and $\mathbb{Z}_n$'s, and indicate the composition with ``$+$'', i.e.~as addition, such that $\beta^{-1}=-\beta$. The automorphism $\triangleright_\ell$ then reduces the second homotopy group into orbits $\{\beta,-\beta\}$, and relaxing the condition on the base point (i.e.~the reduction from based to free homotopy classes) corresponds to replacing $\pi_2[\textrm{Gr}_{p,N}]$ by the set of orbits. This allows us to express the free homotopy classes concisely as
\begin{equation}
\label{eq:free_homotopy}
   [\mathbb{I}^2\!,\mathrm{Gr}_{p,N}]^{(0,0)} 
   \!=\! [\mathbb{S}^2\!,\mathrm{Gr}_{p,N}] 
   \!=\! \pi_2[\mathrm{Gr}_{p,N}]/\{+1,-1\}\;\!.
\end{equation}
However, the last equation reduces simply to $[\mathbb{I}^2,\mathrm{Gr}_{p,N}]^{(0,0)} = \pi_2[\mathrm{Gr}_{p,N}]$ for $p\neq 2$. 

It is worth noting that the non-contractible loop $\ell\subset \mathrm{Gr}_{p,N}$ that appears in the construction is \emph{not} homotopy equivalent to the image of any of the non-contractible loops of the Brillouin zone torus. Rather, the motion of the base point along $\ell$ can be understood as an adiabatic deformation of the Hamiltonian $H(\boldsymbol{k})$ (i.e.~an element of the free homotopy set), while it causes a change of homotopy classes of the associated based map $f_H:(\mathbb{I}^2,\partial \mathbb{I}^2)\rightarrow (\mathrm{Gr}_{p=2,N},[R_0]):\boldsymbol{k}\mapsto H(\boldsymbol{k}) \mapsto [R(\boldsymbol{k})]$ with a fixed base point $[R_0] =[R(\boldsymbol{k}_0)]= f_H(\partial \mathbb{I}^2) $ (i.e.~$[f_H]$ is an element of the based homotopy group $\pi_2[\mathrm{Gr}_{2,N}]$).\footnote{This is most naturally shown in terms of the map lifted to the oriented Grassmannian, i.e.~$f^+_H:(\mathbb{I}^2,\partial \mathbb{I}^2)\rightarrow (\mathrm{Gr}^+_{p=2,N},[R_0]^+)$, see Appendix~\ref{ap:homotopy_grassmannian} and \ref{ap:homotopy_action}.} 

Crucially, we note that $\pi_2[\mathrm{Gr}^+_{p,N}] = \pi_2[\mathrm{Gr}_{p,N}]$ because the sphere is simply connected and because $\mathrm{Gr}^+_{p,N} \rightarrow \mathrm{Gr}_{p,N}$ is a double cover, see Appendix~\ref{ap:homotopy_grassmannian}. We then show in Appendix~\ref{ap:homotopy_action} that the action of $[\ell]$ on an element $[f^+]\in\pi_2[\mathrm{Gr}^+_{p,N}]$ must involve the subframe-orientation reversal, cf.~Eq.~(\ref{eq:orientation_reversal_sr}). Therefore, orientable vector subbundles can be classified in terms of \emph{oriented} subbundles modulo the forgetting of orientation and the discarding of the base point, which, in the case $p=2$, leads to the two-to-one redundancy $\beta \sim -\beta$ for all $\beta\in \pi_2[\mathrm{Gr}^+_{p=2,N}]$. 
Since maps to the oriented Grassmannians are easier to analyze, in the later sections of the manuscript dedicated to the systematic construction of tight-binding models from homotopy, we start with the construction of the \textit{oriented} bundles and then address the effect of forgetting orientation on the homotopy classification of band structures.

More precisely, whenever we are given a \emph{concrete collection of eigenvectors} of a band subspace (rather than just the unoriented vector space they span), the bundle has been equippied with a specific choice of orientation, and as such it can be classified by a unique element $\beta \in \pi_2[\textrm{Gr}^+_{p,N}]$. Then, by dropping the (arbitrary) choice of the eigenvector gauge, the bundle becomes indistinguishable from a bundle with the opposite orientation. This implies that the element $\beta$ becomes indistinguishable from the element $-\beta$ and the system is classified by a unique element $\vert\beta\vert\in [\mathbb{S}^2,\mathrm{Gr}_{p,N}]$. In other words, there exists an adiabatic deformation of the Hamiltonian (nontrivial for $p=2$) which connects the elements $\beta$ and $-\beta$. We give this transformation explicitly in Appendices~\ref{ap_1} and~\ref{ap_3} respectively for the three-band and four-band tight-binding models that are presented in Secs.~\ref{3_bands} and~\ref{4_bands}.

Below, whenever we say that we deal with an \emph{explicit model}, we mean an oriented bundle defined by a single valued function $R(\boldsymbol{k})\in \mathsf{SO(N)}$ for all $\boldsymbol{k}$. In contrast, when we discuss the (free) \emph{homotopy class representative}, we mean an orientable bundle, that is an equivalence class of two explicit models with opposite orientations.

\subsection{Euler class and second Stiefel-Whitney class}\label{subsec:2D-cohomology-classes}

The relevant second homotopy groups for oriented classifying spaces are listed in Table~\ref{Tab::Flaghoms}~\cite{BzduSigristRobust}. The stable limit of the second homotopy group is given by $N -p \geq 3$, for which we have 
\begin{eqnarray}
\pi_2[\mathrm{Gr}^+_{1,N\geq 4}] &=& \mathbb{0}, \nonumber \\ \pi_2[\mathrm{Gr}^+_{2,N\geq 5}] &=& \mathbb{Z}, \\ \textrm{ and}\quad \pi_2[\mathrm{Gr}^+_{p\geq3 , N\geq p+3}] &=& \mathbb{Z}_2. \nonumber
\end{eqnarray}
Notably, the second homotopy invariant characterizing an \textit{oriented} \emph{two-band} vector subbundle $\mathcal{B}^+(p=2)$ in the stable limit corresponds to the Euler class~\cite{Hatcher_2}, $\chi[\mathcal{B}^+_I(p=2)] \in H^2(\mathbb{T}^2,\mathbb{Z}) =  \mathbb{Z}$. 
The Euler class is computed as the integral of the Pfaffian of the two-band Berry-Wilczek-Zee curvature~\cite{Zhao_PT,Ahn2019, bouhon2019nonabelian} over the Brilouin zone. It can also be conveniently computed as a two-band Wilson loop winding \cite{BJY_linking,bouhon2018wilson}. The reversal of subframe orientation [Eq.~(\ref{eq:orientation_reversal_sr})] exchanges the sign of Euler class (see Methods of \cite{bouhon2019nonabelian}), in other words, the chosen orientation of an oriented two-band subbundle is faithfully indicated by the Euler class. (This plays an important role in the derivation of Eq.~(\ref{eq:free_homotopy}) in Appendix~\ref{ap:homotopy_action}.)

In contrast, when the oriented vector subundle under consideration consists of three or more bands, the second homotopy invariant in the stable limit corresponds to the second SW class $w_2[\mathcal{B}^+_I(p\geq 3)] \in H^2(\mathbb{T}^2,\mathbb{Z}_2) = \mathbb{Z}_2$. The second SW class can be conveniently computed as the parity of the number of $\pi$ crossings in the Wilson loop flow~\cite{BJY_nielsen}. Contrary to the Euler class of two-band subbundles, the second SW class is insensitive to the reversal of subframe orientation. Finally, one-band subspaces, i.e.~associated to a real orientable line subbundle, are always \textit{stably trivial}. (We discuss in Sec.~\ref{subsec:triviality_NS} one example of \textit{unstable nontrivial} line bundle.)

Because of the assumed triviality of $\mathcal{E}_{N,N}$, the second SW class satisfies the sum rule $ 0 = w_2[\mathcal{B}^+_I(p) \oplus \mathcal{B}^+_{II}(N-p)] = (w_2[\mathcal{B}^+_I(p)] + w_2[\mathcal{B}^+_{II}(N-p)]) \mod 2$, where we have used the fact that the first SW class is zero for oriented vector bundles. Therefore 
\begin{equation}
    w_2[\mathcal{B}^+_I(p)] = w_2[\mathcal{B}^+_{II}(N-p)] ,\;\label{eqn:Whitney-sum-w2}
\end{equation}
implying that the same element of $H^2(\mathbb{T}^2,\mathbb{Z})$ characterizes both the occupied and the unoccupied vector subbundle, i.e.~it entirely characterizes the total oriented gapped bundle $\mathcal{E}^+_{p,N}$.
For a rank-2 oriented vector subbundle, the second SW class is given as the parity of the Euler class~\cite{BJY_linking},
\begin{equation}
    w_2[\mathcal{B}^+_I(2)] = \chi[\mathcal{B}^+_I(2)] \mod 2\;,
\end{equation}
which implies that the Euler class must also satisfy the sum rule in Eq.~(\ref{eqn:Whitney-sum-w2}) $\mod 2$, i.e. 
\begin{equation}
    \chi[\mathcal{B}^+_I(2)] \mod 2= w_2[\mathcal{B}^+_{II}(N-2)] \label{eqn:Euler-2SW-sum} \;.
\end{equation}
Since the Euler class contains more information than the $\mod 2$ reduction, Eq.~(\ref{eqn:Euler-2SW-sum}) implies that it entirely characterizes the oriented total gapped bundle $\mathcal{E}^+_{2,N}$.

We finally consider the reduction, up to a sign, when dropping the explicit choice of orientation. We find that the topology of orientable gapped band structures is classified by the following stable free homotopy sets
\begin{equation}
\begin{aligned}
    [\mathbb{S}^2,\mathrm{Gr}_{1,N\geq 4}] &= 0 \;,\\
    [\mathbb{S}^2,\mathrm{Gr}_{2,N \geq 5}] &= \mathbb{N} \;,\\
    [\mathbb{S}^2,\mathrm{Gr}_{p\geq 3,N\geq p+3}] &= \mathbb{Z}_2\;, 
\end{aligned}
\end{equation}
where for orientable two-band subspaces we define the \textit{reduced Euler class} $\overline{\chi} $, obtained through the reduction modulo sign of the Euler class of the associated oriented subbundle, i.e.
\begin{equation}
\label{eq:reduced_Euler}
    \overline{\chi} [\mathcal{B}(2)]  = \vert \chi[\mathcal{B}^+(2)] \vert \;. 
\end{equation}
The orientable subspaces with more bands are characterized by the second SW class which, contrary to the Euler class, does not require a definite orientation, 
\begin{equation}
    w_2[\mathcal{B}(p\geq 3)] = w_2[\mathcal{B}^+(p\geq 3)] \in \mathbb{Z}_2 \;.
\end{equation}

\begin{table}[h]
\begin{equation*}
\begin{array}{ c | l | c }
	N = p_1+p_2+\dots & \mathrm{Fl}^+_{p_1,p_2,\dots} & \pi_2  \\
	\hline
	\hline
	2 & \mathrm{Fl}^+_{1,1} = \mathrm{Gr}^+_{1,2} = \mathbb{S}^1 & \mathbb{0} \\
	\hline 
	3& \mathrm{Fl}^+_{2,1} = \mathrm{Gr}^+_{2,3} = \mathbb{S}^2 &  2\mathbb{Z} \\
	& \mathrm{Fl}^+_{1,1,1} & \mathbb{0} \\
	\hline
	4& \mathrm{Fl}^+_{3,1} = \mathrm{Gr}^+_{3,4} = \mathbb{S}^3 & \mathbb{0} \\
	& \mathrm{Fl}^+_{2,2} = \mathrm{Gr}^+_{2,4} = \mathbb{S}^2\times \mathbb{S}^2 &  \mathbb{Z}\oplus\mathbb{Z} \\
	& \mathrm{Fl}^+_{2,1,1} & 2\mathbb{Z} \\
	& \mathrm{Fl}^+_{1,1,1,1} & \mathbb{0} \\
	\hline
	(m\geq 3) & & \\
	1+m &  \mathrm{Fl}^+_{1,m} = \mathrm{Gr}^+_{1,1+m} = \mathbb{S}^m &  \mathbb{0} \\
	2+m &  \mathrm{Fl}^+_{2,m} = \mathrm{Gr}^+_{2,2+m} &  \mathbb{Z} \\
	3+m &  \mathrm{Fl}^+_{3,m} = \mathrm{Gr}^+_{3,3+m}  &  \mathbb{Z}_2 \\
\end{array}
\end{equation*}
\caption{Classification of \textit{oriented} band structures, i.e.~over the simply-connected base space $B = \mathbb{S}^2$ representing the Brillouin zone torus in the absence of Berry phases. Table indicates the second homotopy groups, $\pi_2$, of oriented Grassmannian and flag varieties as discussed in the text. The factor 2 in $2\mathbb{Z}$ is a convention in order to match with the computed value of the Euler class, see Sec.~\ref{3_bands}. By $\pi_2[\mathrm{Gr}_{p,N}] = \pi_2[\mathrm{Gr}^+_{p,N}]$ and Eq.~(\ref{eq:free_homotopy}), the topologically inequivalent \textit{orientable} phases are classified by the reduction, up to a sign, of the second homotopy group.}
\label{Tab::Flaghoms}
\end{table}

\section{Refined band partitioning}\label{sec:partitioning}

\subsection{Multiple gap conditions}\label{sec:flag-1}
The single gap condition is naturally generalized to \emph{multiple} gap conditions when several blocks of bands are separated from each other by energy gaps both from above and from below everywhere in the Brillouin zone $B$, cf.~Fig.~\ref{fragile}. We use $\mathfrak{N}$ to indicate the total number of band subspaces, and we write the subbundle of the $i$-th band subspace ($i=I,II,III,\dots,\mathfrak{N}$) as $\mathcal{B}_i(p_i)$ with $p_i$ its number of bands, and $N=\sum_{i=I}^\mathfrak{N} p_i$ the total number of bands. The total gapped bundle can be expressed as 
\begin{equation}
\mathcal{E}_{p_I,\ldots,p_\mathfrak{N};N}=\mathcal{B}_I(p_I)\cup\ldots \cup \mathcal{B}_I(p_\mathfrak{N})
\end{equation}
where the ordering of the subspaces follows the increasing band energy. Similar to Sec.~\ref{sec:homotopy-groups}, in the present section we assume the stable limit, i.e.~$N-p_{\text{min}}\geq 3$ with $p_{\text{min}}=\min_{i} p_i$. 

Formally, the classifying space of a Hamiltonian with multiple gap conditions generalizes the Grasmannian to a \emph{flag variety}
\begin{equation}
\mathrm{Fl}_{p_I,p_{II},...,p_{\mathfrak{N}}} = \mathsf{O}(N)/[\mathsf{O}(p_I)\times \mathsf{O}(p_{II})\times \cdots \mathsf{O}(p_\mathfrak{N})]\label{eqn:flag-variety-def}
\end{equation}
where the quotient corresponds to the gauge structure obtained by flattening every block of bands separately. The work of Ref.~\cite{Wu1273} revealed non-Abelian band topology of nodal lines in $PT$-symmetric systems by considering the \emph{complete} flag variety $\mathsf{O}(N)/\mathsf{O}(1)^{\times N} = \textrm{Fl}_{1,1,\ldots,1}$, while ideas interpretable in terms of a \emph{partial} flag $\textrm{Fl}_{p-1,2,N-p-1}$ were employed by the work of Ref.~\cite{bouhon2019nonabelian} to analyze the topological properties of principal band nodes in $C_2T$-symmetric models. One can also construct an \emph{oriented} flag variety $\textrm{Fl}^+$ by replacing $\mathsf{O}\mapsto\mathsf{SO}$ in Eq.~(\ref{eqn:flag-variety-def}) for both the total space and the quotients.

\subsection{Homotopy classes of flag varieties}\label{sec:flag-2}

The first homotopy group of the flag variety in Eq.~(\ref{eqn:flag-variety-def}) is easily shown\footnote{In contrast, computing the homotopy classes $[\mathbb{T}^d ,\textrm{Fl}_{p_I,p_{II},\dots,p_{\mathfrak{N}}}]$ is a nontrivial problem. Nevertheless, by restricting to two-dimensional and orientable systems, the topologies of any band structure can be inferred from the second homotopy groups of Grassmannians that are discussed in Sec.~\ref{sec:homotopy-groups}.} to be $\pi_1[\mathrm{Fl}_{p_I,...,p_{\mathfrak{N}}}] = \mathbb{Z}_2^{\mathfrak{N}-1}$. This result is interpretable in terms of the quantized Berry phases of each subbundle (i.e.~by their first SW classes) on a closed path $l$, subject to the contraint $\sum_{i=I}^\mathfrak{N} \gamma_i[l] = 0\;\,(\!\mod 2\pi)\,$ that follows from the Whitney sum formula and from the triviality of the total Bloch bundle. In analogy with the single gap case discussed in Sec.~\ref{subsec:second}, the generators of the first homotopy group are associated with certain paths $\{\ell_i\}_{i=I}^{\mathfrak{N}-1}$ in $\textrm{Fl}_{p_I,...,p_{\mathfrak{N}}}$, such that adiabatically evolving the Hamiltonian along $\ell_i$ reverses the local orientation of subbundles $\mathcal{B}_i(p_i)$ and $\mathcal{B}_{i+1}(p_{i+1})$~\cite{Wojcik:2019,bouhon2019nonabelian}. 

We further consider the topological classification of total multi-gapped bundles in two dimensions. We explicitly consider only the case when all $\mathcal{B}_i(p_i)$ are orientable. For simplicity, we first assume that each subbundle is equipped with an explicit orientation, becoming $\mathcal{B}^+_i(p_i)$, and we implement the effect of dropping the orientations in a second step. Under these assumptions, the discussion in Sec.~\ref{subsec:first} implies that the first homotopy groups play no role (as $\pi_1[\mathrm{Gr}^+_{p_i,N}] = 0$), and according to Sec.~\ref{subsec:second} the homotopy classification of each oriented subbundle is captured by the stable second homotopy group $\pi_2[\textrm{Gr}^+_{p_i,N\geq p_{\text{min}}+3}]$, 
which depends on $p_i$. It follows that 
$(i)$ one-band oriented subspaces ($p_i=1$) have a trivial topology, $(ii)$ two-band oriented subspaces ($p_i=2$) have a $\mathbb{Z}$ topology indicated by the Euler class which sign reflects orientation, and $(iii)$ multiband subspaces ($p_i\geq 3$) have a $\mathbb{Z}_2$ topology indicated by the second SW class which is blind to orientation. We can indicate a generic homotopy equivalence class of total gapped bundles with a prescribed partitioning of bands as $(\beta_I,\beta_{II},\ldots,\beta_\mathfrak{N})$ where the indicators $\beta_i$ is a $\mathbb{0}$, $\mathbb{Z}$ resp. $\mathbb{Z}_2$ number depending on the value of $p_i$. It follows from the Whitney sum formula for orientable subbundles, from the triviality of the total Bloch bundle, and from the discussion in Sec.~\ref{subsec:2D-cohomology-classes} that $\sum_{i=I}^\mathfrak{N} \beta_i = 0 \mod 2$.

Similar to Sec.~\ref{subsec:second}, dropping the orientations of the subbundles reduces the second homotopy groups into orbits under automorphism induced by the first homotopy group of the classifying space. Since adiabatic evolution of the Hamiltonian along $\ell_i$ reverses the orientation of subbundles $\mathcal{B}^{+}_i(p_i)$ and $\mathcal{B}^{+}_{i+1}(p_{i+1})$, the corresponding automorphism reverses $\triangleright_{\ell_i}: (\beta_i,\beta_{i+1})\mapsto (-\beta_i,-\beta_{i+1})$ while keeping the other indicators intact. Similar to the single gap case, $\triangleright_{\ell_i}$ acts non-trivially only when $p_i=2$ or $p_{i+1}=2$. By forming arbitrary compositions of automorphisms $\{\triangleright_{\ell_I}\}_{i=1}^{\mathfrak{N}-1}$, we can flip the sign of any even number of the indicators $\beta_i$. In other words, the orbits (i.e.~the elements of the \emph{free} homotopy set $[\mathbb{T}^2,\textrm{Fl}_{p_I,\ldots,p_\mathfrak{N}}]^{(\vec{0},\vec{0})}$ where $(\vec{0},\vec{0})$ indicates the vanishing Berry phases of each subbundle along the two non-contractible cycles of the Brillouin zone) consists of collections $(\pm \beta_I,\pm \beta_{II},\ldots,\pm\beta_\mathfrak{N})$ that differ from each other by an \emph{even} number of sign reversals. Whenever any of the indicators is $\mathbb{0}$ or $\mathbb{Z}_2$ valued, but also when it is $\mathbb{Z}$ valued but takes the zero value, its sign reversal does not correspond to any change of topology, meaning that the orbits under automorphisms $\{\triangleright_{\ell_i}\}_{i=1}^{\mathfrak{N}-1}$ also admit arbitrary (including \emph{odd}) number of sign reversals.

\subsection{Repartitioning of bands and fragile topology}\label{sec:fragile-topo}
With the obtained understanding of the topology of the generalized flag manifold, let us consider the effect of \emph{repartitioning} the bands 
\begin{equation}
    \mathcal{B}_i(2) \cup \mathcal{B}_{i+1}(1) \rightarrow \mathcal{B}'_i(3) = \mathcal{B}_i(2) \oplus \mathcal{B}_{i+1}(1) \;,
\end{equation}
caused by the closing (or discarding) of the energy gap between band subspaces $\mathcal{B}_i$ and $\mathcal{B}_{i+1}$. The repartitioning induces the following reduction of topological charge 
\begin{equation}
r:\;\;\begin{aligned}
   \mathbb{N} & \rightarrow \mathbb{Z}_2 \\ 
    \overline{\chi}[\mathcal{B}_i(2)] &\mapsto 
    w_2[\mathcal{B}'_i(3)] 
    = \overline{\chi}[\mathcal{B}_i(2)] \mod 2 \;.
\end{aligned}
\end{equation}
We thus conclude that whenever a two-band subspace has an \textit{even} (\textit{odd}) Euler class, the effect of adding an extra  \textit{trivial band trivializes} (respectively \textit{reduces}) the topology of the combined 3-band subspace. For this reason, Euler class is described as a \textit{fragile} topology~\cite{Ft1,bouhon2018wilson}. Fragile topology is thus weaker than the stable topology known from Chern insulators where the nontrivial topology is robust under the addition of trivial bands. However, fragile topology must be sharply contrasted from the unstable topology of Hopf insulators that only exists in strictly two-level systems ($\pi_3[\mathrm{Gr}_1(\mathbb{C}^2)] = \mathbb{Z}$)~\cite{Hopf_1,Hopf_2, Unal2019,Alex:2019}. Indeed, in Hopf insulators the embedding of the two-level Hamiltonian into three-(or more)-band Hamiltonian destroys the whole topology, while the nontrivial fragile topology of a few-band \textit{subspace} is conserved as long as the energy gaps separating it from the other bands are maintained.

\subsection{Nodal points}\label{nodal_points}
The principal observable linked to the reduced Euler class of an \textit{orientable two-band subbundle} $\mathcal{B}_i(2)$ is the number of stable nodal points formed between the two bands, i.e.~there is a minimal number of nodal points $\#_{\mathrm{NP}} = 2  \overline{\chi}[\mathcal{B}_{i}(2)] = 2 \vert \chi[\mathcal{B}^+_{i}(2)]\vert $ that cannot be annihilated as long as the gaps with the adjacent bands, $\mathcal{B}_{i-1}$ and $\mathcal{B}_{i+1}$, remain open  \cite{BJY_nielsen,bouhon2019nonabelian}. We emphasize that this result is only valid in the orientable case~\cite{BJY_nielsen}, otherwise the Euler class cannot be defined~\cite{Montambaux_2018}.

We emphasize that \textit{stable} nodal points here indicates those that cannot be removed within a two-band subspace as long as the adjacent gaps remain open. Band structures may host additional pairs of nodal points within a two-band subspace that \textit{can} be annihilated when the nodes are collapsed onto each other. This may however require a \textit{large} deformation of the band structure (similarly to generic Weyl points), i.e.~\textit{unstable} nodal points in this context are topologically robust relatively to small local perturbations of the band structure. In that sense, the ``stability'' of unstable nodal points can be measured, crudely, by the shortest distance that separates them in the Brillouin zone.\footnote{A more detailed analysis would be needed to obtain the measure of stability in terms of the deformation of the tight-binding parameters, which will be studied in an upcoming work.}

By allowing band inversions of the two principal bands with a third band, additional nodal points can be generated or annihilated in pairs within the two adjacent (below and above) energy gaps~\cite{BJY_nielsen, bouhon2019nonabelian}. This facilitates the braiding of principal and adjacent band nodes which is accompanied by non-Abelian phase factors~\cite{Wu1273}. The $\mathbb{Z}_2$ second SW class of the three-band subspace then indicates the stable \emph{parity} of the minimal number of pairs of nodal points, i.e.~$w_2[\mathcal{B}_{i}(p_i)] = 1$ indicates that at least one pair of nodal points cannot be annihilated within the $p_i$-band subspace.

\section{Geometric construction}\label{geo}

\subsection{Strategy}

We now embark on employing the notions developed in the previous sections to construct a general geometrical framework. Before we turn to the topic, we emphasize that out strategy is again to first develop \emph{explicit models} equipped with a specific orientation of each subbundle. We subsequently drop the orientation and arrive at \emph{homotopy class representatives} of orientable bands.

Accordingly, we note that all real vector bundles over a sphere are orientable since all base loops can be contracted to a point, i.e.~$ e^{\imi \gamma[l]} = +1$ for all $l\subset B= \mathbb{S}^2$. Inversely, all orientable topological phases can be effectively modeled over the sphere since $[\mathbb{T}^2, \mathrm{Gr}_{p,N}]^{(0,0)} = [\mathbb{S}^2, \mathrm{Gr}_{p,N}]$. This motivates the strategy~\cite{BzduSigristRobust} to generate representative tight-binding models for all the homotopy classes. 

The general framework is then presented as follows. After setting up some definitions and identifying the appropriate universal bundle structure, we then describe how this structure can be pulled back to the torus to obtain specific coordinates to facilitate the desired maps. To achieved this we make use of the so-called Pl\"{u}cker embedding into more manageable exterior product spaces that allows us to paramaterize the map into the Grassmannians in a tractable manner. After having discussed this embedding, we close with the homotopy aspects of our construction.

\subsection{The tautological bundle}
The \emph{tautological bundle} of the oriented Grassmannian, $\mathcal{F}^+_{p,N} \rightarrow \mathrm{Gr}^+_{p,N}$, is defined as the vector bundle obtained by taking the oriented $p$-dimensional hyperplane $V_I = \mathrm{Span}\{u_1,\dots, u_p\}$ at \textit{every} point $[R]^+ $ of the oriented Grassmannian, where $R=(R_I R_{II}) = (u_1\cdots u_p u_{p+1} \cdots   u_N)\in \mathsf{SO}(N)$\footnote{\unexpanded{ Alternatively, the tautological vector bundle can be defined in the following way. First, the band vector subspaces can be defined as the range (i.e.~image) $V_{i} = \mathrm{ran} \mathbb{P}_{i}$ with the projectors $\mathbb{P}_{i} = R_i R_i^T$, for $i=I,II$. Then, $V_{II} = \mathrm{ran}\, \mathbb{P}_{II} = \mathrm{ran}\, \mathbb{Q}_{I} $ with $\mathbb{Q}_i = \mathbb{1}_N - \mathbb{P}_i$. Thus, $\mathcal{F}^+_{p,N} = \bigcup_{[R] \in \mathrm{Gr}^+_{p,N}} \mathrm{ran}\, \mathbb{P}_I$ and $\mathcal{F}^+_{N-p,N} = \bigcup_{[R]\in \mathrm{Gr}^+_{p,N}} \mathrm{ran}\,(\mathbb{1}_N -\mathbb{P}_I)$.}}. As mentioned in Sec.~\ref{sec:orientability}, and as more carefully elaborated in Sec.~\ref{sec:Plucker} below, the oriented $p$-plane can also be expressed in an $\mathsf{SO}(p)$-invariant fashion as the wedge product $u_1 \wedge \cdots \wedge u_p$. The tautological bundle is canonical, in the sense that its structure follows directly (without extra assumptions) from the construction of the Grassmannian. 

We note that by fixing an oriented $p$-dimensional hyperplane in $\mathbb{R}^N$, we implicitly but uniquely also define the complementary oriented $(N-p)$-dimensional hyperplane $V_{II} = \mathrm{Span}\{u_{p+1},\dots, u_N\}$ such that $\mathbb{R}^N = V_I \oplus V_{II}$. This can be also seen as the Hodge dual of $u_1\wedge \cdots \wedge u_p$. However, in general for the equivalence classes we have $[\mathcal{F}^+_{p,N}] \neq  [\mathcal{F}^+_{N-p,N}]$ since they do not need to have equal ranks, while $\mathrm{Gr}^+_{p,N} = \mathrm{Gr}^+_{N-p,N}$. For this reason we introduce the notion of oriented \textit{tautological total gapped bundle}, in analogy with our definition of $\mathcal{E}^{+}_{p,N}$ in Sec.~\ref{sec:orientability}, as  $\mathcal{T}^+_{p,N} = \mathcal{F}^+_{p,N} \cup \mathcal{F}^+_{N-p,N}$.

We now define a \textit{reference total gapped bundle} from which all the phases can be generated. This is achieved through a map $f_1 : \mathbb{S}^2 \rightarrow \mathrm{Gr}^+_{p,N}$ such that $f_1(\mathbb{S}^2)$ belongs to the homotopy class that generates $\pi_2[\mathrm{Gr}^+_{p,N}]$ (by abuse of language, we will say that $f_1(\mathbb{S}^2)$ \emph{is} the generator of the second homotopy group). The reference bundle is then defined as the pullback $\mathcal{R}^+_{p,N} = f_1^*\mathcal{T}^+_{p,N}$, i.e.~the restriction of the tautological total gapped bundle $\mathcal{T}^+_{p,N}$ induced by $f_1$. Now, in analogy with the way every oriented vector (sub)bundle can be obtained as a pullback of the tautological bundle,~i.e.~$\mathcal{B}^+(p) = f^*_\mathcal{B} \mathcal{F}^+_{p,N}$ with a suitable $f_\mathcal{B}$, an explicitly constructed $f_1$ allows us to express an arbitrary 
oriented total gapped bundle $\mathcal{E}^+_{p,N}$ 
as a pullback of $\mathcal{T}^+_{p,N}$.

\subsection{Pullback to the Brillouin zone torus}

\begin{figure*}[t]
\centering
\begin{tabular}{c}
	\includegraphics[width=0.7\linewidth]{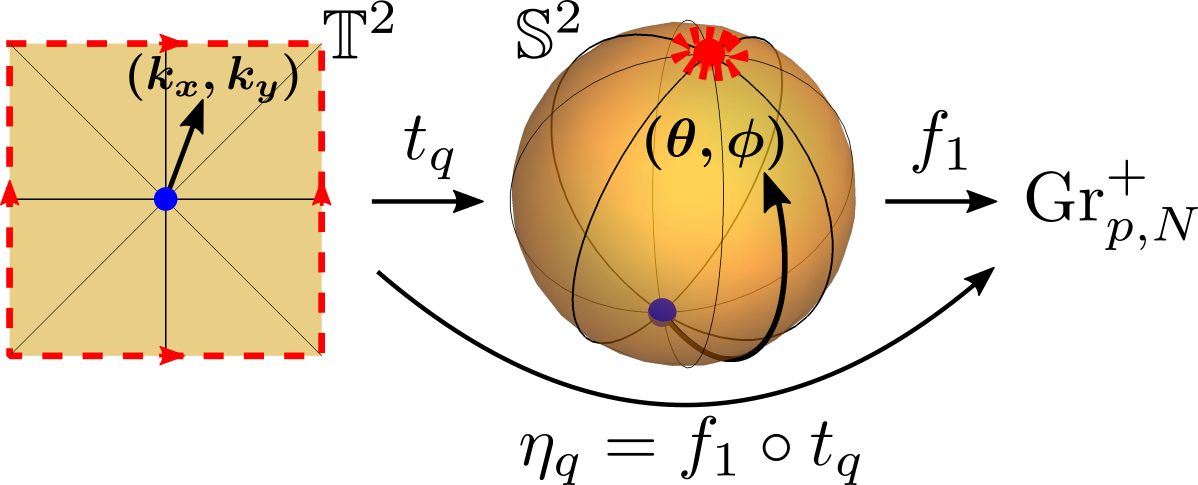}
\end{tabular}
\caption{\label{pullback} Composition map $\eta_q = f_1 \circ t_q$ through which the pullback bundle $\mathcal{E}^+_{p,N}=\eta_q^*\mathcal{T}^+_{p,N}$ is built. We define the map $t_q$ such that the Brillouin zone center is mapped to the ``blue pole'' of the sphere, and the Brillouin boundary to the ``red pole''. The points with the same distance from the Brillouin zone center, $\max\{\left|k_x\right|\left|k_y\right|\}$, are mapped to the same polar angle $\theta$ on the sphere. The map $f_1$ is then constructed such that its image $f_1(\mathbb{S}^2)$ induces the generator(s) of $\pi_2[\mathrm{Gr}^+_{p,N}]$. As a result, windings producing non-trivial Euler class can be imposed. In the text we also refer to the center of the Brillouin zone, $(k_x,k_y)=(0,0)$, as the $\Gamma$ point, and the corner of the Brillouin zone, $(k_x,k_y)=(\pi,\pi)$, as the M point.}
\end{figure*}

In order to connect with tight-binding models, we define a continuous function $t_{q} :\mathbb{T}^2 \rightarrow \mathbb{S}^2$ that maps the Brillouin zone torus onto the sphere with $\mathrm{deg}\,t_{q} = q \in \mathbb{Z}$, i.e.~$t_{q}$ wraps $\vert q\vert$ times over the sphere with the orientation $\sgn \, q=\pm 1$. Parametrizing the sphere $\mathbb{S}^2$ with the usual spherical coordinates, a simple choice of $t_q$ is obtained by taking inside the first Brillouin zone $\mathbb{I}^2 \simeq \left|k_{x,y}\right|\leq \pi$ the mapping 
\begin{subequations}\label{eqn:tq-map}
\begin{equation}
	t_{q} : \boldsymbol{k} \mapsto (\theta_q(\boldsymbol{k}),\phi_q(\boldsymbol{k}) ) \;,
\end{equation} 
with 
\begin{equation}
\begin{aligned}
    \theta_q(\boldsymbol{k}) &=  \max  \left(\vert k_x \vert , \vert k_y \vert \right)  \;,\\
	\phi_q(\boldsymbol{k}) &=  q\, \mathrm{arg}\,(k_x + \mathrm{i} k_y)\, \;,
\end{aligned}
\end{equation} 
\end{subequations}
where we set $\phi_q(0,0) = 0$. Note that $\phi_q$ has a branch cut on $\{k_x \leq 0, k_y=0\}$ and that it is discontinuous at $\boldsymbol{k}=(0,0)$. However, these discontinuities disappear in the Cartesian coordinates of a point of the sphere $e_r = (\cos \phi_q  \sin \theta_q,\sin \phi_q  \sin \theta_q,\cos \theta_q)$. Furthermore, although $\theta_q$ is not differentiable at $\vert k_x\vert=\vert k_y\vert$, the map is continuous. (The differentiability of the resulting Bloch Hamiltonians representatives of each topological phase will be easily restored in Secs.~\ref{3_bands} and~\ref{3_bands}.). The composition map $\eta_q = f_1 \circ t_q$ thus sends each point $\boldsymbol{k}\in \mathbb{T}^2$ of the Brillouin zone  to a point $\eta_{q}(\boldsymbol{k}) = f_1(\theta_{q}(\boldsymbol{k}),\phi_{q}(\boldsymbol{k}))$ of a sphere inside the Grassmannian, cf.~Fig.~\ref{pullback}.

A generic oriented total gapped bundle over the Brillouin zone torus is obtained as a pullback by the composition map~$\eta_q$, 
\begin{equation}
\mathcal{E}^{q+}_{p,N}  = t_q^*\mathcal{R}^+_{p,N} = t_q^{*}(f_1^* \mathcal{F}^+_{p,N})  
\equiv \eta_q^{*}\mathcal{F}^+_{p,N},
\end{equation}
and according to the diagram 
\begin{equation}\label{eqn:total-bundles-diagram}
\begin{tikzcd}[]
\mathcal{E}^{q+}_{p,N} \arrow[r, "h' "] \arrow[d, "\pi_{\mathbb{T}^2}"] &
			\mathcal{R}^+_{p,N}  \arrow[r, "h"] \arrow[d, "\pi_{\mathbb{S}^2}"]
						& \mathcal{T}^+_{p,N} \arrow[d, "\pi_{\mathrm{Gr}}"] & \\
\mathbb{T}^2 \arrow[r, "t_q"] &
			\mathbb{S}^2 \arrow[r, "f_1" ]
						& \mathrm{Gr}^+_{p,N} \arrow[r, hook,"\iota"] &  \bigwedge^p\mathbb{R}^{N}\;.
\end{tikzcd}
\end{equation}
where $h$, and $h'$, are bundle maps obtained as the inverse of the pullbacks $f_1^*$, and $t_q^*$, respectively, and the mapping $\iota$ on the bottom-right is the Pl\"{u}cker embedding of the Grassmannian explained in Sec.~\ref{sec:Plucker}. 
It follows from the described construction, that the integer $q$ determines the isomorphism class of the oriented total gapped bundle, such that $[\mathcal{E}^{q+}_{p,N}] = \beta_q \in \pi_2[\mathrm{Gr}^+_{p,N}] $. This defines the homomorphism of groups 
\begin{equation}
\beta:\;
\begin{aligned}
 \mathbb{Z} &\rightarrow \pi_2[\mathrm{Gr}^+_{p,N}] \\
 q &\mapsto \beta_q
\end{aligned}
\end{equation}
with $\beta_q^{-1} = \beta_{-q} $, $ \beta_{q'+q} = \beta_q' + \beta_{q}$, and $\beta_0 = 0$. 

Finally, writing the oriented total gapped bundle as $\mathcal{E}^{q+}_{p,N} = \mathcal{B}^{q+}_I(p)\cup  \mathcal{B}^{q+}_{II}(N-p)$, the corresponding diagram for the vector subbundle $\mathcal{B}^{q+}_I(p)$ is 
\begin{equation}
\begin{tikzcd}[]
\mathcal{B}^{q+}_I(p) \arrow[r, "h' "] \arrow[d, "\pi_{\mathbb{T}^2}"] &
			\mathcal{B}^{+}_{p,N}  \arrow[r, "h"] \arrow[d, "\pi_{\mathbb{S}^2}"]
						& \mathcal{F}^{+}_{p,N} \arrow[d, "\pi_{\mathrm{Gr}}"] & \\
\mathbb{T}^2 \arrow[r, "t_q"] &
			\mathbb{S}^2 \arrow[r, "f_1" ]
						& \mathrm{Gr}^+_{p,N}\;,
\end{tikzcd}\label{eqn:reference-bundle-diagram}
\end{equation}
[and similarly for $\mathcal{B}^{q+}_{II}(N-p)$], where $\mathcal{B}^+_{p,N}$ is the vector subbundle of the bundle $\mathcal{R}^+_{p,N} = f_1^* \mathcal{T}^+_{p,N}$ defined over $\mathbb{S}^2$.
We emphasize that since the associated total gapped bundle $\mathcal{E}^{q+}_{p,N}$ is generated from the pullback by a map to the \textit{oriented} Grassmannian ($\eta_q:\mathbb{T}^2\rightarrow \mathrm{Gr}^+_{p,N}$) it has no nontrivial weak invariants. 
In other words, the Berry phases over either the occupied and the unoccupied band subspaces are all trivial, even though the base space ($\mathbb{T}^2$) contains non-contractible loops.

\subsection{Pl\"{u}cker embedding}\label{sec:Plucker}

We now motivate how to represent the image $f_1(\mathbb{S}^2) \subset  \mathrm{Gr}^+_{p,N}$ via a general procedure that would be discussed more explicitly for three-band and four-band models in Secs.~\ref{3_bands} and~\ref{4_bands}. To achieve this, we employ the \emph{Pl\"ucker embedding} $\iota :\mathrm{Gr}^+_{p,N} \xhookrightarrow{}  \bigwedge^p(\mathbb{R}^{N}) $ which represents the oriented Grassmannian as a $p(N-p)$-dimensional submanifold of the $p$-th exterior power of $\mathbb{R}^{N}$ (i.e.~the $(\substack{N \\ p})$-dimensional Euclidean vector space spanned by $N$-component and fully anti-symmetric $p$-vectors). 
As a result, the image of the Pl\"{u}cker embedding can be generally represented by~\cite{Kozlov_Gr} 
\begin{equation}
\iota (\mathrm{Gr}^+_{p,N}) = \mathbb{K}_p \cap \mathbb{S}^{\left(\substack{N \\ p}\right)-1}.\label{eqn:Plucker-image}
\end{equation}
Here, $\mathbb{K}_p$ is the cone of \textit{simple} (or \emph{decomposable}) $p$-vectors, i.e.~those of the form $\bigwedge_{i=1}^p v_i$ for some collection of vectors $\{v_i\}_{i=1}^p$ in $\mathbb{R}^N$ (not necessarily pairwise orthogonal), and $\mathbb{S}^{\left(\substack{N \\ p}\right)-1}$ is the unit sphere in $\bigwedge^p(\mathbb{R}^{N})$ with respect to the linear inner product defined on simple $p$-vectors as\footnote{Equivalently, the right-hand side of Eq.~(\ref{eqn:inner-on-exterior-power}) is the determinant of a matrix $A$ with elements $A_{ij} = \left<v_{i},v'_j\right>$.}
\begin{equation}
\begin{aligned}
\Big\langle\bigwedge_{i=1}^p v_i,\bigwedge_{i=1}^p v'_i\Big\rangle
=\sum_{\sigma\in\mathsf{S}_p}(-1)^{\mathrm{sign}\,\sigma}\prod_{i=1}^p \left<v_{i},v'_{{\sigma (i)}}\right>\label{eqn:inner-on-exterior-power}
\end{aligned}
\end{equation}
where $\mathsf{S}_p$ is the permutation group of $p$ elements. 

The Pl\"{u}cker embedding is defined explicitly as follows. For a given point $[R]^+\in \mathrm{Gr}^+_{p,N}$ we take a representative $R=(R_I R_{II})$, and we construct $\iota([R]^+)$ as the wedge product of the columns of subframe $R_I$, that is 
\begin{equation}
\iota([R]^+) = u_{1} \wedge \dots \wedge u_{p} \equiv \omega_p.\label{eqn:Plucker-map}
\end{equation}
Crucially, the $p$-vector in Eq.~(\ref{eqn:Plucker-map}) is invariant under the $\mathsf{SO}(p)\times\mathsf{SO}(N-p)$ gauge transformations of the frame $R$, meaning that all choices of the representative of $[R]^+$ result in the same image $\iota([R]^+)$. Furthermore, note that $\omega_p$ is by definition a simple $p$-vector, and it is easy to check that it has a unit norm, implying $\iota (\mathrm{Gr}^+_{p,N}) \subseteq \mathbb{K}_p \cap \mathbb{S}^{\left(\substack{N \\ p}\right)-1}$. The validity of the \emph{equality} in Eq.~(\ref{eqn:Plucker-image}) is less obvious\footnote{The inverse map, i.e.~the one that assigns to any simple $p$-vector with unit norm a unique oriented $p$-plane, can be obtained by considering the $QR$ decomposition of the matrix $(v_1,\ldots,v_p)$ representing the $p$-vector.}, but can be proved~\cite{Kozlov_Gr}.

Note that one can similarly define $\iota (\mathrm{Gr}^+_{N-p,N})$ as the exterior product $\omega_{N-p}$ of the $N-p$ eigenvectors of subframe $R_{II}$. Although in general $\omega_p \neq \omega_{N-p}$ (in a similar way as generically $\mathcal{F}^+_{p,N} \neq \mathcal{F}^+_{N-p,N}$), the two objects are canonically related as Hodge duals, namely $*(\omega_p)=\omega_{N-p}$. 
The invariance of $\omega_p$ and of $\omega_{N-p}$ under gauge transformations $R\mapsto R [G_I \oplus G_{II}] = (R_I G_I~R_{II} G_{II})$ with $G_I \in \mathsf{SO}(p)$ and $G_{II} \in \mathsf{SO}(N-p)$, and the fact that the images of the two subframes are uniquely related as Hodge duals, together imply that Eq.~(\ref{eqn:Plucker-map}) is a faithful representation of the oriented Grassmannian.

In the following sections, we obtain an explicit parametrization of $\mathrm{Gr}^+_{p,N}$ for $N=3$ and $N=4$ through the Pl\"{u}cker embedding.

Starting from the general parametrization of an element $R\in \mathsf{SO}(N)$, the task is to find the restriction to the parametrization of an element $\omega_p$ of the image $ (\iota \circ f_1)(\mathbb{S}^2) $ (corresponding to the generator of the second homotopy group of the Grassmannian), as this provides the parametrization of the subframes $R_I$ and $R_{II}$, which directly encode the Hamiltonian $H(\bs{k})$. By explicitly solving this problem for $\mathsf{SO}(3)$ and $\mathsf{SO}(4)$, we derive explicit three-band and four-band tight-binding models for all the homotopy classes.

\subsection{Homotopy classes of total gapped bundles}\label{sec:homotopy-reference-bundle}
We argued in Sec.~\ref{subsec:2D-cohomology-classes} that two-band subspaces are characterized by the Euler class $\chi[\mathcal{B}^{q+}(2)] $, while $p$-band subspaces with $p\geq 3$ are characterized by the second SW class, $w_2[\mathcal{B}^{q+}(p)] $. We will see in Secs.~\ref{3_bands} and~\ref{4_bands} for several concrete examples that the element of the second homotopy group for the oriented total gapped bundle determines the Euler and the second SW classes of both vector subbundles $\mathcal{B}^{q+}_{I}(p)$ and $\mathcal{B}^{q+}_{II}(N-p)$, depending on~$q$ for the map $t_q : \mathbb{T}^2 \rightarrow \mathbb{S}^2$ [Fig.~(\ref{pullback})].

When the classifying space is a product space, i.e. $\mathrm{C} = \prod_{j\in A} X_j$ ($A $ being some indexing set), the second homotopy group splits as a direct sum $\pi_2[\mathrm{C}] = \bigoplus_{j} \pi_2[X_{j}] $. Thus the map $f_1$ splits accordingly into the components $\{f_{1,j}(\mathbb{S}^2) \}_{j\in A'}$
where the indexing set $A'$ contains the components $X_j$ with a non-trivial $\pi_2[X_j]$. 
This scenario notably occurs for $\mathrm{Gr}^+_{2,4}$ that is discussed in Sec.~\ref{4_bands}. 
Under these circumstances, one needs to replace the base space of the reference total gapped bundle in Eq.~(\ref{eqn:reference-bundle-diagram}) by a product $\mathscr{S}^{\vert A'\vert}=\bigtimes_{j=1}^{\vert A'\vert} \mathbb{S}^2_j$, i.e.~one copy of $\mathbb{S}^2$ for each generator of $\pi_2[\mathrm{C}]$, and the maps relate $f_{1,j}:\mathbb{S}^2_j\to X_j$. 
The map from $\mathbb{T}^2$ to $\mathscr{S}_{A'}$ is characterized by a vector of integers $\boldsymbol{q}=(q_1,\ldots,q_{A'})$, with each element encoding the map in Eq.~(\ref{eqn:tq-map}) to the respective $\mathbb{S}^2_j$. 
The composition map $\eta_{\boldsymbol{q}} = f_1\circ t_{\boldsymbol{q}}$ then determines the homotopy class (and therefore also the Euler resp.~the second SW class of both band subspaces) of the mapping $\eta_q:\mathbb{T}^2\rightarrow \mathrm{Gr}^+_{p,N}$.

\section{Three-band models}\label{3_bands}

We now turn the attention to the specific case of $N=3$ bands, and we deploy the machinery developed in the previous sections in a concrete context. From a physical perspective, we point out that the $N=3$ topology has appeared in numerous physical settings. In particular, for nematic systems the
associated topology has been extensively studied~\cite{volovik2018investigation,Beekman20171,Kamienrmp,Genqcs2016, Machon2016}, as well as non-Hermitian band topology has been related to it~\cite{Wojcik:2019}.

\subsection{The three-band classifying space}\label{3_bands_A}

Nontrivial topology can be achieved in a three-band system when a band gap separates a two-band (occupied) subspace and a single-band (unoccupied) subspace (see Table \ref{Tab::Flaghoms}). The classifying space of the orientable phases is $\mathbb{R}\mathrm{P}^2$ leading to the homotopy classification of all the topologically nonequivalent phases $[\mathbb{S}^2,\mathbb{R}\mathrm{P}^2] = 2\mathbb{N}$ (see Sec.~\ref{sec:homotopy-groups}). The factor ``$2$'' here is a convention made such that the value of the topological invariant agrees with the computed Euler class, as we elaborate below. 

As explained in Sec.~\ref{subsec:second}, the associated bundle of the Hamiltonian of an orientable phase can be given an orientation. Therefore, in the following we work with oriented bundles as a mean to build explicit tight-binding models. The classifying space of the oriented phases is $\mathrm{Gr}^+_{2,3} = \mathsf{SO}(3)/[\mathsf{SO}(2)\times \{1\}] \cong \mathbb{S}^2$ (cf.~Sec.~\ref{subsec:projective_plane} and Fig.~\ref{double_cover_RP2}), leading to the homotopy classification $\pi_2[\mathrm{Gr}^+_{2,3}] = 2\mathbb{Z}$ with the topological invariant given by an even integer Euler class $\chi$. The homotopy classes of the orientable phases are then obtained by dropping the orientation, resulting in the $2\mathbb{N}$ topological classification with the reduced Euler class $\overline{\chi} = \vert \chi\vert$ as the topological invariant. We give an explicit example of such a reduction~\cite{Wojcik:2019} in Appendix~\ref{ap_0}.

We start from an arbitrary element 
\begin{equation}
R = (u_1u_2u_3)\in \mathsf{SO}(3) \label{eqn:3-frame}
\end{equation}
that is parametrized by three continuous angles (e.g.~the Euler angles). Choosing a Cartesian frame $(e_1 e_2 e_3)$ of $\mathbb{R}^3$ to decompose the eigenvectors, i.e.~$u_i = u_i^1 e_1 +u_i^2 e_2+ u_i^3 e_3$ for $i=1,2,3$, the Pl\"ucker embedding $\iota(\mathrm{Gr}^+_{2,3})$ of the two-band subspace is given through the bivectors 
\begin{equation}
\begin{aligned}
u_1\wedge u_2 = \;\; &(u_1^2 u_2^3-u_1^3 u_2^2) e_2 \wedge e_3 \\ 
+ &(u_1^3 u_2^1-u_1^1 u_2^3) e_3 \wedge e_1 \\ 
+ &(u_1^1 u_2^2-u_1^2 u_2^1) e_1 \wedge e_2.
\end{aligned}
\end{equation}
Note that the expressions in the parentheses correspond to components of $u_3$, by virtue of the property in Eq.~(\ref{eqn:3-frame}).
By formally identifying the basis of the three-dimensional Euclidean vector space $\bigwedge^2\mathbb{R}^3 $ with the Cartesian frame of $\mathbb{R}^3$ via the Hodge dual, i.e.~$*(e_2\wedge e_3,e_3\wedge e_1,e_1\wedge e_2) = (e_1,e_2,e_3)$, we get $*(u_1\wedge u_2) = u_3 \in \mathbb{S}^2$, which is a specific instance of the duality  
discussed in Sec.~\ref{sec:Plucker}. 

We thus infer that spherical coordinates for three-dimensional orthonormal frames defined on $\mathbb{S}^2$ provide the desired mapping $f_1$ onto  the non-trivial sphere inside the Grassmannian, i.e.~we use 
\begin{equation}
\;f_1(\theta,\phi) = [(u_1 \, u_2 \, u_3)]^+\;,
\end{equation}
with
\begin{equation}
\begin{aligned}
u_3 &= e_r = (\cos\phi\sin\theta,\sin\phi\sin\theta,\cos\theta), \\ 
u_1 &= e_{\theta} = \frac{\partial_{\theta} e_r}{\vert \partial_{\theta} e_r\vert} = (\cos\phi\cos\theta,\sin\phi\cos\theta,-\sin\theta), \\ 
u_2 &= e_{\phi} = \frac{\partial_{\phi} e_r}{\vert \partial_{\phi} e_r\vert} = (-\sin\phi,\cos\phi,0),
\end{aligned}\label{eqn:up-south-east}
\end{equation}
which correspond respectively to the directions ``up'', ``south'', and ``east'' at every point $(\theta,\phi)$ of the sphere. Indeed, $ u_1\wedge u_2 = e_{\theta} \wedge e_{\phi}$, which is invariant under any $G_I \in \mathsf{SO}(2)$ gauge transformation $R_I \rightarrow R_I G_I$, represents the oriented plane perpendicular to $u_3 = e_r$ and, by definition, an element of $\mathrm{Gr}^+_{2,3}$.

\subsection{Three-band reference total gapped bundle}

Note that $(e_{\theta},e_{\phi})$ is an oriented orthonormal frame of $T_{(\theta,\phi)}\mathbb{S}^2$, i.e.~of the tangent space at the point $(\theta,\phi)$ of the sphere, and $e_r$ is the basis of the normal space to the point of the sphere. Therefore, these vectors span the tangent bundle $T\mathbb{S}^2$, and the normal bundle $N\mathbb{S}^2$, respectively. Furthermore, since $\mathrm{Gr}^+_{2,3} \cong \mathbb{S}^2$, the total gapped tautological bundle $\mathcal{T}^+_{2,3}\rightarrow \mathrm{Gr}^+_{2,3} $ is topologically equivalent to the couple $T\mathbb{S}^2\cup N\mathbb{S}^2 $, so we write \begin{equation}
\mathcal{T}^+_{2,3} \sim T\mathbb{S}^2\cup N\mathbb{S}^2.
\end{equation}
Then, since $f_1(\mathbb{S}^2) \cong \mathbb{S}^2$, the same is true for the pullback bundle 
\begin{equation}
\mathcal{R}^+_{2,3} = f_1^* \mathcal{T}^+_{2,3} = \mathcal{B}^+(2) \cup \mathcal{B}^+(1)\label{eqn:ref-tot-gap-bundle-2-3}
\end{equation}
with $\mathcal{B}^+(2) \sim T\mathbb{S}^2$ and $\mathcal{B}^+(1) \sim  N\mathbb{S}^2 $.

\begin{figure*}[t]
\centering
\begin{longtable}{cc}
    \multicolumn{1}{l}{(a)} & \\
    & \multirow{2}{*}{
            \includegraphics[width=0.35\linewidth]{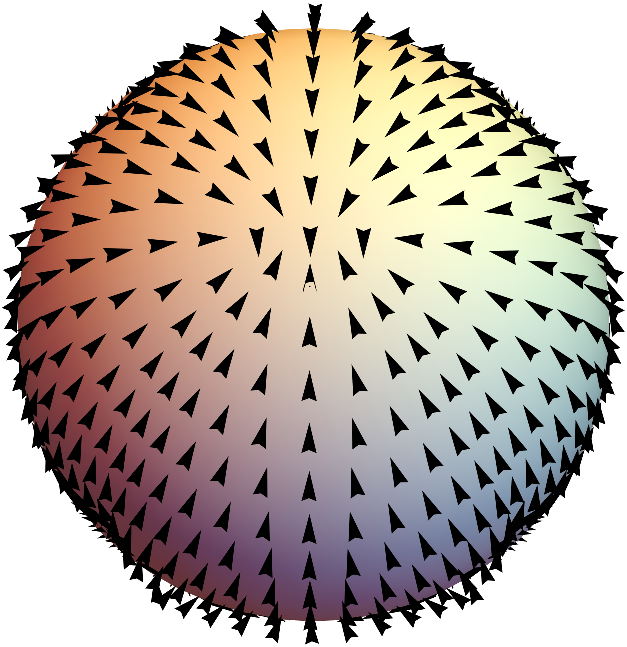} 
        } \\
    \includegraphics[width=0.33\linewidth]{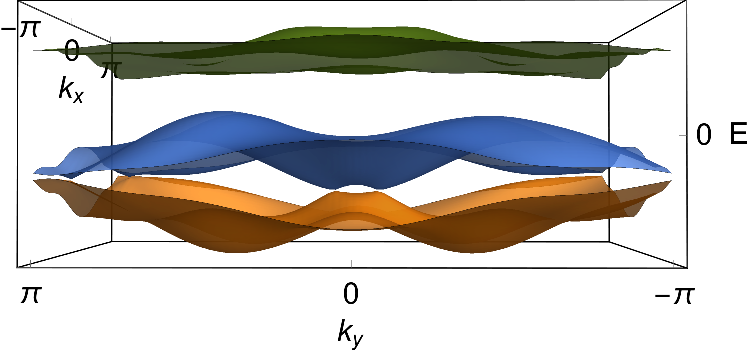}  & \\
    \includegraphics[width=0.27\linewidth]{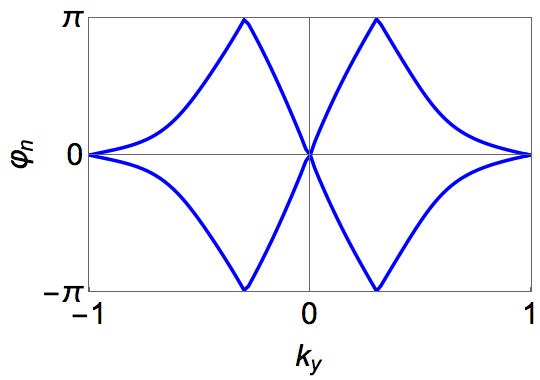} & \\
    \multicolumn{1}{l}{(b)} &  \\
    & \multirow{2}{*}{
            \includegraphics[width=0.35\linewidth]{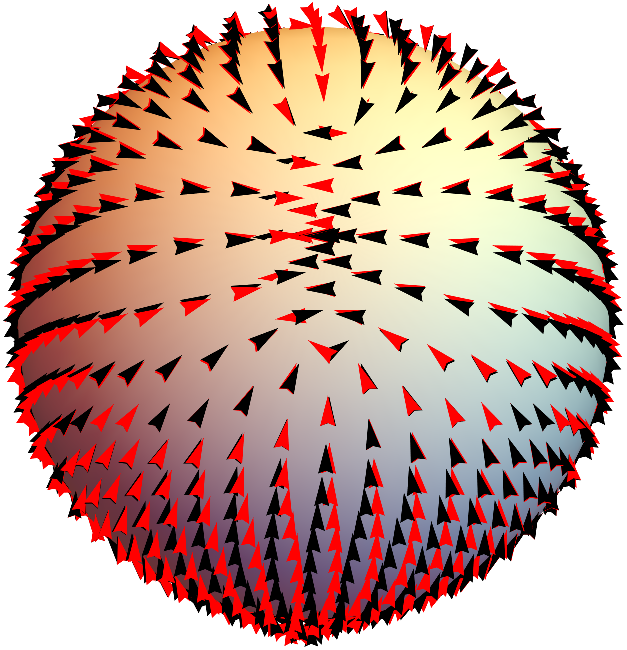}
        } \\
    \includegraphics[width=0.33\linewidth]{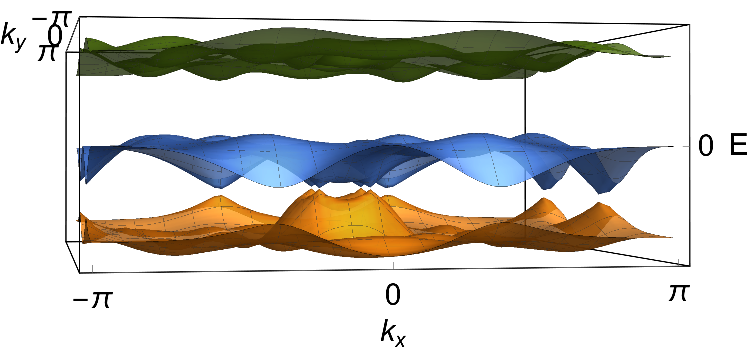}  & \\
    \includegraphics[width=0.27\linewidth]{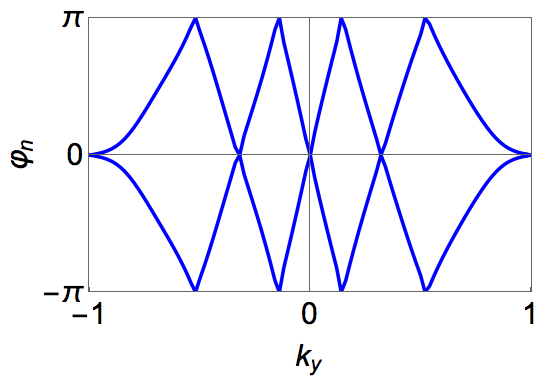} &
\end{longtable}
\caption{\label{VF} Band structure and tangent vector field realization of fragile topology of $\mathrm{Gr}^+_{2,3}$, together with the Wilson loop winding of occupied two-band subspace indicating the Euler class. Panel (a) shows $\mathcal{E}^{q=1,+}_{2,3}$ with Euler class $\vert\chi\vert=2q=2$. The mapping $t_{q=1}$ of Fig.~\ref{pullback} from the Brillouin zone covers the sphere once. We show one vector field directly given by the eigenvectors of lowest energy of the two-band subspace $\mathcal{B}^{q=1,+}_I(2)$. As a global section of the tangent bundle of the sphere, it is characterized through the Poincar\'e-Hopf theorem with the Euler characteristic $\vert\chi\vert=2$, see Eq.~(\ref{eqn:Poincare}). Similarly, panel (b) shows $\mathcal{E}^{q=2,+}_{2,3}$ with Euler class $\vert\chi\vert=2q=4$. The mapping $t_{q=2}$ of Fig.~\ref{pullback} from the Brillouin zone covers the sphere two times. We show one vector field given by the eigenvectors of lowest energy, over the halves $-\pi \leq k_x \leq 0$ (black), and $0 \leq k_x \leq \pi$ (red), of the Brillouin zone. In both cases the sphere is shown on the side of the image of $\Gamma$, i.e.~$t_q(0,0)$, that is the blue pole in Fig.~\ref{pullback}. We thus see in both cases that the vortex structures of the tangent vector fields directly reflects the nodal points of the eigenvalues band structure, with $\#_{\mathrm{NP}} = 2\vert \chi\vert$ globally. Although these nodes come in two pairs that are pairwise close in momentum space, making them hard to distinguish visually, inspecting the nodes in more detail, as shown in Fig. \ref{twonodes} for the panel (a), confirms their presence in the anticipated number.
The tight-binding models have been generated with the Mathematica code available at Ref.~\cite{abouhon_EulerClassTightBinding}.} 
\end{figure*}

It is well known through the hairy ball theorem that the tangent bundle of the 2-sphere is non-trivial, namely any global section, (i.e.~any smooth tangent vector field) must have zeros associated with a vortex structre. We illustrate this known fact on an example in Fig.~\ref{VF}(c), which displays the eigenvectors of a tight-binding model presented in the next section. The hairy ball theorem is formalized by the \emph{Poincar\'e-Hopf theorem}, which states 
\begin{equation}
\sum_{j} \mathrm{index}_{x_j} (v) = \chi[\mathbb{S}^2] = 2, \label{eqn:Poincare}
\end{equation}
where $x_j$ is the location of a zero of the vector field $v$, $\mathrm{index}_{x_j} (v)$ is the winding number of $v/\vert v\vert$ around the zero $x_j$, and $\chi[\mathbb{S}^2]=2$ is the Euler characteristic of the sphere \cite{Frankel}. Thus, any tangent vector field must have two sources of vorticity 1. In the more general context of our classification scheme,
the Euler characteristic of the sphere is substituted by the Euler class of the rank-2 subspace of our reference bundle, which however is still~$ \chi[\mathcal{B}^+_{2,3}(2)] = 2$ by virtue of Eq.~(\ref{eqn:ref-tot-gap-bundle-2-3}) [c.f.~also to the diagram in Eq.~(\ref{eqn:reference-bundle-diagram})]. We show below that the vortices of the tangent vector field correspond to nodal points within the two-band subspace. 

\begin{figure*}[t]
\centering
\begin{tabular}{cc}
	\includegraphics[width=0.33\linewidth]{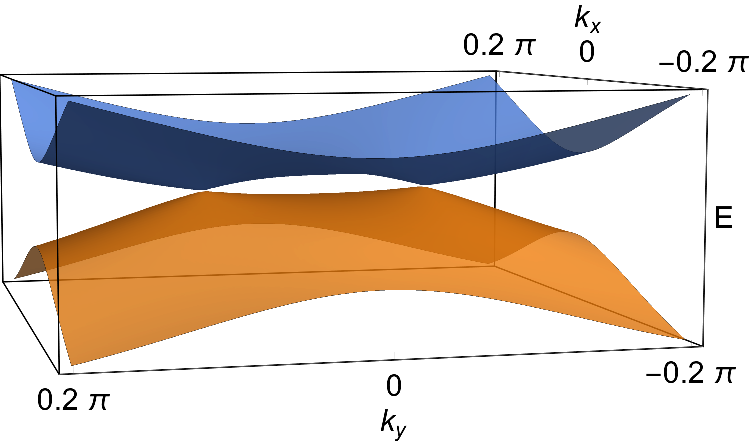} &
	\includegraphics[width=0.396\linewidth]{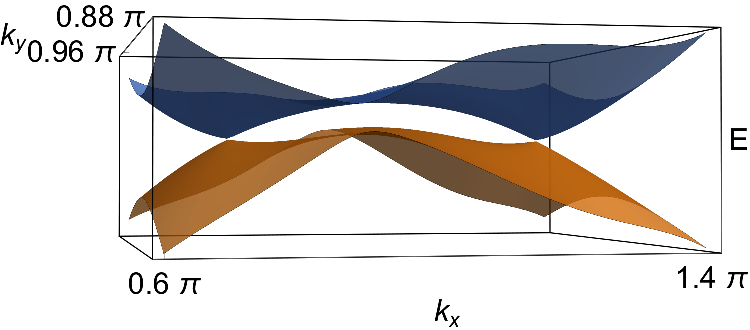} \\
	$\Gamma$ & M
\end{tabular}
\caption{\label{twonodes} Detailed image of the band node structure of Fig.~\ref{VF}, confirming that $\Gamma$ (the center of the Brillouin zone) and $M$ (the corner of the Brillouin zone) indeed host a pair of nodes. 
As a result, we confirm the presence of the anticipated $\#_{\mathrm{NP}} = 2\vert \chi\vert=4$ number of nodes.
} 
\end{figure*}

In this context, we emphasize the \textit{stable triviality} of the tangent bundle of the sphere, i.e.~that it becomes trivial upon the direct sum with a trivial bundle. The normal bundle of the sphere is trivial, and the direct sum gives $T\mathbb{S}^2 \oplus N\mathbb{S}^2 \cong \mathbb{S}^2 \times \mathbb{R}^{3}$. In other words, the nontrivial tangent bundle (with the non-zero characteristic Euler class $\chi[T\mathbb{S}^2] = 2$) is trivialized by the trivial normal bundle \cite{Hatcher_1,Hatcher_2} (resulting in the vanishing characteristic second SW class $w_2[\mathbb{S}^2 \times \mathbb{R}^{3}]=0$). The trivialization (or \emph{reduction}) of the fragile topology of two-band subspaces  upon closing the adjacent energy gap as discussed in Sec.~\ref{sec:fragile-topo} can thus be perceived as a fingerprint of the stable triviality of the tangent bundle of the sphere. 

\subsection{Simple model generation}
Importantly, we 
make use of the presented machinery to generate explicit models of fragile topological phases over the Brillouin zone torus for an arbitrary homotopy class $[\mathcal{E}^{q+}_{2,3}] = \beta_q = 2q \in 2\mathbb{Z}$ (corresponding respectively to Euler class $\chi[\mathcal{B}^{q+}_I(2)] = 2q$). As elaborated previously, we first construct explicit models with a specific orientation that are classified by $2\mathbb{Z}$, and then we drop the orientation resulting in a $2\mathbb{N}$ topological classification. The continuous deformations that relate oriented tight-binding Hamiltonians with Euler class $\pm q$ are explicitly presented in Appendix~\ref{ap_0}.

For $q=1$ the mapping $t_{q=1}$ (cf.~Fig.~\ref{pullback}) wraps the Brillouin zone torus around the sphere once, and we have $\mathcal{E}^{q=1,+}_{2,3} \cong \mathcal{R}^+_{2,3} \cong T\mathbb{S}^2 \cup N\mathbb{S}^2$. The Hamiltonian representative is then readily given by~\cite{BzduSigristRobust}
\begin{equation}
\label{eq:H_3_bands}
H(\boldsymbol{k}) = R(\boldsymbol{k}) [(-\mathbb{1}_2)\oplus 1] R(\boldsymbol{k})^T,
\end{equation} 
where $\mathbb{1}_2$ is a $2\times 2$ identity matrix, $1$ is the identity in the single-band unoccupied subspace, and the frame $R(\bs{k}) = t_{q=1}^*f_1(\theta,\phi) = (e_{\theta},e_{\phi},e_r)$ with 
\begin{equation}
 e_{a}(t_{q=1}(\boldsymbol{k})) = e_{a}(\theta_{q=1}(\boldsymbol{k}),\phi_{q=1}(\boldsymbol{k}))
\end{equation}
for $a = \theta, \phi, r$, with $f_1(\theta,\phi)$ defined in Eq.~(\ref{eqn:up-south-east}) and $t_q(\bs{k})$ from Eqs.~(\ref{eqn:tq-map}). Sampling $H(\boldsymbol{k})$ over a grid and performing an inverse discrete Fourier transform, we obtain the hopping parameters of a tight-binding model, which we truncate to a few neighbors (see Appendix~\ref{app:tb} and Ref.~\cite{abouhon_EulerClassTightBinding} for more details)
without affecting the topological features. Accordingly, we note that while the bands of the Hamiltonian $H(\boldsymbol{k})$ constructed above are flat and the two occupied bands are fully degenerate, these features are lost in our tight-binding models after performing the truncation. Indeed, imposing the perfect flatness and degeneracy results after performing the inverse Fourier transform in an infinite-range hopping amplitudes.

We show in Fig.~\ref{VF}(a) the band structure of the obtained tight-binding model (truncated at two nearest neighbors in both directions of the square lattice, see Appendix~\ref{app:tb} and Ref.~\cite{abouhon_EulerClassTightBinding})
that is a representative Hamiltonian for $\mathcal{E}^{q=1,+}_{2,3}$ with Euler class $\chi[\mathcal{B}_I^{q=1,+}(2)]=2$. In agreement with the rule $\#_{\mathrm{NP}} = 2 \vert \chi \vert$, we find $4$ nodal points between the two lower bands of Fig.~\ref{VF}(a), two around $\Gamma=(0,0)$ and two around $\textrm{M}=(\pi,\pi)$. Since the pairs of nodes appear very close to each other, we zoom in the neighborhood of the points $\Gamma$ and $\textrm{M}$ in Fig.~\ref{twonodes} to properly resolve them.

We now demonstrate explicitly the equivalence of vector bundles $\mathcal{B}^{q=1,+}_{I}(2) \sim T\mathbb{S}^2$ and $\mathcal{B}^{q=1,+}_{II}(1) \sim N\mathbb{S}^2$ mentioned below Eq.~(\ref{eqn:ref-tot-gap-bundle-2-3}). To achieve this, we transfer the eigenvectors $\{u_1,u_2,u_3\}$ of the band structure in Fig.~\ref{VF}(a) defined over the Brillouin zone torus to two tangent vectors and one normal vector over each point of the sphere. This is done through the mapping $t_{q=1}$ as a ``pushforward'' of vector fields, i.e.~$t_{1*} u_i(\boldsymbol{k}) \mapsto u_i(t_1(\boldsymbol{k})) = u_i (\theta_1(\boldsymbol{k}),\phi_1(\boldsymbol{k}))$ for $i=1,2,3$. We thus plot on the right Fig.~\ref{VF}(a) the tangent vector field given by the eigenvectors of the \emph{lower} energy band of the two-band subspace of the band structure shown on the left of Fig.~\ref{VF}(a). As a global section of the tangent bundle $T\mathbb{S}^2$, it can be characterized by invoking the Poincar\'e-Hopf theorem from Eq.~(\ref{eqn:Poincare}). 
This indicates the equivalence between the Euler characteristic $\chi[\mathbb{S}^2]=2$ and the Euler class $\chi[\mathcal{B}^{q=1,+}_{I}(2)] = 2$. We also observe that the nodal points of the band structure with fragile topology [left upper panel of Fig.~\ref{VF}(a)] correspond to vortices of the section of $T\mathbb{S}^2$ [right Fig.~\ref{VF}(a)].

We can repeat the exercise for arbitrary $q\in\mathbb{Z}$. We show an example of band structure for $\mathcal{E}^{q=2,+}_{2,3}$ in Fig.~\ref{VF}(b) with the Euler class $\chi=2q =4$. We find $\#_{\mathrm{NP}} = 2\vert \chi \vert = 2\cdot 4 =8$ nodal points in the two occupied bands, namely 4 on the Brillouin zone boundary and 4 around the $\Gamma$ point. Owing to the pullback construction, we can force a geometric picture of $\mathcal{E}^{q>1,+}_{2,3}$ as the tangent and normal bundles of a generalized surface (not a manifold). Since $t_q:\mathbb{T}^2\rightarrow \mathbb{S}^2$ wraps the sphere $q$-times, we can divide the Brillouin zone into $q$ cells that are each mapped onto $\mathbb{S}^2$, i.e.~we get a tangent vector field over the sphere for each of the $q$ cells. Taken together we can think of it as a tangent field over a surface that wraps on itself with $q$ sheets and with the two $q$-fold ramification points $t_q(0,0)$ and $t_q(\boldsymbol{k})\vert_{\boldsymbol{k}\in \partial \mathrm{BZ}}$, where the former is the image of $\Gamma$ (blue point in Fig.~\ref{pullback}), and the latter is the image of the boundary of the first Brillouin zone $\partial \mathrm{BZ} = \{\boldsymbol{k} \vert \max(\vert k_x\vert,\vert k_y\vert)=\pi \}$ (red point in Fig.~\ref{pullback}).

\subsection{Comment on the ``triviality'' of \texorpdfstring{$N\mathbb{S}^2$}{NS2}}\label{subsec:triviality_NS}

We address here a possible source of confusion concerning the triviality of vector bundles, taking the normal bundle of the sphere, and all the line bundles that are pulled back from it, as an example. Further examples in other dimensions are discussed in Sec.~\ref{subsec:higher_dimensional}.

The homotopy classification of the total gapped bundle characterizes the (stable or unstable) homotopy classes of all the bands at the same time. In the case of the sphere we have seen that the total bundle $T\mathbb{S}^2 \cup N\mathbb{S}^2$ is characterized by the second homotopy invariant $\beta = 2 \in 2 \mathbb{Z}$. The homotopy method is in contrast with the cohomology invariants that characterize the \textit{stable} topology of a single subbundle at a time, i.e.~$\chi[T\mathbb{S}^2] = 2$ and $w_1[N\mathbb{S}^2] = w_2[N\mathbb{S}^2] = 0$. The trivial cohomology class of the normal bundle of the sphere reflects the fact that the vector field generated by $e_r$ over the sphere defines a smooth global section, i.e.~$N\mathbb{S}^2$ is a trivial vector bundle. 

However, the same vector field defines a \emph{hedgehog structure} over the sphere, and as such it cannot be continuously deformed into a constant section, e.g.~$v(\theta,\phi) = e_r(0,0) = \textrm{const.}$ for all $(\theta,\phi)$. This is precisely captured by the homotopy invariant of the total gapped bundle which can be obtained as the Skyrmion number of the normal vector field (see Appendix~\ref{ap_0} and Ref.~\cite{Wojcik:2019}). From this perspective, one concludes that the normal bundle of the sphere is non-trivial as it belongs to a different homotopy class than the constant vector field. 

We emphasize, crucially, that the statement about the non-triviality of $N \mathbb{S}^2$ is only true as long as we keep the underlying dimension of Euclidean space $\mathbb{R}^3$ (or equivalently the rank of the total bundle $N=3$) fixed. The trivial cohomology indicates that if we embed the $2$-sphere in an higher dimensional Euclidean space the normal bundle can be fully trivialized, i.e.~the extra dimensions allow a continuous deformation of the hedgehog structure into the constant vector field. The nontrivial homotopy invariant of the normal bundle must thus be interpreted as an \textit{unstable} homotopy which refines the cohomology classification. The same conclusion applies to all line bundles $\mathcal{B}^{q+}(1)$ pulled back from $N\mathbb{S}^2$.

\section{Four-band models}\label{4_bands}
In this section we study the $\mathrm{Gr}_{2,4}$ case, i.e.~the classifying space of an orientable band structure with four bands and with a single gap condition that separates an ``occupied'' two-band subspace from an ``unoccupied'' two-band subspace. As in the previous section we start from the classification of the oriented phases and then we address the effect of dropping the orientation. We first discuss the parametrization of the classifying space obtained from the Pl\"ucker embedding from which representative tight-binding models of all the homotopy classes can be derived. We then study in detail several explicit tight-binding models and we establish the complete list of all topologically nonequivalent phases for $0\leq \vert\chi\vert \leq 3$. We finally briefly address the stability of the topological invariants under the repartitioning of the bands.

\subsection{Parametrization}
The diagonalizing matrix, $R=(u_1\,u_2\,u_3\,u_4)$, underlying the Hamiltonian belongs to $\mathsf{SO}(4)$ and thus can be parametrized by six continuous angles. The exterior product of the occupied states $\omega = u_1\wedge u_2$ is now a point of the image of the Pl\"ucker embedding $\iota(\mathrm{Gr}^+_{2,4})$, i.e.~a simple unit bivector. Notably, since the number of unoccupied bands is the same, the Hodge dual $*\omega = u^3\wedge u^4$ also is a unit bivector in \emph{the same} space $\iota(\mathrm{Gr}^+_{2,4})$. We show in Appendix \ref{ap_1} that their linear combinations $v_{\pm} = \frac{1}{\sqrt{2}}(u_1\wedge u_2\pm u_3\wedge u_4)\in V_{\pm},$ where $V_{\pm}$ are two complementary three-dimensional vector spaces of bivectors which partition the second exterior power of $\mathbb{R}^4$ (a space of dimension ${4  \choose 2} = 6$) into two halves, i.e.~$\bigwedge^2\mathbb{R}^4 = V_+ \oplus V_-$. Furthermore $\vert v_{+}\vert = \vert v_{-}\vert = 1$, thus $v_{+}$ ($v_{-}$) belongs to a unit sphere in $V_+$ ($V_-$). Therefore, the image of the Pl\"ucker embedding is the four-dimensional submanifold $\mathbb{S}^2_{+} \times \mathbb{S}^2_{-}$ with  points $\omega = v_+ + v_-$. The inverse embedding induces a diffeomorphism $\mathrm{Gr}^+_{2,4} \cong \mathbb{S}^2 \times \mathbb{S}^2$, implying $\pi_2[\mathrm{Gr}^+_{2,4}] = \mathbb{Z}\oplus \mathbb{Z}$. We observe that $f_1$ splits into two generators $\{f_1^{(j)}(\mathbb{S}^2) =  \mathbb{S}^2_{j}\}_{j=+,-}$ of $\pi_2[\mathrm{Gr}^+_{2,4}]$ parametrized by $\{(\theta_{j},\phi_{j})\}_{j=+,-}$.

Let us now consider the image $\mathcal{M} = \iota(\mathrm{Gr}^+_{2,4}) = \mathbb{S}^2_{+} \times \mathbb{S}^2_{-}$ whose points are bivectors parametrized by a pair of angles
$(\theta_+,\phi_+,\theta_-,\phi_-)$ (see Appendix \ref{ap_1} for more details). The inverse embedding induces a parametrization $P$ of the four-band diagonalizing matrices 
\begin{subequations}\label{eqn:Gr24-to-SO4}
\begin{equation}
P: \mathcal{M} \rightarrow \mathrm{Gr}^+_{2,4} \hookrightarrow  \mathsf{SO}(4)
\end{equation}
through the assignment 
\begin{equation}
P:\omega(\theta_+,\phi_+,\theta_-,\phi_-) \mapsto [R]^+ \mapsto R(\alpha_1,\alpha_2,\alpha_3,\alpha_4),
\end{equation}
\end{subequations}
i.e.~there is a reduction from six continuous angles for a generic element $R\in\mathsf{SO}(4)$ to four angles for the representatives $[R]^+\in \textrm{Gr}^+_{2,4}$. It is worth noting that $P$ is nothing but a section of the tautological total gapped bundle $\mathcal{T}^+_{2,4}\rightarrow \mathrm{Gr}^+_{2,4}$.  
The explicit parametrization $P$ 
depends on the chosen encoding
of elements $R \in \mathsf{SO}(4)$. This is done in Appendix~\ref{ap_1} in terms of the Lie algebra of real and anti-symmetric matrices.

\subsection{Generating the models}

Once the parametrization in Eqs.~(\ref{eqn:Gr24-to-SO4})
is found, we can readily apply our machinery to generate a tight-binding model corresponding to any homotopy class of $\pi_2[\mathrm{Gr}_{2,4}^+] = \mathbb{Z} \oplus \mathbb{Z}$. Adapting the discussion from Sec.~\ref{sec:homotopy-reference-bundle} to the present situation, we replace the base space of the reference total gapped bundle by $\mathbb{S}^2_1\times \mathbb{S}^2_2$, modifying the scheme in Eq.~(\ref{eqn:total-bundles-diagram}) into
\begin{equation}
\begin{tikzcd}[]
\mathcal{E}^{\boldsymbol{q}+}_{p,N} \arrow[r, "h' "] \arrow[d, "\pi_{\mathbb{T}^2}"] &
			\mathcal{R}^+_{p,N}  \arrow[r, "h"] \arrow[d, "\pi_{\mathbb{S}^2}"]
						& \mathcal{T}^+_{p,N} \arrow[d, "\pi_{\mathrm{Gr}}"] & \\
\mathbb{T}^2 \arrow[r, "t_{\boldsymbol{q}}"] &
			\!\mathbb{S}^2_1\!\times\! \mathbb{S}^2_2\! \arrow[r, "f_1" ]
						& \mathrm{Gr}^+_{p,N} \arrow[r, hook,"\iota"] &  \!\mathbb{S}^2_+\!\times\!\mathbb{S}^2_-\!\;
\end{tikzcd}
\end{equation}
where the pair of integers $\boldsymbol{q}=(q_+,q_-)$ dictates how many times $\eta_{\boldsymbol{q}}:\mathbb{T}^2\rightarrow \mathbb{S}^2_{+} \times \mathbb{S}^2_{-}$ wraps around each of the two target spheres as we cover the Brillouin zone torus. 
The map $f_1$ splits into $(f_{1,+},f_{1,-})$ such that $\iota\circ f_{1,+}$ wraps $\mathbb{S}^2_1$ around $\mathbb{S}^2_+$ (and $\iota\circ f_{1,-}$ wraps $\mathbb{S}^2_2$ around $\mathbb{S}^2_-$) exactly once. 
Then we generate all the topological phases through the pullback of the tautological total gapped bundle, $\mathcal{E}^{\boldsymbol{q}+}_{2,4} = (\eta_{\boldsymbol{q}})^* \mathcal{T}^+_{2,4}$, where $\eta_{\boldsymbol{q}} = f_1 \circ t_{\boldsymbol{q}}$. 
The pair $(q_+,q_-)\in \mathbb{Z}\oplus\mathbb{Z}$ then determines the homotopy invariant $(\beta_1(q_+,q_-),\beta_2(q_+,q_-))\in \pi_2[\mathrm{Gr}^+_{2,4}] = \mathbb{Z}\oplus \mathbb{Z}$, and we simply take $(\beta_1,\beta_2) = (q_+,q_-)$.

\begin{figure*}[t!]
\centering
\begin{tabular}{ccc} 
    \multicolumn{1}{l}{(a) \quad $(q_+,q_-)=(1,0)$} & 
    $(\chi_I,\chi_{II}) = (1,1)$ & 
    $\#_{\mathrm{NP}}[\mathcal{B}_I] = 2\chi_I = 2$\\
    \includegraphics[width=0.33\linewidth]{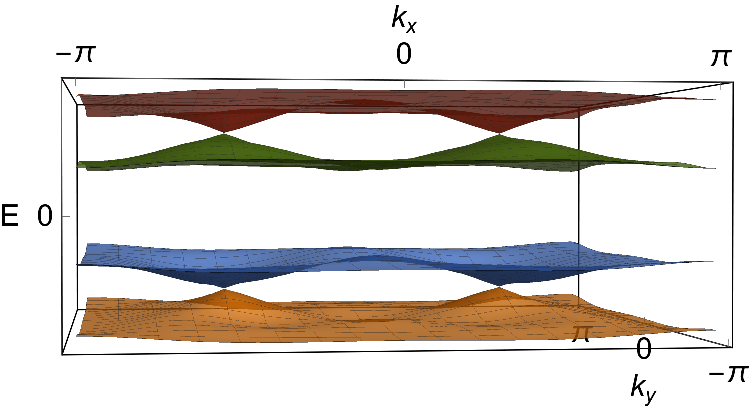} & 
    \includegraphics[width=0.27\linewidth]{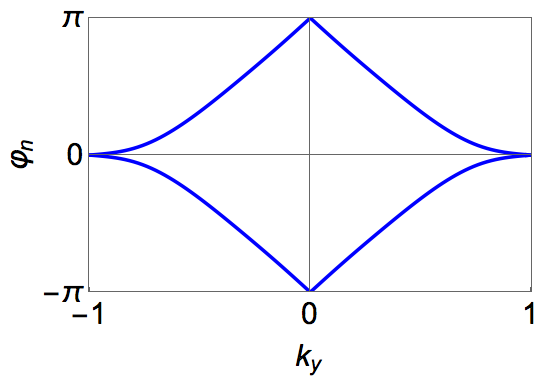} &
    \includegraphics[width=0.19\linewidth]{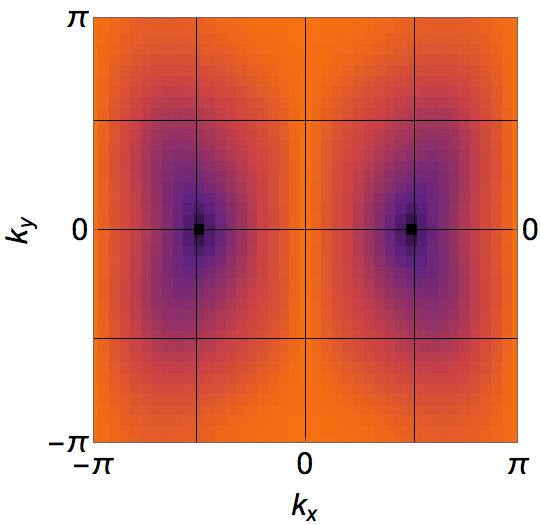} \\
   \multicolumn{1}{l}{(b)\quad $(q_+,q_-)=(2,0)$} &
   $(\chi_I,\chi_{II}) = (2,2)$ & 
   $\#_{\mathrm{NP}}[\mathcal{B}_I] = 2\chi_I = 4$ \\
    \includegraphics[width=0.33\linewidth]{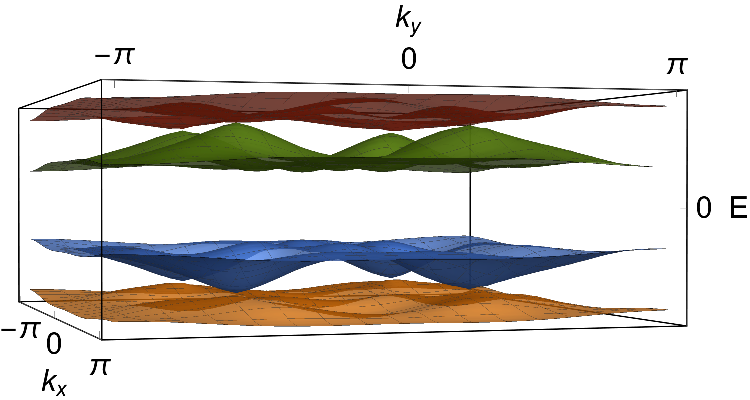} &
    \includegraphics[width=0.27\linewidth]{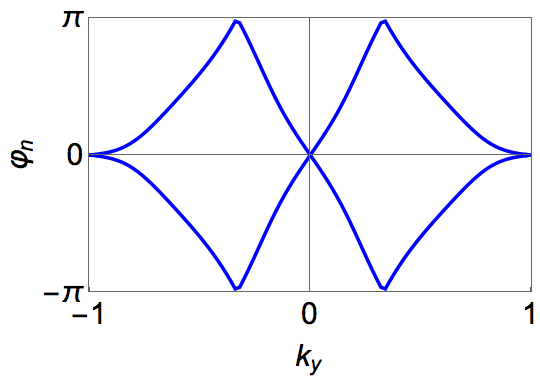} &
    \includegraphics[width=0.19\linewidth]{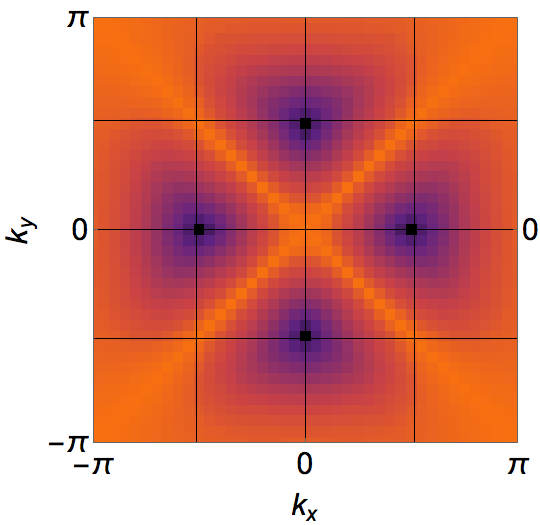} \\
    \multicolumn{1}{l}{(c)\quad $(q_+,q_-)=(3,0)$} &
    $(\chi_I,\chi_{II}) = (3,3)$ & 
    $\#_{\mathrm{NP}}[\mathcal{B}_I]= 2\chi_I = 6$\\
    \includegraphics[width=0.33\linewidth]{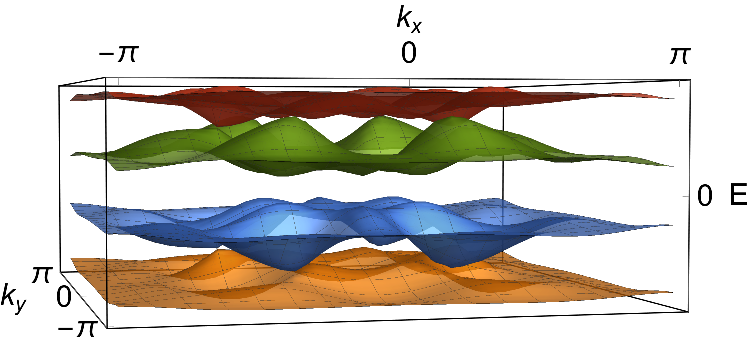} &
    \includegraphics[width=0.27\linewidth]{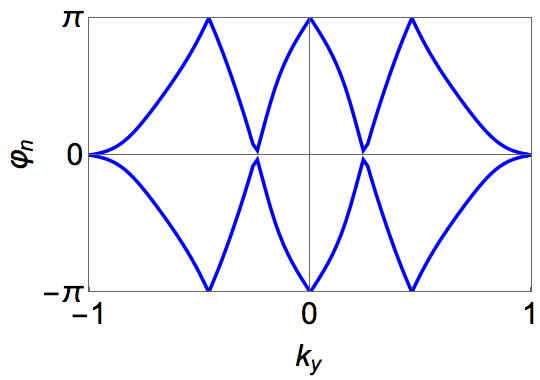} &
    \includegraphics[width=0.19\linewidth]{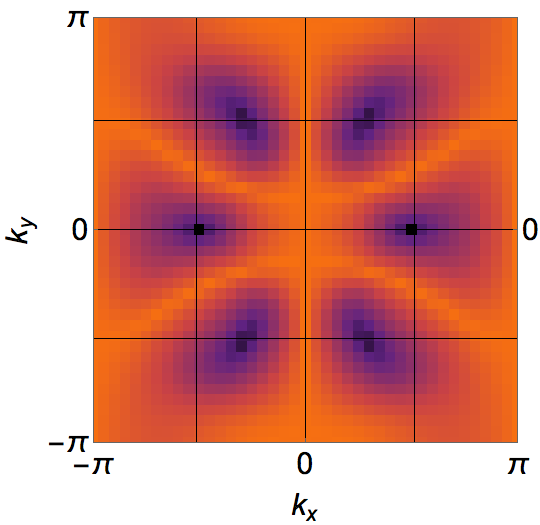} \\
    \multicolumn{1}{l}{(d)\quad $(q_+,q_-)=(1,-1)$} &
    $(\chi_I,\chi_{II}) = (2,0)$ & 
    $\#_{\mathrm{NP}}[\mathcal{B}_I]= 2\chi_I = 4$\\
    \includegraphics[width=0.33\linewidth]{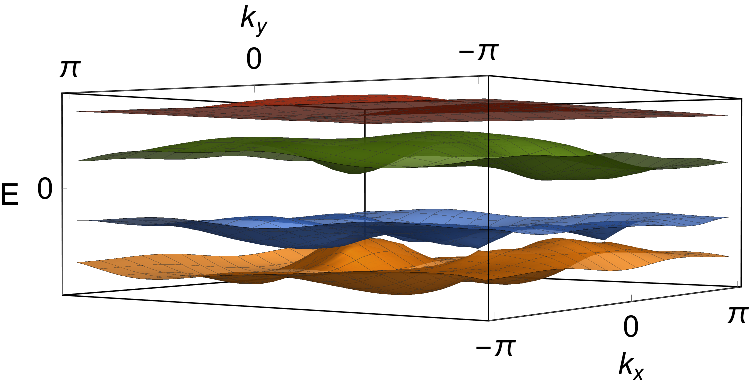} &
    \includegraphics[width=0.27\linewidth]{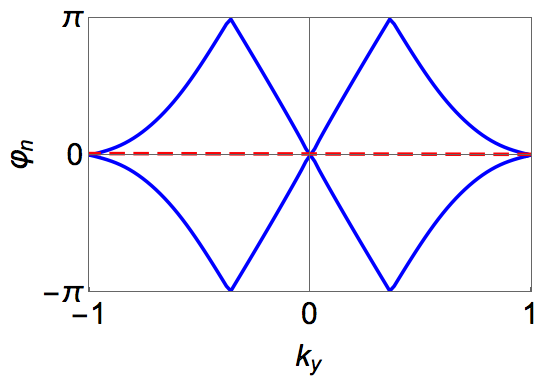} &
    \includegraphics[width=0.19\linewidth]{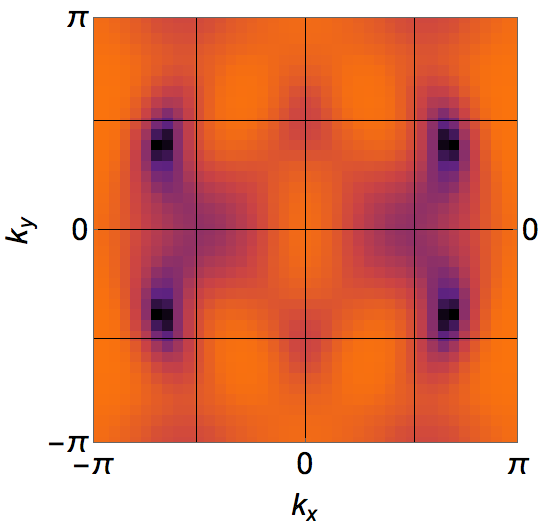}\\
    \multicolumn{1}{l}{(e)\quad $(q_+,q_-)=(2,1)$} &
    $(\chi_I,\chi_{II}) = (1,3)$ & 
    $\#_{\mathrm{NP}}[\mathcal{B}_{II}] = 2\chi_{II}= 6$\\
    \includegraphics[width=0.33\linewidth]{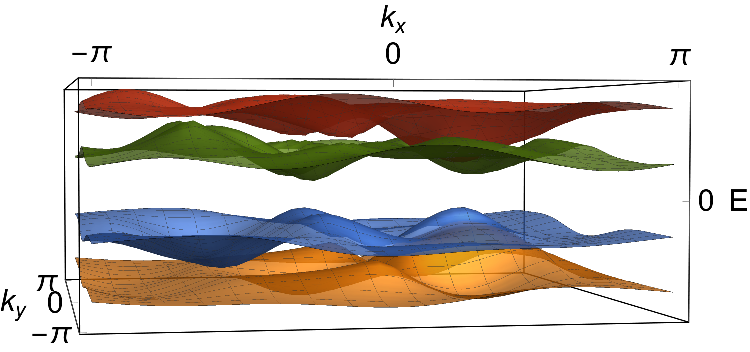} &
    \includegraphics[width=0.27\linewidth]{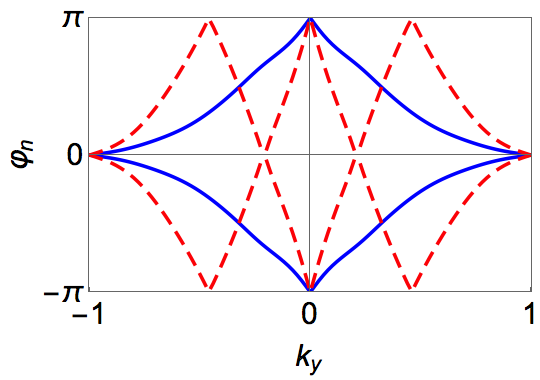} &
    \includegraphics[width=0.19\linewidth]{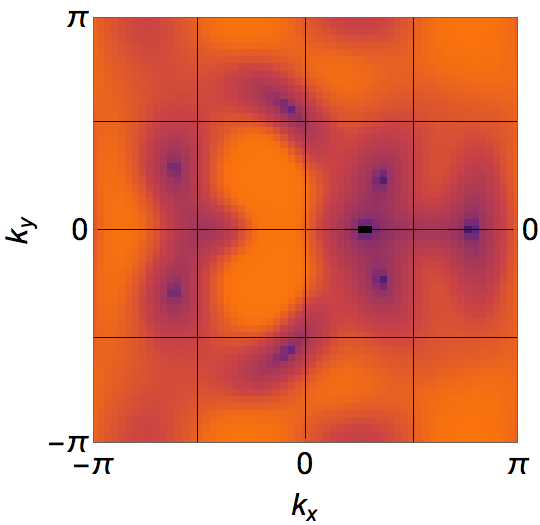}
\end{tabular}
\caption{\label{4bands} Band structures (left column) associated to the real oriented vector bundles $\mathcal{E}^{(q_+,q_-),+}_{2,4} $ based on the Grassmannian $\mathrm{Gr}^+_{2,4}$. Wilson loop flow (middle column) of the lower two-band subspaces (blue) and, when different, of the higher two-band subspaces (dashed red). The Wilson loop winding gives the reduced Euler class $\vert \chi_{I,II}\vert$. The correspondence between the geometric construction and the topology follows the bijection $(\chi_I,\chi_{II}) = (q_+ - q_- , q_+ + q_-)$. Location of the $\#_{\mathrm{NP}} = 2\chi_i$ nodal points of the $i$-th two-band subspace, $i=I$ in panels (a-d) and $i=II$ in panel (e) (right column, dark for nodes and orange for large band separation within the subspace). The tight-binding models have been generated with the \texttt{Mathematica} code downloadable from Ref.~\cite{abouhon_EulerClassTightBinding}.} 

\end{figure*}

We show in the left column of Fig.~\ref{4bands} band structures of truncated tight-binding models (for details see Appendix~\ref{app:tb} and Ref.~\cite{abouhon_EulerClassTightBinding}) associated to $\mathcal{E}^{(q_+,q_-)}_{2,4}$ for different combinations of $(q_+,q_-)$, which were generated in a similar way as the three-band models discussed in Sec.~\ref{3_bands}. By construction, each band structure is composed of two two-band subspaces, $\mathcal{B}^+_I(2)$ and $\mathcal{B}^+_{II}(2)$. The topology of each oriented subspace is characterized by the Euler class, i.e.~$\chi_{I} = \chi[\mathcal{B}^+_I(2)] \in \mathbb{Z}$ and $\chi_{II} = \chi[\mathcal{B}^+_{II}(2)] \in \mathbb{Z}$, which we compute here as a two-band Wilson loop winding~\cite{bouhon2018wilson} (middle column of Fig.~\ref{4bands})\footnote{\unexpanded{The Wilson loop winding gives the reduced Euler class $\vert \chi\vert$. The \emph{integer} invariant can be obtained from the winding of the Pfaffian of the Wilsonnian Hamiltonian Ref.~\cite{bouhon2019nonabelian}.}}.

Both Euler classes are directly determined by the numbers $(q_+,q_-)$. While $(q_+,q_-)$ takes value in a free group, the pair of Euler classes $(\chi_I,\chi_{II})$ must satisfy the sum rule $(\chi_I+\chi_{II})\mod 2 =0$~\cite{BzduSigristRobust, BJY_nielsen}.\footnote{This follows from the sum rule for the second Stiefel-Whitney class of the total bundle, namely $w_2[\mathcal{B}_I(2)\oplus \mathcal{B}_{II}(2)] = (w_2[\mathcal{B}_I(2)] + w_2[\mathcal{B}_{II}(2)])\mod 2 = (\chi[\mathcal{B}_I(2)] + \chi[\mathcal{B}_{II}(2)] ) \mod 2 = 0$.} From the data presented in Fig.~\ref{4bands}, we conclude that there is the following homomorphism of groups from the homotopy invariants to the cohomology invariants,  
\begin{equation}
   m : (q_+,q_-) \rightarrow 
   \left\{
   \begin{array}{rcl} 
        \chi_I &=& q_+ - q_- \, \\
        \chi_{II} &=& q_+ + q_- \,.
   \end{array}\right.
\end{equation} 
We explain in Appendix~\ref{ap_1} how the homomorphism $m$ follows from the chosen parametrization of $\mathsf{SO}(4)$. We emphasize that while the parity of the Euler classes $(\chi_I,\chi_{II})$ must be equal (by virtue of the Whitney sum formula), their sum \emph{does not} need to vanish.

As anticipated, we again observe that the number of stable nodal points within each two-band subspace follows $\#_{\mathrm{NP}}[\mathcal{B}^+_i(2)] = 2\vert\chi_{i}\vert$, $i=I,II$, see the right column of Fig.~\ref{4bands} showing the location of nodal points (black points) of one to the two-band subspaces. This does not prevent accidental pairs of nodal points as is found in Fig.~\ref{4bands}(e) which exhibits 8 nodal points in the unoccupied two-band subspace while the minimum of only 6 stable nodal points is expected.

Beyond the phases that are represented in Fig.~\ref{4bands}, all the other topologically nontrivial phases within $0 \leq  \vert \chi_{I}\vert,\vert\chi_{II}\vert  \leq 3$ can readily be obtained through the transformations 
\begin{equation}
    \begin{array}{c rcr c rcr}
        (i) & (q'_+,q'_-) &=& (q_+,-q_-) & \mathrm{for} &  (\chi'_{I},\chi'_{II}) & = & (\chi_{II},\chi_{I}), \\
        (ii) & (q'_+,q'_-) &=& -(q_+,q_-) & \mathrm{for} & (\chi'_{I},\chi'_{II}) &=& -(\chi_{I},\chi_{II}) ,\\
        (ii) & (q'_+,q'_-) &=& (q_-,q_+) & \mathrm{for} & (\chi'_{I},\chi'_{II}) &=& (-\chi_{I},\chi_{II}) .
    \end{array}
\end{equation}
(The topologically trivial case with $\chi_{I,II}=0$ is easily obtained as a constant Hamiltonian, and therefore not listed in Fig.~\ref{4bands}.) 

\subsection{Dropping of orientation}

The topology of orientable band structures is classified by the free homotopy classes (see Sec.~\ref{subsec:second}) for which there is no canonical definition of an orientation. Therefore, the orientation assumed so far must be dropped. This has the effect of the following reduction of the homotopy classification of band structures (see Sec.~\ref{sec:flag-2})
\begin{equation}
    \;  \mathbb{Z}\oplus \mathbb{Z} \rightarrow [\mathbb{Z}\oplus \mathbb{Z}]/\triangleright_\ell \;,
\end{equation} 
where the quotient set is defined through the equivalence relation given by the automorphism $\triangleright_\ell (\beta_+,\beta_-) =  -(\beta_+,\beta_-)$ that reverses the orientations of both subbundles $\mathcal{B}^+_I(2)$ and $\mathcal{B}^+_{II}(2)$. We give in Appendix \ref{ap_3} an explicit example of a continuous deformation of the Hamiltonian that reverses the Euler class of oriented subbundles of models with two occupied and two unoccupied bands.

This implies that any two phases with, on the one hand, $(\chi_I,\chi_{II})$ and, on the other hand, $(-\chi_I,-\chi_{II})$, belong to the same homotopy class, which we write \begin{equation}
    \; (\chi_I,\chi_{II}) \sim (-\chi_I,-\chi_{II}) \;.
\end{equation} 
On the contrary, 
\begin{equation}
  \;   (\chi_I,\chi_{II}) \nsim (\chi_I,-\chi_{II}) \sim (-\chi_I,\chi_{II}) \;.
\end{equation} 

Given the sum rule of Euler classes, and given the above reduction, we readily obtain the following list of all equivalence classes of topologically nonequivalent phases that are bounded by $0 \leq  \vert \chi_{I}\vert,\vert\chi_{II}\vert  \leq 3$, and written in terms of Euler class, i.e.
\begin{equation}
    (\chi_I, \chi_{II}) : \left\{
        \begin{array}{rcl}
            (0,0) & & \\
             (1,1) & \sim &(-1,-1) \\
             (2,0) & \sim &(-2,0) \\
             (0,2) & \sim &(0,-2) \\
             (1,3) & \sim & (-1,-3) \\
             (3,1) & \sim & (-3,-1) \\
             (-1,3) & \sim & (1,-3) \\
             (-3,1) & \sim & (3,-1) \\
             (3,3) & \sim & (-3,-3) \\
             (-3,3) & \sim & (3,-3) 
        \end{array}
    \right. \,,
\end{equation}
of which Fig.~\ref{4bands} presents only a subset.

\subsection{Fragile topology of four-band models}\label{sec:4-fragile}

We conclude this section by commenting on the fragility through a repartitioning of the bands. Let us start from the band structure with $(\chi_I,\chi_{II}) = (2,0)$ [Fig.~\ref{4bands}(d)]. As indicated by their Euler class, the two higher bands are connected by a minimum of \emph{zero} nodal points, and as such they can be separated by an energy gap [as readily visible in Fig.~\ref{4bands}(d)] thus enabling a finer partitioning $\mathcal{B}_I(2)\cup \mathcal{B}_{IIa}(1)\cup\mathcal{B}_{IIb}(1) $. Then we can lower the band $IIa$ in energy and close the gap with the block $I$, leading to the new partitioning $\mathcal{B}'_I(3)\cup \mathcal{B}'_{II}(1)\cong \mathcal{B}_I(2)\oplus \mathcal{B}_{IIa}(1)\cup \mathcal{B}_{IIb}(1) $. The (oriented) classifying space corresponding to the repartitioned bands
is thus $\mathrm{Gr}^+_{3,4} = \mathbb{S}^3$. Then $\pi_2[\mathbb{S}^3] = 0$ tells us that the nontrivial fragile topology of the occupied two-band subspace ($\chi_{I} =2$) has been trivialized, in agreement with $w_2[\mathcal{B}'_I(3)] = w_2[\mathcal{B}_I(2) \oplus \mathcal{B}_{IIa}(1)] = (w_2[\mathcal{B}_I(2)] + w_2[\mathcal{B}_{IIa}(1)])\mod 2 = \chi_{I} \mod 2+ 0 =0$. 

The same conclusion can alternatively be obtained by considering the tangent bundle to the 3-sphere which is associated to the classifying space $\mathbb{S}^3$. We explain in Sec.~\ref{sec:generalizations} that oriented vector subbundles pulled back from $\mathcal{F}^+_{3,4} \cong T\mathbb{S}^3 $, i.e.~the tangent bundle of the 3-sphere, are classified by $\pi_d[\mathbb{S}^3]$ where $d$ is the dimensionality of the system. Then, since $\pi_2[\mathbb{S}^3] = 0$, the two-dimensional restriction of any vector subbundle pulled back from $\mathcal{F}^+_{3,4}$ must be fully trivial.

\section{Generalizations}\label{sec:generalizations}
We now briefly highlight how our construction can be extended in certain directions, underpinning the generality of the presented geometric framework.

\subsection{Band structures with multiple gaps}\label{multigap}
As a first direction, we can generalize  to systems with more bands and with multiple energy gaps, as has been briefly foreshadowed in Secs.~\ref{sec:flag-1} and~\ref{sec:flag-2}. We discuss here the concrete example of $\mathrm{Fl}_{2,2,2}$ for the orientable phases. In this case the total gapped bundle is composed of three vector subbundles, $\mathcal{E}_{2,2,2;6} = \mathcal{B}_I(2) \cup \mathcal{B}_{II}(2) \cup \mathcal{B}_{III}(2) $. First lifting the problem to the oriented bundle, we have 
\begin{equation}
\pi_2[\mathrm{Fl}^+_{2,2,2}] = \mathbb{Z}\oplus \mathbb{Z} \oplus \mathbb{Z}\ni (c_1,c_2,c_3).
\end{equation}
The two-band subspaces are characterized by an Euler class, $\chi_i=\chi[\mathcal{B}_i(2)] \in \mathbb{Z}$, $i=I,II,III$, which must satisfy the sum rule $[\chi_I + \chi_{II} + \chi_{III}] \;\mod 2 = 0$, which follows directly from the Whitney sum formula and from the triviality of the total Bloch bundle. There is then a homomorphism from the homotopy invariants to the cohomology invariants,
\begin{equation}
\chi_I = c_2+c_3\; , \quad \chi_{II} = c_3+c_1 \;, \quad \chi_{III} = c_1+c_2\;. 
\end{equation}

The classification of orientable phases, as opposed to oriented ones, is then obtained by dropping the orientation. This is obtained by taking the quotient
\begin{equation}
   \; [\mathbb{S}^2,\mathrm{Fl}_{2,2,2}] = \pi_2[\mathrm{Fl}_{2,2,2}]/\pi_1[\mathrm{Fl}_{2,2,2}]\;, 
\end{equation}
where the first homotopy group
is $\pi_1[\mathrm{Fl}_{2,2,2}] = \mathbb{Z}_2 \oplus \mathbb{Z}_2$ with elements corresponding to the four Berry phase configurations \begin{multline}
(\gamma_I,\gamma_{II},\gamma_{III}) \in \\ \{(0,0,0),(\pi,\pi,0),(\pi,0,\pi),(0,\pi,\pi)\}\mod 2\pi.
\end{multline}
After dropping the orientation, we hence obtain the following homotopy equivalence expressed through the Euler class of each two-band subspace, i.e.
\begin{equation}
\begin{aligned}
    (\chi_I,\chi_{II},\chi_{III}) &\sim (-\chi_I,-\chi_{II},\chi_{III}) \\
    &\sim (-\chi_I,\chi_{II},-\chi_{III}) \\
  &\sim  (\chi_I,-\chi_{II},-\chi_{III}).
  \end{aligned}
\end{equation}
While the above results constitute a complete homotopy classification of orientable band structures associated to $\mathrm{Fl}_{2,2,2}$, the flag manifold $\mathrm{Fl}^+_{2,2,2}$ has a dimension of $12$ which makes explicit parametrization challenging. 

\subsection{Higher-dimensional fragile topologies}\label{subsec:higher_dimensional}

Our construction, relating the nontrivial tangent bundle of the sphere and the nontrivial fragile topological band structures of three-level systems, can be straightforwardly generalized to other dimensions. Indeed, any band structure with $p$ occupied bands and one unoccupied band has the classifying space $\mathbb{R}\mathrm{P}^p$, or $\mathrm{Gr}^+_{p,p+1} \cong \mathbb{S}^p$ when the orientation is artificially fixed. The total tautological bundle is then $\mathcal{T}^+_{p,p+1} = \mathcal{F}^+_{p,p+1} \cup \mathcal{F}^+_{1,p+1} \cong T\mathbb{S}^p \cup N\mathbb{S}^p$, where the tautological rank-$p$ subbundle is the tangent bundle of the $p$-sphere, $\mathcal{F}^+_{p,p+1} \cong T\mathbb{S}^p$, and the tautological line bundle is the normal bundle of the $p$-sphere, $\mathcal{F}^+_{1,p+1} \cong N\mathbb{S}^p$. 

Fixing $d$ as the dimensionality of the system, we focus on the orientable phases (note that the definition of orientability of Sec.~\ref{sec:orientability} generalizes to an arbitrary dimension of the base space). Then, generalizing the two-dimensional case discussed in Sec.~\ref{3_bands} ($d=p=2$), in general the Hamiltonian defines a map $\eta_q^{(d,p)}= f^{(d,p)} \circ t^{(d)}_q  :\mathbb{T}^d \rightarrow \mathbb{S}^d \rightarrow \mathbb{S}^p$ where $t^{(d)}_q$ wraps $\mathbb{T}^d$ on $\mathbb{S}^d$ $q$-times, and $f^{(d,p)}(\mathbb{S}^d) $ is an element of $\pi_d[\mathbb{S}^p]$. Any associated total gapped bundle is thus obtained as the pullback bundle $\mathcal{E}^{q+,(d)}_{p,p+1} = (f^{(d,p)}\circ t^{(d)}_q)^* \mathcal{T}^+_{p,p+1}$ with a topology that is classified both by $\pi_d[\mathbb{S}^d]$ (corresponding to the classification of maps $t_q^{(d)}: \mathbb{T}^d \to \mathbb{S}^d$) and by $\pi_d[\mathbb{S}^p]$ (for the classification of maps $f^{(d,p)}: \mathbb{S}^d \to \mathbb{S}^p$).
We conclude that in general the topology is captured by the direct sum
\begin{equation}
    [\mathcal{E}^{q+}_{p,p+1}] \cong \pi_d[\mathbb{S}^d] \oplus \pi_d[\mathbb{S}^p] \;.
\end{equation}
In the special case $d=p$, the topology is classified by the single homotopy group $\pi_d[\mathbb{S}^d]$, similar to Sec.~\ref{geo}. When $d<p$, we have $\pi_d[\mathbb{S}^p] = 0$ and all associated total gapped bundles have a trivial topology. This explains the triviality achieved by the repartitioning of bands as discussed in Sec.~\ref{sec:4-fragile}. On the contrary when $d>p$, we predict a richer classification than the examples studied so far in this work. One well-known example is the Hopf bundle and the associated Hopf insulators obtained for $d=3$ and $p=2$ \cite{Hopf_1,Hopf_2, Unal2019,Alex:2019}. 

We conclude this section by a generalization of Sec.~\ref{subsec:triviality_NS} that addressed the unstable non-trivial homotopy of the normal bundle of the sphere $N\mathbb{S}^2$. It is a classical result of algebraic topology that the tangent bundles $T \mathbb{S}^d$ of the $d$-sphere for $d=1,3,7$ are fully parallelizable, i.e.~smooth global sections (vector fields) can be formed over these spheres. Nevertheless, since $\pi_d[\mathbb{S}^d] = \mathbb{Z}$ for all $d$'s, it readily follows from our construction that topologically nontrivial bundles can be obtained as the pullback of parallelizable tangent bundles. This again points to the finer topological content of the homotopy classification as compared to the cohomology (i.e.~stable) characterization of vector bundles.

The higher dimensional topologies we are alluding to here will be studied comprehensively elsewhere.

\section{Conclusion and discussion}\label{sec:conclusions}

We have provided a geometric perspective on fragile topological phases beyond symmetry indicators, while also addressing structures that emerge when a refined partitioning of bands is taken into account. This topological analysis also underlies the novel braiding properties found in \cite{Wu1273, bouhon2019nonabelian} when nodes of different bands are considered.
The framework rests on direct homotopy evaluations of the relevant Grassmannians using a geometrical construction, which involves the so-called Pl\"ucker embedding into the more manageable projective exterior product spaces. This construction does not only provide descriptive power, enumerating the possible topologies on a generic footing, but in reverse also allows for a direct construction of rather tractable models displaying the desired topological features. These explicit Hamiltonian models provide a valuable platform to investigate the possible experimental signatures of fragile topology beyond symmetry indicators. In this regard, we point out the very recent Ref.~\cite{Unal_quenched_Euler}, in which some of the present authors investigated the expected fingerprints of the 3-band Euler-insulator models of Sec.~\ref{3_bands} in cold-atoms setups.

We conclude with outlining prospective theoretical generalizations. Firstly, there are the extensions already outlined in Sec.~\ref{sec:generalizations}, namely the generalization to fragile topology in the presence of multiple bands gaps on the one hand (Sec.~\ref{multigap}), and the generalization to higher-dimensional spaces on the other hand (Sec.~\ref{subsec:higher_dimensional}). Although in the present work we only consider real-symmetric Hamiltonians, such as in the presence of $C_2\mathcal{T}$ or $\mathcal{PT}$ symmetry, the finer repartitioning of bands discussed in Sec.~\ref{sec:flag-1} can similarly be considered for complex Hamiltonians classified by complex flag varieties.

Related to the generalization to higher dimensions, when the third momentum is played by the time direction of a periodic drive, it appears that the language of flag varieties may provide a natural language to describe other classes of topological systems, especially in the case of periodically driven Floquet systems~\cite{Floquet1,Floquet2,Floquet3}. Here, the periodicity of the quasienergy implies that there is no canonical choice of chemical potential. One therefore often assigns the same importance to \emph{all} spectral gaps of Floquet systems~\cite{Sun:2018b}, suggesting a very natural application for the language of refined band partitioning developed in Sec.~\ref{sec:partitioning}. 

Another promising application of the refined band partitioning and of fragile topology arises in the context of non-Hermitian models~\cite{modes, kawabatanonhermitian, zhounonhermitian}, where non-standard gap conditions were recently investigated using homotopy theory~\cite{Sun:2019,li:2019}. Indeed, as noted in Sec.~\ref{3_bands}, the topology of a generic two-band non-Hermitian Hamiltonian has been shown by Ref.~\cite{Wojcik:2019} to be essentially equivalent to the fragile topology of $3$-band real-symmetric Hamiltonians discussed here. Furthermore, in analogy with the non-Abelian reciprocal braiding of band nodes in real-symmetric Hamiltonians~\cite{Wu1273,bouhon2019nonabelian}, refined band partitioning in non-Hermitian models has been shown to also facilitate noncommutative exchange of exceptional points inside momentum space~\cite{ZhongNOnHerm}.

The final extension, which is of particular importance for the study of materials, concerns the interplay with crystalline symmetries. We have shown in Ref.~\cite{bouhon2018wilson} that a point group of crystalline symmetries can lead to an obstruction on the Wilson loop winding (Euler class) of two-band subspaces. This was proved to be directly rooted in the representation theory of the Wilson loop. All the observations we made here should have a similar natural explanation from the exhaustive topological classification of band structures and their explicit realization. We will report on this in due time at another occasion. 

\section{Acknowledgements}
R.-J.~S.~acknowledges funding from Trinity college, the Marie Curie programme under EC Grant agreement No.~842901 and the Winton programme at the University of Cambridge.
T.~B.~was supported by an Ambizione Grant No.~185806 from the Swiss National Science Foundation.

\appendix

\section{Geometric and topological properties of Grassmannians}\label{ap:homotopy_grassmannian}

We review here a few basic facts about Grassmannians that motivate the results of their homotopy groups and that allow us to formulate in a rigorous way the action of $\pi_1[\mathrm{Gr}_{p,N}]$ on $\pi_2[\mathrm{Gr}_{p,N}]$ in Appendix \ref{ap:homotopy_action}. 

Let us start from the definition of the Grassmannian in Sec.~\ref{subsec:un_oriented_grass}. $\mathsf{O}(N)$ is a Lie group and $\mathsf{O}(p) \times \mathsf{O}(N-p)$ is a Lie subgroup (by Cartan's closed subgroup theorem \cite{Lee_SM}), so that the unoriented Grassmannian is a smooth (and closed \cite{Hatcher_2}) manifold. Then, representing $\mathrm{Gr}_{p,N}$ as the space of (non-oriented) $p$-dimensional hyperplanes passing through the origin in $\mathbb{R}^N$, it readily follows that the Grassmannian is path-connected, and thus connected, as any two hyperplanes (i.e.~two points of the Grassmannian) can be smoothly rotated into each other. The same conclusions hold similarly for the oriented Grassmannian. 

There is a continuous two-to-one surjective map from the oriented Grassmannian to the unoriented Grassmannian,  
\begin{equation}
\label{eq:covering_map}
    \bar{q} : \mathrm{Gr}^+_{p,N}\rightarrow \mathrm{Gr}_{p,N}: \{[R]^{+},[R^\mathrm{sr}]^{+}\} \mapsto [R] \;,
\end{equation}
where we assume $R\in \mathsf{SO}(N)$, in particular with $\bar{q}([\mathbb{1}_N]^+) = \bar{q}([\mathbb{1}_N^\mathrm{sr}]^+) = [\mathbb{1}_N]$ (see Eqs.~(\ref{eq:leftcoset}) and (\ref{eq:leftcoset_oriented})). The map $\bar{q}$ is a covering map and $\mathrm{Gr}^+_{p,N}$ a covering space of the base space $\mathrm{Gr}_{p,N}$ \cite{Lee_SM}.\footnote{\unexpanded{We remark the relative freedom in defining the covering map as we could have shifted the image as $\bar{q}([\mathbb{1}_N]^+) = \bar{q}([\mathbb{1}_N^\mathrm{sr}]^+) = [R]$ with an arbitrary $R \in \mathsf{SO}(N)$, since there is no favored choice of origin for a Grassmannian. This freedom does not play any role in the topology of Grassmannian though.}} Therefore, the oriented Grassmannian $\mathrm{Gr}^+_{p,N}$ is the \textit{orientable double cover} of the unoriented Grassmannian $\mathrm{Gr}_{p,N}$. It is interesting to note that the non-orientability of the connected $\mathrm{Gr}_{p,N}$ is a necessary and sufficient condition for the connectedness of $\mathrm{Gr}^+_{p,N}$, i.e.~the orientable double cover is made of a single piece, as it can be readily seen in the example of $\mathbb{S}^2\rightarrow \mathbb{R}\mathrm{P}^2$ shown in Fig.~\ref{double_cover_RP2} (that is to be contrasted e.g.~with the double cover group $\mathsf{O}(N)$ of $\mathsf{SO}(N)$). Furthermore, we take for granted that the oriented Grassmannian is simply connected (i.e.~$\pi_1[\mathrm{Gr}^+_{p,N}]=0$ \cite{Hatcher_1,Hatcher_2}) which makes it the universal cover of the unoriented Grassmannian \cite{Lee_SM}. 

Further important information on the homotopy of Grassmannians is readily obtained from the lifts of continuous maps through the covering map $\bar{q}$. The \textit{lifting criterion} says that every continuous map $f:\mathbb{S}^2\rightarrow \mathrm{Gr}_{p,N}$ lifts to a continuous map $f^+:\mathbb{S}^2\rightarrow \mathrm{Gr}^+_{p,N}$ with $f = \bar{q}\circ f^+$, such that the lift $f^+$ is uniquely defined once a reference image point is fixed, e.g.~$f^+(k_0) = [R_0]^+$ \cite{Lee_SM,Hatcher_1}. It further follows that every map $f^+ :\mathbb{S}^2\rightarrow \mathrm{Gr}^+_{p,N}$ projects to a map $\bar{q}\circ f^+ : \mathbb{S}^2\rightarrow \mathrm{Gr}_{p,N}$. 

A lift can be composed with an automorphism of the covering map $\bar{q}$, called a \textit{deck transformation}, defined by an homeomorphism $\varphi_{\bar{q}} : \mathrm{Gr}^+_{p,N} \rightarrow \mathrm{Gr}^+_{p,N}$ such that $\bar{q}\circ\varphi_{\bar{q}} = \bar{q}$ \cite{Lee_SM}. There are only two possible choices for $\varphi_{\bar{q}}$ given a fixed $\bar{q}$, namely the identity, i.e.~with $\varphi_{\bar{q}} ([\mathbb{1}_N]^+) = [\mathbb{1}_N]^+$, and the orientation reversal [Eq.~(\ref{eq:orientation_reversal_sr})], i.e.~with $\varphi_{\bar{q}} ([\mathbb{1}_N]^+) = [\mathbb{1}^\mathrm{sr}_N]^+$. (This freedom plays the role of a gauge symmetry of the gapped Hamiltonians.) Let us summarize the above definitions in one diagram
\begin{equation}\label{diagram:lift_automorphism}
\begin{tikzcd}[]
            &
			\mathrm{Gr}^+_{p,N}  \arrow[r, "\mathrm{sr}"] \arrow[d, "\bar{q}"]	& \arrow[dl, "\bar{q}"]
			\mathrm{Gr}^+_{p,N}  \\
\mathbb{S}^2 \arrow[ru, "f^+"] \arrow[r, "f"] &
			\mathrm{Gr}_{p,N}
						& 
\end{tikzcd}.
\end{equation}
Since $\mathrm{sr}^2 = \mathrm{id}$, the set of automorphisms of $\bar{q}$ form the \textit{automorphism group of the covering} $\mathrm{Aut}_{\bar{q}}[\mathrm{Gr}_{p,N}] =\{\mathrm{id},\mathrm{sr}\}$.

The lift of one point, say $[R(\boldsymbol{k}_0)]=[R_0]$, is called the fiber and is here given by $\bar{q}^{-1}([R_0]) = \{[R_0]^+,[R^{\mathrm{sr}}_0]^+\}$. Let us transfer the lifting property to the homotopy classes $[f]_{[R_0]} \in \pi_2[\mathrm{Gr}_{p,N}]$ and $[f^+]_{[R_0]^+} \in \pi_2[\mathrm{Gr}^+_{p,N}]$ with the respective base points $[R_0]$ and $[R_0]^+$. We have $\bar{q}_*[f^+]_{[R_0]^+} = [\bar{q}(f^+)]_{[R_0]} = [f]_{[R_0]}$ and $\bar{q}^{-1}_*[f]_{[R_0]} = \{[f^+]_{[R_0]^+}, [\text{sr}(f^+)]_{[R^{\mathrm{sr}}_0]^+}\}$. Fixing the base point of the lifted map, say $[R_0]^+$, we thus have a bijection of homotopy classes leading to the isomorphism \cite{Hatcher_1}
\begin{equation}\label{eq:iso_pi2}
    \pi_2[\mathrm{Gr}_{p,N}] \cong \pi_2[\mathrm{Gr}^+_{p,N}] \;.
\end{equation}
This isomorphism lies at the core of our strategy to built homotopy-based tight-binding models in Sec.~\ref{geo}.

We define the action of a nontrivial element $[\ell] \in \pi_1[\mathrm{Gr}_{p,N}]$ (with the base point $[R_0]$) on the fiber $\bar{q}^{-1}([R_0])$ in terms of the end point of the lifted paths $\ell^+_{a,b}$, i.e.~$[R_0]^+ \cdot [\ell] = \ell^+_a(1) = [R_0^{\mathrm{sr}}]^+$ and $[R_0^{\mathrm{sr}}]^+ \cdot [\ell] = \ell^+_b(1) = [R_0]^+$, called the \textit{monodromy action} of the fundamental group on the fiber of the covering map. In our case, it is equivalent to the restriction of the deck transformation `sr' to the fiber over a point. 

Let us now consider the first homotopy group of the unoriented Grassmannian. Since $\mathrm{Gr}^+_{p,N}$ is connected, there exists a continuous path $\ell^+_a:[0,1]\rightarrow \mathrm{Gr}^+_{p,N}:s\mapsto \ell^+_a(s)$ connecting any element $[R]^+ = \ell^+_a(0)$ to its reversal partner $[R^\mathrm{sr}]^+=\ell^+_a(1)$. By projecting onto $\mathrm{Gr}_{p,N}$, the path defines a non-contractible loop $\ell = \bar{q}(\ell^+_a)  \subset \mathrm{Gr}_{p,N}$. Inversely, the lift of a non-contractible loop $\ell\subset \mathrm{Gr}_{p,N}$ produces an open path in $\mathrm{Gr}^+_{p,N}$ which end points have reversed (subframe) orientation. This directly captures the non-orientability of $\mathrm{Gr}_{p,N}$. (That also characterizes non-orientable vector subbundles $\mathcal{B}_I\vert_{l}$ over a base loop  $l$ as discussed in Sec.~\ref{sec:homotopy-groups}.) Furthermore, the lift can be composed with the nontrivial deck transformation `$\mathrm{sr}$', from which we get the path $\ell^{+}_b =\mathrm{sr}(\ell^+_a)$ with $\ell^{+}_b(0) = \ell^{+}_a(1) $ and $\ell^{+}_b(1) = \ell^{+}_a(0) $. The open path $\ell^{+}_b$ then projects as $\ell = \bar{q}(\ell^{+}_b)$. If we now take the composed loop $\ell\cdot\ell \subset \mathrm{Gr}_{p,N}$ (contrary to the composition of functions, we read the composition of paths/loops from left to right), it lifts to a closed loop $(\ell\cdot\ell)^+ = \ell^+_a \cdot \ell^+_b = \ell^+_a  \cdot \mathrm{sr}(\ell^+_a)\subset \mathrm{Gr}^+_{p,N}$ which is contractible to a point since $\pi_1[\mathrm{Gr}^+_{p,N}]=0$. Thus $\ell \cdot \ell$ is itself contractible by continuity of the covering map $\bar{q}$. We conclude that the unoriented Grassmannian has a non-trivial fundamental group $\pi_1[\mathrm{Gr}_{p,N\geq 3}] = \mathbb{Z}_2$ \cite{BzduSigristRobust}. It is actually true in general that the automorphism group of a universal covering is isomorphic to the fundamental group of the space being lifted, i.e.~$\mathrm{Aut}_{\bar{q}}[\mathrm{Gr}^+_{p,N}] \cong \pi_1[\mathrm{Gr}_{p,N}]$ \cite{Lee_SM}.

We illustrate these properties with the example of the projective plane in Sec.~\ref{subsec:projective_plane}.

\section{Action of $\pi_1[\mathrm{Gr}_{p,N}]$ on $\pi_2[\mathrm{Gr}_{p,N}]$}\label{ap:homotopy_action}

\begin{figure*}[t!]
\centering
\begin{tabular}{c}
	\includegraphics[width=0.85\linewidth]{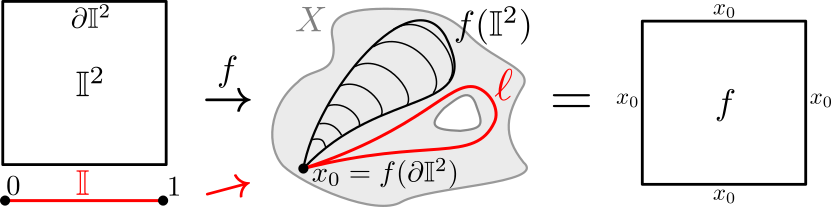}
\end{tabular}
\caption{\label{homotopy_square} Graphic representation of the second homotopy group $\pi_2[X]$, following \cite{Hatcher_1}, in terms of the continuous maps $f:(\mathbb{I}^2,\partial \mathbb{I}^2) \rightarrow (X,x_0)$ with the base point $x_0 = f(\partial \mathbb{I}^2 ) $ ($\mathbb{I} = [0,1]$ is the unit interval, and $\mathbb{I}^2$ the unit square). The homotopy equivalence classes of such maps $[f]$ constitute the elements of $\pi_2[X]$. The elements of of the first homotopy group $\pi_1[X]$ are defined as the homotopy equivalence classes $[\ell]$ of the maps (red) $\ell: (\mathbb{I},\partial \mathbb{I})\rightarrow (X,x_0)$ with the base point $x_0 = \ell(0) = \ell(1)$. }
\end{figure*}

This appendix gives the precise definition of the action of $\pi_1[\mathrm{Gr}_{p,N}]$ on $\pi_2[\mathrm{Gr}_{p,N}]$ which then leads to the results of Sec.~\ref{subsec:second}. For this we use the definition of the first, resp.~the second, homotopy groups in terms of continuous functions $\ell:(\mathbb{I},\partial \mathbb{I})\rightarrow (X,x_0)$ from the unit interval $\mathbb{I}=[0,1]$, resp.~continuous functions $f:(\mathbb{I}^2,\partial\mathbb{I}^2) \rightarrow (X,x_0)$ from the unit square $\mathbb{I}^2=[0,1]\times[0,1]$, to a topological space $X$, such that the boundaries $\partial \mathbb{I} = \{0,1\}$, resp.~$\partial \mathbb{I}^2$, are mapped to the base point $ x_0 = \ell(0) = \ell(1) = f(\partial \mathbb{I}^2) \in X$, see Fig.~\ref{homotopy_square} that follows Ref.~\cite{Hatcher_1}. Importantly, the base point is kept constant for all the maps and the orientation of the unit interval (square), seen as a subspace of the real vector space $\mathbb{R}^{1(2)}$, transfers to an orientation of the image $ \ell(\mathbb{I})$ ($f(\mathbb{I}^2)$) within the target space $X$. These allow the definition of composition of homotopies which then induces a group structure on the homotopy equivalence classes $[\ell] \in \pi_1[X]$ ($[f]\in \pi_2[X]$) \cite{Hatcher_1}. By removing the constraint of a fixed base point we obtain the free homotopy set $[\mathbb{I}^{d},X]\cong [\mathbb{S}^{d},X]$ which may lack a group structure. We show below that this is the case for the unoriented Grassmannian. 

\begin{figure*}[t]
\centering
\begin{tabular}{c}
	\includegraphics[width=0.85\linewidth]{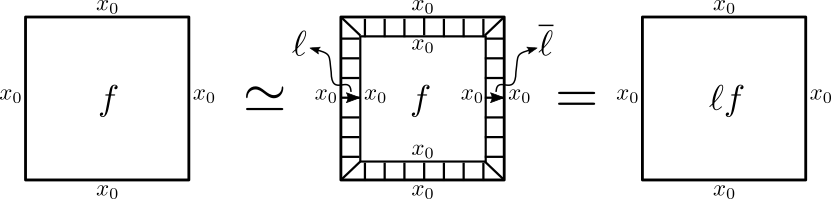}
\end{tabular}
\caption{\label{homotopy_base_point} Graphic representation of the homotopy $F_{\ell}:f\rightarrow \ell f $ which generates the displacement of the base point $x_0=\ell(0)=\ell(1)$ along the loop $\ell\subset \mathrm{Gr}_{p,N}$. While this is a homotopy equivalence within the free homotopy set $[\mathbb{S}^2,\mathrm{Gr}_{p,N}]$ ($f\simeq \ell f$), it can lead to a change of homotopy classes of the based homotopy group $\pi_2[\mathrm{Gr}_{p,N}]$.
} 
\end{figure*}

In the following we take $X=\mathrm{Gr}_{p,N}$ and $x_0 = [R(\boldsymbol{k}_0)] = [R_0]$.\footnote{\unexpanded{We can now motivate the lifting criterion presented in Appendix \ref{ap:homotopy_grassmannian}. By taking a one-dimensional cross-section of the mapping $f$ [Fig.~\ref{homotopy_square}], i.e.~$f\vert_l$ with $l\subset \mathbb{I}^2$ and $\partial l \in \partial\mathbb{I}^2$, we find $[f\vert_l] = [0] \in \pi_1[\mathrm{Gr}_{p,N}]$. We thus conclude that the image $f(\mathbb{I}^2)$ for any $[f] \in \pi_2[\mathrm{Gr}_{p,N}]$ is orientable, and $f$ can be lifted to a map into the orientable double cover, i.e.~$f^+:(\mathbb{I}^2,\partial \mathbb{I}^2)\rightarrow (\mathrm{Gr}^+_{p,N},x_0)$. (The gapped total bundle associated to $f$, interpreted as the classifying map, is thus orientable as well.)}} Let us define the homotopy \cite{Hatcher_1} $F_{\ell} : f \rightarrow \ell f$ as the pre-composition of $f$ with the displacement of the base point $x_0$ along the loop $\ell$ as shown in Fig.~\ref{homotopy_base_point}. $F_{\ell}$ induces the (right) action of $[\ell] $ on $[f]$ through the group automorphism
\begin{equation}
    \triangleright_{\ell}:\pi_2[\mathrm{Gr}_{p,N}]\rightarrow \pi_2[\mathrm{Gr}_{p,N}] :[f]\mapsto [f]\cdot [\ell] = [\ell f] \;.
\end{equation} 
While the homotopy $F_{\ell}$ preserves the homotopy classes in the free homotopy set $[\mathbb{S}^2,\mathrm{Gr}_{p,N}]$ (we write $f\simeq \ell f$), it can lead to a change of homotopy classes in $\pi_2[\mathrm{Gr}_{p,N}]$. It is however not obvious to see it from the above definition. It turns out that we can derive the effect of the action $\triangleright_{\ell}$ from the action of $[\ell]$ on the homotopy class $[f^+]\in \pi_2[\mathrm{Gr}^+_{p,N}]$ of the lifted map $f^+$, as done below. 

\begin{figure*}[t]
\centering
\begin{tabular}{c}
	\includegraphics[width=0.85\linewidth]{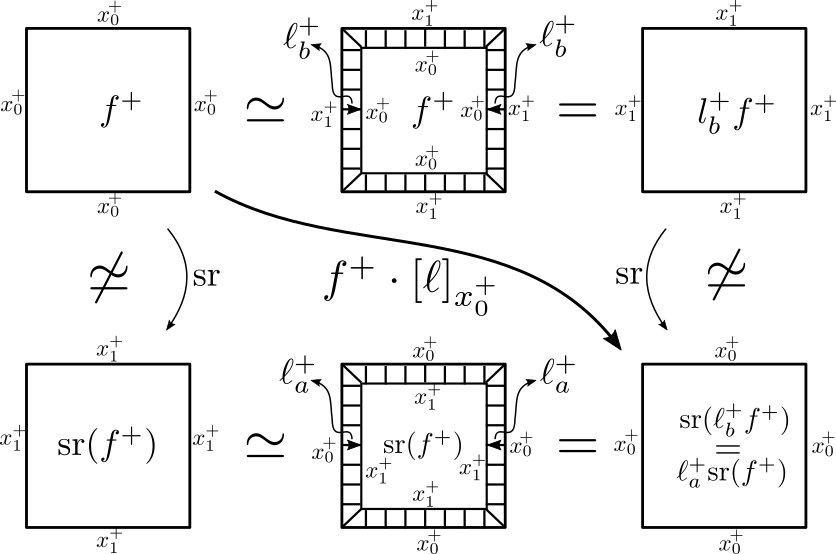}
\end{tabular}
\caption{\label{homotopy_base_point_lifted} Action of $[\ell]\in\pi_1[\mathrm{Gr}_{p,N}]$ on the lifted map $f^+:(\mathbb{I}^2,\partial \mathbb{I}^2) \rightarrow (\mathrm{Gr}^+_{p,N},x_0^+)$ with base point $x_0^+$, which we write $f^+\cdot [\ell]_{x_0^+}$. It is given as the composition of the based-point-changing homotopy $F_{\ell^+_b} : f^+ \rightarrow l^+_bf^+$ (corresponding to the monodromy action $ x_0^+ \cdot [\ell] = x_1^+$, see Appendix \ref{ap:homotopy_grassmannian}) and the nontrivial deck transformation `$\mathrm{sr}$' that reverses orientation. This induces the action of $[\ell]\in\pi_1[\mathrm{Gr}_{p,N}]$ on $[f^+]\in \pi_2[\mathrm{Gr}^+_{p,N}]$, given by $[f^+] \cdot [\ell]_{x_0^+} = [\text{sr}(l_b^+f^+)]_{x_0^+} = [l_a^+ \text{sr}(f^+)]_{x_0^+} = [\text{sr}(f^+)]_{x_1^+} $. We show in the text that it is a nontrivial automorphism of $\pi_2[\mathrm{Gr}^+_{p,N}]$ for $p=2$, and trivial otherwise.  
} 
\end{figure*}

The loop $\ell$ and the sphere image $f(\mathbb{I}^2)$ lift respectively to a path $\ell^+ \subset \mathrm{Gr}^+_{p,N}$ and a sphere image $f^+(\mathbb{I}^2) \subset \mathrm{Gr}^+_{p,N}$ [Appendix~\ref{ap:homotopy_grassmannian}]. We now can define the lifted homotopy $F^+_{\ell}$ as the homotopy of the lifted maps, i.e.~$F^+_{\ell} = F_{\ell^+} : f^+ \rightarrow \ell^+f^+ $ which is unique by the \textit{homotopy lifting property} \cite{Hatcher_1}. If $\ell_0$ is a contractible loop in $\mathrm{Gr}_{p,N}$, the lifted loop $\ell_0^+$ is then also contractible in $\mathrm{Gr}^+_{p,N}$ (see Appendix~\ref{ap:homotopy_grassmannian}), in which case there is a homotopy $f^+ \simeq \ell_0^+ f^+$ and $[f^+] = [\ell_0^+ f^+]$ in $\pi_2[\mathrm{Gr}^+_{p,N}]$. On the contrary, for a non-contractible loop $\ell$, we have seen in Appendix~\ref{ap:homotopy_grassmannian} that it lifts to a path $\ell^+_b$ connecting end points of opposite orientation in $\mathrm{Gr}^+_{p,N}$, which we write $x_0^+=\ell^+_b(0)$ ($\equiv [R(\boldsymbol{k}_0)]^+$) and $x_1^+ = \ell^+_b(1)$ ($ \equiv [R(\boldsymbol{k}_0)^{\text{sr}}]^+$). We thus have a based-point-changing homotopy $F_{\ell^+_b}:f^+\rightarrow \ell^+_bf^+$ according to the first row of Fig.~\ref{homotopy_base_point_lifted}. Since $\ell^+_bf^+$ has the base point $x_1^+$, it cannot be compared within $\pi_2[\mathrm{Gr}^+_{p,N}]$ to $f^+$ which has the base point $x_0^+$ [Fig.~\ref{homotopy_base_point_lifted}]. Applying the orientation reversal (the deck transformation `sr') gives $\text{sr}(\ell^+_bf^+) = \ell^+_a \text{sr}(f^+)$ with the base point $x_0^+$. Therefore, $f^+$ and $\ell^+_a\text{sr}(f^+)$ can now be compared within $\pi_2[\mathrm{Gr}^+_{p,N}]$. 

We thus define the action of $[\ell]$ on $f^+$ through 
\begin{equation}
    f^+\cdot [\ell]_{x_0^+} = \text{sr}\circ F_{\ell^+_b}(f^+) = \text{sr} (\ell^+_b f^+) = \ell^+_a \text{sr} ( f^+) \;,
\end{equation}
according to Fig.~\ref{homotopy_base_point_lifted}, which then induces the action of $[\ell]$ on $[f^+]_{x_0^+}$ as 
\begin{multline}
    \triangleright_{\ell}^+ : \pi_2[\mathrm{Gr}^+_{p,N}]\rightarrow \pi_2[\mathrm{Gr}^+_{p,N}] :\\ [f^+]_{x_0^+} \mapsto [f^+] \cdot [\ell]_{x_0^+} = [\ell^+_a \text{sr} ( f^+) ]_{x_0^+} \;.
\end{multline}
Projecting back onto $\mathrm{Gr}_{p,N}$, we get $\bar{q}(f^+\cdot [\ell]_{x_0^+}) = \bar{q}(\ell^+_a \text{sr}(f^+)) = \ell f$, since $\bar{q}(\text{sr}(f^+)) = \bar{q}(f^+) = f$, and thus $\bar{q}_*\triangleright_{\ell}^+([f^+]_{x_0^+}) = \triangleright_{\ell}([\bar{q}(f^+)]_{x_0}) = \triangleright_{\ell}([f]_{x_0}) $ as expected from Eq.~(\ref{eq:iso_pi2}).  

We now show that $\triangleright_{\ell}$ acts non-trivially in the case $p=2$. Before doing so though it is useful to briefly review the cohomology invariants introduced in Sec.~\ref{subsec:2D-cohomology-classes} that can be used as indicators of the stable homotopy classes of the lifted map $f^+$. On the one hand, the Euler class $\chi_{f^+_{p=2}} \in \mathbb{Z}$ can be used as an indicator of the (stable) homotopy class\footnote{The Euler class is defined for the rank-2 oriented subbundle $\mathcal{B}^+(2)$ of the gapped total bundle $\mathcal{E}^+_{2,N}$ associated to $f^+$, i.e.~the pullback by $f^+$ of the total tautological bundle $\mathcal{E}^+_{2,N} = (f^{+})^*\mathcal{T}^+_{2,N}$, see Sec.~\ref{geo}.} $[f^+_{p=2}]\in \pi_2[\mathrm{Gr}^+_{2,N\geq 5}]$, such that the sign $\sgn\{\chi_{f^+_{p=2}}\} \in \pm1$ defines faithfully the orientation of $f^+_{p=2}(\mathbb{I}^2)$. In other words, the reversal of the orientation of $f^+_{p=2}$ must also flip the sign of the Euler class, i.e.
\begin{equation}
\label{eq:Euler_reversal}
    \chi_{\text{sr}(f^+_{p=2})} =- \chi_{f^+_{p=2}} \;.
\end{equation}
On the other hand, the case $p=1$ is stably trivial (i.e.~$\pi_2[ \mathrm{Gr}^+_{1,N\geq 3}] =0$), and the case $p\geq 3$ is indicated by the $\mathbb{Z}_2$ second Stiefel-Whitney class which forgets orientation, see Sec.~\ref{subsec:2D-cohomology-classes}.

The second row of Fig.~\ref{homotopy_base_point_lifted} gives $\ell^+_a \text{sr}(f^+) \simeq \text{sr}(f^+)$, and from Eq.~(\ref{eq:Euler_reversal}) we find
\begin{align}
    \chi_{f^+_{p=2}\cdot [\ell]_{x_0^+}} = \chi_{\ell^+_a \text{sr}(f^+_{p=2})} = \chi_{\text{sr}(f^+_{p=2})} = -\chi_{f^+_{p=2}} \;.
\end{align}
Since the Euler class is an indicator of the stable homotopy class $[f^+_{p=2}]\in \pi_2[\mathrm{Gr}^+_{2,N}]$ and from Eq.~(\ref{eq:iso_pi2}) there is a one-to-one correspondence $[f^+_{p=2}]_{x_0^+} \rightarrow [f_{p=2}]_{x_0} $, we finally conclude 
\begin{equation}
\label{eq:pi1_action}
    \triangleright_{\ell}([f_{p=2}]) = [\ell f_{p=2}] = [f_{p=2}]^{-1} \;.  
\end{equation}

For $p\neq 2$, we either have $[f_{p=1}] = [0]$ or $[f_{p\geq 3}] \in \mathbb{Z}_2 $ that is indicated by the second Stiefel-Whitney class which is invariant under orientation reversal. Thus, $\triangleright_{\ell}$ acts nontrivially only when $p=2$. Since $[f_{p\neq 2}] = [f_{p\neq 2}]^{-1}$ though we can generalize Eq.~(\ref{eq:pi1_action}) to every $p$.

\section{Tight-binding models}\label{app:tb}
In order to get an explicit tight-binding model, we first sample $H(\boldsymbol{k})$ over a grid $\Lambda^*$ in the Brillouin zone and perform an inverse discrete Fourier transform ($FT$). This gives us the hopping matrix elements $t_{\mu\nu}(\boldsymbol{R}_j-\boldsymbol{0}) = FT[ \{H_{\mu\nu}(\boldsymbol{k}_m)\}_{m\in \Lambda^*}](\boldsymbol{R}_j-\boldsymbol{0}) $. Typically, the hopping elements decay exponentially and we can truncate them beyond a finite support including a few neighbors $\boldsymbol{R}_j$ around the center $\boldsymbol{0}$. The three-band example with Euler class $2$, and all the four-band examples shown in Sec.~\ref{4_bands} are truncated beyond the second neighbors in both directions, i.e.~$t_{\mu\nu}(\boldsymbol{R}_j-\boldsymbol{0}) = 0$ for all  $\boldsymbol{R}_j \in \{n_1 \boldsymbol{a}_1 + n_2 \boldsymbol{a}_2 \}_{n_1,n_2\neq 0,1,2}$, while the three-band example with Euler class $4$ has been truncated beyond the third neighbors in both directions, i.e.~$t_{\mu\nu}(\boldsymbol{R}_j-\boldsymbol{0}) = 0$ for all $\boldsymbol{R}_j \in \{n_1 \boldsymbol{a}_1 + n_2 \boldsymbol{a}_2 \}_{n_1,n_2\neq 0,1,2,3}$.  

The implementation in Wolfram Mathematica of the above algorithm is available at Ref.~\cite{abouhon_EulerClassTightBinding}. The codes generate three-band, and four-band, tight-binding models for any fixed Euler class through the use of the Pl\"ucker embeddings presented in Sec. \ref{3_bands}, and Sec.  \ref{4_bands} and Appendix \ref{ap_1}, respectively. 

\section{Euler class reversal in $\mathbb{R}\mathrm{P}^2$}\label{ap_0}
 
For completeness, in this Appendix we reproduce from Ref.~\cite{Wojcik:2019} the continuous and adiabatic transformation that reverses the Euler class of the two-band oriented subbundle of an orientable gapped three-band model classified by $\mathbb{R}\mathrm{P}^2$, hence realizing the automorphism $\triangleright_\ell :\chi \rightarrow -\chi$ of Sec.~\ref{subsec:second} (and Appendix \ref{ap:homotopy_action}).  

Our representative Hamiltonians of orientable gapped three-band systems [cf.~Eqs.~(\ref{eqn:tq-map}) and~(\ref{eq:H_3_bands})] can conveniently be parametrized as \cite{BzduSigristRobust}
\begin{equation}
\label{eq:H_para}
    H[\boldsymbol{n}](\theta,\phi) = 2 \boldsymbol{n}(\theta,\phi)\cdot \boldsymbol{n}(\theta,\phi)^T - \mathbb{1}_3 \;,
\end{equation}
with $\boldsymbol{n}(\theta,\phi) = u_3 \in \mathbb{S}^2$ the unit eigenvector of the single unoccupied band, and $\boldsymbol{n} = n_1 e_1 +n_2 e_2+n_3 e_3$.

Since $H[\boldsymbol{n}]$ is explicitly invariant under the inversion $\boldsymbol{n} \rightarrow -\boldsymbol{n}$ there is not canonical signed Euler class associated with the Hamiltonian. The indeterminacy can be lifted though by assigning a smooth structure to the vector field $\{ \boldsymbol{n}(\theta,\phi) \vert (\theta,\phi)\in \mathbb{S}^2\}$, which is allowed by virtue of the triviality of any real line bundle defined over the sphere (see Sec.~\ref{subsec:2D-cohomology-classes}). 

As in Sec.~\ref{3_bands} this is achieved by setting \cite{BzduSigristRobust} $\boldsymbol{n}(\theta,\phi) = e_r = (\cos\phi_q \sin\theta_q, \sin \phi_q \sin \theta_q ,\cos\theta_q) \in \mathbb{S}^2 $, where $q\in \mathbb{Z}$ defines the number of times $\boldsymbol{n}$ wraps around the sphere [cf.~Eq.~(\ref{eqn:up-south-east})]. The Euler class of the oriented occupied two-band subbundle is then $\chi_0 = 2q \in 2\mathbb{Z}$. We have thereby promoted the Hamiltonian Eq.~(\ref{eq:H_para}) to an \textit{oriented} total gapped bundle with a well defined Euler class. 

Since we are interested in an automorphism of the based homotopy group $\pi_2[\mathbb{R}\mathrm{P}^2]$ it is crucial to specify a chosen base point that will serve as a reference for comparing any two elements of the group. Let us fix $\boldsymbol{n}(\theta=0,\phi=0) = e_3$ at the blue pole of the sphere [Fig.~\ref{pullback}].

Defining the rotation 
\begin{equation}
    S(s) = \left(\begin{array}{ccc}
        \cos s & 0 & -\sin s \\
        0 & 1 & 0 \\
        \sin s & 0 & \cos s
    \end{array}\right)\;.
\end{equation}
for $s\in[0,\pi]$, we obtain a smooth deformation of the Hamiltonian through 
\begin{equation}
    H[S(s)\cdot \boldsymbol{n}] = 2 (S(s)\cdot\boldsymbol{n})\cdot (S(s)\cdot\boldsymbol{n})^T - \mathbb{1}_3 \;,
\end{equation}
which is adiabatic (i.e.~it preserves the gap between the eigenvalues) since it can be rewritten as the change of basis, i.e.
\begin{equation}
    H[S(s)\cdot \boldsymbol{n}] = S(s)\cdot H[\boldsymbol{n}] \cdot S(s)^T \;.
\end{equation}
Exploiting the gauge freedom of the Hamiltonian (i.e.~$H[\boldsymbol{n}] = H[-\boldsymbol{n}]$), we eventually find 
\begin{equation}
    H[-S(s)\cdot \boldsymbol{n}] = 2 (-S(s)\cdot\boldsymbol{n})\cdot (-S(s)\cdot\boldsymbol{n})^T - \mathbb{1}_3 \;,
\end{equation}
which at $s=\pi$ preserves the base point of the original Hamiltonian, i.e.~$-S(\pi)\cdot \boldsymbol{n}(0,0) = \boldsymbol{n}(0,0) = e_3 $. Furthermore, we find
\begin{equation}
    -S(\pi)\cdot \boldsymbol{n} = n_1 e_1 - n_2 e_2 + n_3 e_3 \;,
\end{equation}
such that the Euler class of $H[-S(\pi)\cdot \boldsymbol{n}]$ is $ \chi_{\pi} = -\chi_0 = -2q$. Therefore, at $s=\pi$ the transformation realizes the automorphism $\triangleright_{\ell} : \chi_0 \mapsto \chi_{\pi} = - \chi_0$.

As a conclusion, the above construction defines the continuous deformation of Hamiltonian 
\begin{equation} 
    \triangleright(s) : \mathbb{R}\mathrm{P}^2 \rightarrow \mathbb{R}\mathrm{P}^2 :  H[\boldsymbol{n}] \rightarrow H[-S(s)\cdot \boldsymbol{n}] .
\end{equation}
with $\triangleright(0) = \mathrm{id}$ and $\triangleright(\pi) = \triangleright_{\ell}$. Then, keeping track of $H[\boldsymbol{n}]$ at the base point $(\theta,\phi)=(0,0)$ through the deformation, i.e. $\{\triangleright(s) H[\boldsymbol{n}](0,0) \vert s\in [0,\pi]\}$, this defines a non-contractible loop within $\mathbb{R}\mathrm{P}^2$, i.e.~a generator of $\pi_1[\mathbb{R}\mathrm{P}^2] = \mathbb{Z}_2$ \cite{Wojcik:2019}.

\section{Pl\"ucker embedding for $\mathrm{Gr}^+_{2,4}$}\label{ap_1}

In this Appendix we derive the explicit Pl\"ucker embedding for $\mathrm{Gr}^+_{2,4} = \mathsf{SO}(4)/[\mathsf{SO}(2)\times \mathsf{SO}(2)] \cong \mathbb{S}^2 \times \mathbb{S}^2$. We do it starting from the parametrizations of $\mathsf{SO}(4)$ in terms of the Lie algebra of real and anti-symmetric matrices. 

\subsection{Parametrization of $\mathsf{SO}(4)$}

A matrix $R \in \mathsf{SO}(4)$ can be decomposed as \cite{Haber_antisym} $R = Q_c U_r$ with 
\begin{equation}
	Q_c = \exp\left( \begin{array}{cc} C & D \\ D & -C	\end{array}\right) \,,\quad
	U_r = \left( \begin{array}{cc} \mathrm{Re} U & -\mathrm{Im} U \\ \mathrm{Im} U & \mathrm{Re} U	\end{array}\right) \,,
\end{equation}
where $C$ and $D$ are arbitrary real antisymmetric matrices, and $U_r$ is a generic matrix in $\mathsf{U}(2)$. $Q_c$ can be parametrized as \cite{Haber_antisym}
\begin{multline}
	Q_c = \\
	\left(\begin{array}{cccc} \cos\rho & \sin\rho\sin\xi & 0 & \sin\rho\cos\xi \\
			-\sin\rho\sin\xi  &  \cos\rho  & 	- \sin\rho\cos\xi  & 0 \\
				0 & 	 \sin\rho\cos\xi & \cos\rho & -  \sin\rho\sin\xi \\
				- \sin\rho\cos\xi & 0 &  \sin\rho\sin\xi & \cos\rho
	 \end{array}\right)\,,
\end{multline}
with the angle $\rho = \sqrt{c^2+d^2}$, where $c=\mathrm{Pf}[C]$ and $d=\mathrm{Pf}[D]$, and an other angle defined through $\cos \xi = c/\rho$ and $\sin \xi = d/\rho$. The range of these angles are $\rho, \xi \in [0,2\pi)$. A generic matrix $U_r \in \mathsf{U}(2)$ can be decomposed as
\begin{equation}
	U_r = e^{\imi \varphi/2} \left(\begin{array}{cc} e^{\imi \phi_1} \cos\psi & e^{\imi \phi_2} \sin\psi \\
			-e^{-\imi \phi_2} \sin\psi & e^{-\imi \phi_1} \cos\psi
		\end{array}\right)\;,
\end{equation}
with the angles $\varphi,\phi_1,\phi_2 \in [0,2\pi)$ and $\psi \in [0,\pi)$. 
This results in
\begin{equation}
\label{param_SO4}
\!\!\!	R(\rho,\xi,\varphi, \psi, \phi_1,\phi_2) \!=\! Q_c(\rho,\xi) U_r(\varphi, \psi, \phi_1,\phi_2) \!\in\! \mathsf{SO}(4).\!\!\!
\end{equation}
We now need the constraints among the six angles, $\{\rho,\xi,\varphi,\psi,\phi_1,\phi_2\}$, as to only cover the quotient space $\mathsf{SO}(4)/[\mathsf{SO}(2)\times \mathsf{SO}(2)] \cong \mathbb{S}^2 \times \mathbb{S}^2$. Before doing so, we first review the diffeomorphism of spaces $\mathsf{SO}(4)/[\mathsf{SO}(2)\times \mathsf{SO}(2)] \cong \mathbb{S}^2 \times \mathbb{S}^2$. The readers familiar with the Pl\"ucker embedding may jump to the solution Eq.~(\ref{constrained_SO4}).

\subsection{$\mathsf{SO}(4)/[\mathsf{SO}(2)\times \mathsf{SO}(2)] \cong \mathbb{S}^2 \times \mathbb{S}^2$}

We review here the standard result $\mathsf{SO}(4)/[\mathsf{SO}(2)\times \mathsf{SO}(2)] \cong \mathbb{S}^2 \times \mathbb{S}^2$ obtained through the Pl\"ucker embedding~\cite{mathSE:2215495orig}. This section follows the argument of~\cite{mathSE:2215495orig} with a few more steps.

The Pl\"ucker embedding $\iota:\mathrm{Gr}^+_2(\mathbb{R}^4)\xhookrightarrow{} \Lambda^2\mathbb{R}^4 $ represents the points of the oriented Grassmannian as elements of the second exterior power of $\mathbb{R}^4$, $\Lambda^2\mathbb{R}^4$, that is a vector space of dimension $(\substack{4\\2})=6$ spanned by bivectors, i.e. the exterior product ($\cdot\wedge\cdot$) of two vectors of $\mathbb{R}^4$. More specifically, for $x\in \Lambda^2\mathbb{R}^4$ the image of the Pl\"ucker embedding is defined by the solutions to the system 
\begin{equation}
\label{embedding}
	x\wedge x = 0, \quad \vert x\vert^2_{\wedge} = 2\;,
\end{equation}  
where the norm $\vert\cdot \vert_{\wedge} =\sqrt{\langle \cdot,\cdot \rangle_{\wedge}}$ is defined in terms of a inner product in $\Lambda^2\mathbb{R}^4$, see below.

Let us take $(u_1,u_2,u_3,u_4)$ an oriented orthonormal frame of $\mathbb{R}^4$. There is a bijection between any oriented plane $V \subset \mathbb{R}^4$ and an element $u_1\wedge u_2 \in \Lambda^2\mathbb{R}^4$, given that $V$ is spanned by the orthonormal frame $(u_1,u_2)$. The orthogonal complement $V^c = \{u\in \mathbb{R}^4 \vert \langle u , v \rangle = 0,\; \forall v \in V\}$ is then represented by the Hodge dual $*(u_1\wedge u_2) = u_3\wedge u_4$. 

We have $*( \alpha u_1\wedge u_2 \pm \beta u_3\wedge u_4) = \pm (\beta u_1\wedge u_2 \pm  \alpha u_3\wedge u_4) \in \Lambda^2\mathbb{R}^4, \; \alpha, \beta \in \mathbb{R}$. Thus, the $\pm1$-eigenspaces of the Hodge star $*$, which we note $\Lambda^2_+$ and $\Lambda^2_-$, are composed of elements of the form $v_{\pm} = \alpha (u_1\wedge u_2 \pm  u_3\wedge u_4)$. These are perpendicular with respect to the exterior and the inner products, i.e. $v_+ \wedge v_- = \langle v_+ , v_-\rangle_{\wedge} =  0$ for $v_+\in \Lambda^2_+$ and $v_-\in\Lambda^2_-$, where the inner product of two elements of $\Lambda^2\mathbb{R}^4$ is defined through $\langle a\wedge b, c\wedge d\rangle_{\wedge} = \langle a,c \rangle \langle b,d \rangle - \langle a,d \rangle \langle b,c \rangle$ with $a,b,c,d \in \mathbb{R}^4$. 

Setting $x = v_+ + v_-$, the equation $x\wedge x = 0$ gives $\vert v_+\vert = \vert v_- \vert $, and the equation $\vert x\vert^2 =2$ gives $\vert v_+\vert^2 + \vert v_- \vert^2 = 2$. Combining these we get the relation $\vert v_+ \vert = \vert v_- \vert = 1$. Thus, the system Eq.~(\ref{embedding}) is readily satisfied for $v_{\pm} = \alpha (u_1\wedge u_2 \pm  u_3\wedge u_4)$ with $\alpha = 1/\sqrt{2}$. We conclude that an element of $\mathrm{Gr}^+_{2,4}$ is represented by an element $x = u_1 \wedge u_2 = v_+ + v_- \in \Lambda^2\mathbb{R}^4$ with $v_+\in \Lambda^2_+$ and $v_-\in\Lambda^2_-$.

For $v_\pm\in \Lambda^2_{\pm} \subset \Lambda^2\mathbb{R}^4$ we have $\langle v_+,v_-\rangle_{\wedge} = 0 $, such that $v_+$ and $v_-$ split $\Lambda^2\mathbb{R}^4$ into two orthogonal components each of dimension 3, i.e. $\Lambda^2\mathbb{R}^4 = V_+ \oplus V_-$. Since $v_{\pm}$ are unit bivectors, the spaces $\Lambda^2_{\pm} $ are the unit spheres in $V_{\pm}$, i.e. $(v_+,v_-) \in \Lambda^2_{+} \oplus \Lambda^2_{-} \cong \mathbb{S}^2_+ \times \mathbb{S}^2_-$. Since every point of the oriented Grassmannian is represented through the Pl\"ucker embedding by a bivector $x = v_+ + v_-$, we conclude that the image of the embedding is $\iota(\mathrm{Gr}^+_{2,4}) \cong \mathbb{S}^2_+ \times \mathbb{S}^2_-$.

\subsection{From $\mathsf{SO}(4)$ to $\mathsf{SO}(4)/[\mathsf{SO}(2)\times \mathsf{SO}(2)]$}

The previous section provides the guidelines for the derivation of the constraints Eq.~(\ref{constrained_SO4}) that map the elements of $\mathsf{SO}(4)$ to the elements of $\mathsf{SO}(4)/[\mathsf{SO}(2)\times \mathsf{SO}(2)]$.  

Choosing a Cartesian frame for $\mathbb{R}^4$, each column vector of $R  = (u_1u_2u_3u_4) \in \mathsf{SO}(4)$ reads

\begin{widetext}
\begin{equation}
	u_i = u^1_{i} \hat{e}_1 + u^2_{i} \hat{e}_2 + u^3_{i}\hat{e}_3 + u^4_{i}\hat{e}_4, \quad \mathrm{for}~i=1,2,3,4,  \quad \mathrm{and~with}~\hat{e}_i^j = \delta_{ij}\;.
\end{equation}

We then choose a reference basis for $\Lambda^2\mathbb{R}^4$,
\begin{equation}
	\{ \check{e}_1,\check{e}_2, \check{e}_3, \check{e}_4, \check{e}_5, \check{e}_6  \}
	= \left\{
	\hat{e}_3 \wedge \hat{e}_2 , 
	\hat{e}_3 \wedge \hat{e}_1 , 
	\hat{e}_1 \wedge \hat{e}_2 ,
	\hat{e}_4 \wedge \hat{e}_1 ,
	\hat{e}_2 \wedge \hat{e}_4 ,
	\hat{e}_3 \wedge \hat{e}_4 \right\} \;,
\end{equation}
and compute the elements $v_+ = u_1\wedge u_2 + u_3\wedge u_4 \in \Lambda^2_+$ and $v_- =u_1\wedge u_2 - u_3\wedge u_4 \in \Lambda^2_-$. 

For the parametrization $R(\rho,\xi,\varphi,\psi,\phi_1,\phi_2)$ derived in Eq.~(\ref{param_SO4}), we find
\begin{align}
	v_+ &= 
	\left(\check{e}_1~\check{e}_2~\check{e}_3~\check{e}_4~\check{e}_5~\check{e}_6\right) \cdot \left(a~b~c~a~b~c \right)^T\;, \\
	v_-&= 
	\left(\check{e}_1~\check{e}_2~\check{e}_3~\check{e}_4~\check{e}_5~\check{e}_6\right) \cdot  \left(d~e~f~-d~-e~-f \right)^T\;,
\end{align}
with 
\begin{equation}
	\begin{aligned}
		a &= \cos(\psi)^2 \sin(2\phi_1) + \sin(\psi)^2 \sin(2\phi_2) \,,\\
		b &= \sin(2\psi)\sin(\phi_1-\phi_2) \,,\\
		c &= \cos(\psi)^2 \cos(2\phi_1) + \sin(\psi)^2 \cos(2\phi_2)\,, 
		\\
	\end{aligned} 
	\hspace{1.5cm}
	\begin{aligned}
		d &=\cos(\rho)^2 \sin(\varphi) + \sin(\rho)^2 \sin(\varphi+2\xi) \,,\\
		e &= -\cos(\varphi+\xi)\sin(2 \rho)\,, \\
		f &= \cos(\rho)^2 \cos(\varphi) - \sin(\rho)^2 \cos(\varphi+2\xi)\,,
	\end{aligned}
\end{equation}
Note that $\langle v_+ , v_- \rangle_{\wedge} = ad+be+cf-ad-be-cf \equiv 0$. 

Let us make the following change of basis for $\Lambda^2\mathbb{R}^4$, 
\begin{equation}
\begin{array}{cc}
	\begin{array}{rcl}
		\check{e}'_1 &=& \check{e}_1+\check{e}_4\,,\\
		\check{e}'_2 &=& \check{e}_2+\check{e}_5\,,\\
		\check{e}'_3 &=& \check{e}_3+\check{e}_6\,,
	\end{array} &
	\begin{array}{rcl}
		\check{e}'_4 &=& \check{e}_1-\check{e}_4\,,\\
		\check{e}'_5 &=& \check{e}_2-\check{e}_5\,,\\
		\check{e}'_6 &=& \check{e}_3-\check{e}_6\,,
	\end{array}
\end{array}
\end{equation}
which we rewrite as
\begin{align}
	\left(\check{e}_1~\check{e}_2~\check{e}_3~\check{e}_4~\check{e}_5~\check{e}_6\right) = \left(\check{e}'_1~\check{e}'_2~\check{e}'_3~\check{e}_4'~\check{e}_5'~\check{e}'_6\right)\cdot S\;,\quad\mathrm{with}~
	S = \dfrac{1}{2}\left(\begin{array}{cc} \mathbb{1}_{3\times3} & \mathbb{1}_{3\times3} \\ \mathbb{1}_{3\times3}& -\mathbb{1}_{3\times3} \end{array}\right)\;.
\end{align}	

In the new basis, we then get 
\begin{align}
	v_+ &= \left(\check{e}'_1~\check{e}'_2~\check{e}'_3~\check{e}'_4~\check{e}'_5~\check{e}'_6\right)\cdot S \cdot  (a~b~c~a~b~c)^T \;, \nonumber\\
		&= \left(\check{e}'_1~\check{e}'_2~\check{e}'_3~\check{e}_4'~\check{e}_5'~\check{e}'_6\right)\cdot   (a~b~c~0~0~0)^T = (a,b,c,0,0,0) \;,\\
	v_-  &= \left(\check{e}'_1~\check{e}'_2~\check{e}'_3~\check{e}'_4~\check{e}'_5~\check{e}'_6\right)\cdot S \cdot  (d~e~f~-d~-e~-f)^T \;, \nonumber\\
		&= \left(\check{e}'_1~\check{e}'_2~\check{e}'_3~\check{e}_4'~\check{e}_5'~\check{e}'_6\right) \cdot  (0~0~0~d~e~f)^T = (0,0,0,d,e,f) \;,
\end{align}
i.e. this basis emphasizes that $v_+$ and $v_-$ live in three-dimensional orthogonal complements $V_{\pm}$ of the six-dimensional vector space $\Lambda^2\mathbb{R}^4 = V_+\oplus V_- $.

The restriction of $v_+ = (a,b,c)$ and $v_- = (d,e,f)$ to unit spheres is then obtained through
\begin{align}
\label{constrained_SO4}
	\phi_1 = -\phi_2 = \theta_+/2, \quad \psi = \phi_+/2,  \quad \rho = \theta_-/2, \quad \varphi = -\xi = \phi_- \;, 
\end{align}
with the spherical angles $(\phi_{+(-)},\theta_{+(-)}) \in [0,2\pi) \times [0,\pi] \cong \mathbb{S}^2_{+(-)}$. Indeed, substituting Eq.~(\ref{constrained_SO4}) we find in the basis $\{\check{e}'_i\}_{i=1,\dots,6}$,
\begin{align*}
	v_+ &= (a,b,c,0,0,0) = (\cos \phi_+ \sin \theta_+ ,\sin \phi_+ \sin \theta_+,\cos \theta_+,0,0,0)\\
	& \in \mathbb{S}^2_+ \subset \mathrm{Span}\{ \check{e}'_1 , \check{e}'_2, \check{e}'_3\} \;,\\
	v_-  &= (0,0,0,d,e,f)= (0,0,0,\sin \phi_{-} \cos \theta_{-} ,-\sin \theta_{-} , \cos \phi_{-} \cos\theta_{-}) \\
	&\in \mathbb{S}^2_{-} \subset \mathrm{Span}\{  \check{e}'_4 , \check{e}'_5, \check{e}'_6\} \;.
\end{align*}

The inverse Pl\"ucker embedding then gives a bijection,
\begin{equation}
	\iota^{-1} : \mathbb{S}^2_{+}\times \mathbb{S}^2_{-} \rightarrow \mathsf{SO}(4)/[\mathsf{SO}(2)\times \mathsf{SO}(2)] : \left(\phi_{+} , \theta_{+}, \phi_{-} , \theta_{-}\right)   \mapsto [R(\phi_{+} , \theta_{+}, \phi_{-} , \theta_{-})]   \;,
\end{equation}
\end{widetext}
that we use in Sec.~\ref{4_bands} for building explicit tight-binding models for all homotopy classes.

\section{Euler class reversal in $\mathrm{Gr}_{2,4}$}\label{ap_3}

Let us consider the following transformation for $R=(u_1\,u_2\,u_3\,u_4)$,
\begin{equation}
   \;\triangleright(s) :  R \mapsto R_s = S(s) \cdot R \cdot G \;,
\end{equation}
with the rotation
\begin{equation}
   \; S(s) = \left(\begin{array}{cccc}
        1 & 0 & 0 & 0 \\
        0 & \cos s & -\sin s & 0  \\
        0 & \sin s & \cos s & 0  \\
        0 & 0 & 0 & 1
   \end{array}
   \right)  \;,
\end{equation}
for $s\in[0,\pi]$, and the gauge transformation
\begin{equation}
  \; G = \left(\begin{array}{cccc}
        1 & 0 & 0 & 0 \\
        0 & -1 & 0 & 0 \\
        0 & 0 & 1 & 0 \\
        0 & 0 & 0 & -1
   \end{array}
   \right)  \;.
\end{equation}

We then readily find that applying the Pl\"ucker embedding of the previous section to $R_{s}$ and at $s=\pi$, gives 
\begin{equation}
    \tilde{u}_1 \wedge \tilde{u}_2 = \tilde{v}_+ + \tilde{v}_- = (-a,b,c,-d,e,f),
\end{equation}
in the basis $(\check{e}_1',\check{e}_2',\check{e}_3',\check{e}_4',\check{e}_5',\check{e}_6')$. Therefore, the winding numbers for $R_{\pi}$ are $(\tilde{q}_+,\tilde{q}_-) = (-q_+,-q_-)$, corresponding to a reversal of Euler class $
(\widetilde{\chi}_I,\widetilde{\chi}_{II}) = (-\chi_I, -\chi_{II})$. Choosing the base point $(\theta_{\pm},\phi_{\pm}) = (0,0)$ at which $(a,b,c,d,e,f) = (0,0,1,0,0,1)$, the path of left cosets $ \{\triangleright(s) [ R_s(\theta_{\pm}=0,\phi_{\pm}=0) ] \vert s\in[0,\pi]\}$ defines a non-contractible loop in $\mathrm{Gr}_{2,4}$, i.e. the generator of $\pi_1[\mathrm{Gr}_{2,4}] = \mathbb{Z}_2$.


\begin{thebibliography}{100}%
\makeatletter
\providecommand \@ifxundefined [1]{%
 \@ifx{#1\undefined}
}%
\providecommand \@ifnum [1]{%
 \ifnum #1\expandafter \@firstoftwo
 \else \expandafter \@secondoftwo
 \fi
}%
\providecommand \@ifx [1]{%
 \ifx #1\expandafter \@firstoftwo
 \else \expandafter \@secondoftwo
 \fi
}%
\providecommand \natexlab [1]{#1}%
\providecommand \enquote  [1]{``#1''}%
\providecommand \bibnamefont  [1]{#1}%
\providecommand \bibfnamefont [1]{#1}%
\providecommand \citenamefont [1]{#1}%
\providecommand \href@noop [0]{\@secondoftwo}%
\providecommand \href [0]{\begingroup \@sanitize@url \@href}%
\providecommand \@href[1]{\@@startlink{#1}\@@href}%
\providecommand \@@href[1]{\endgroup#1\@@endlink}%
\providecommand \@sanitize@url [0]{\catcode `\\12\catcode `\$12\catcode
  `\&12\catcode `\#12\catcode `\^12\catcode `\_12\catcode `\%12\relax}%
\providecommand \@@startlink[1]{}%
\providecommand \@@endlink[0]{}%
\providecommand \url  [0]{\begingroup\@sanitize@url \@url }%
\providecommand \@url [1]{\endgroup\@href {#1}{\urlprefix }}%
\providecommand \urlprefix  [0]{URL }%
\providecommand \Eprint [0]{\href }%
\providecommand \doibase [0]{http://dx.doi.org/}%
\providecommand \selectlanguage [0]{\@gobble}%
\providecommand \bibinfo  [0]{\@secondoftwo}%
\providecommand \bibfield  [0]{\@secondoftwo}%
\providecommand \translation [1]{[#1]}%
\providecommand \BibitemOpen [0]{}%
\providecommand \bibitemStop [0]{}%
\providecommand \bibitemNoStop [0]{.\EOS\space}%
\providecommand \EOS [0]{\spacefactor3000\relax}%
\providecommand \BibitemShut  [1]{\csname bibitem#1\endcsname}%
\let\auto@bib@innerbib\@empty
\bibitem [{\citenamefont {Qi}\ and\ \citenamefont {Zhang}(2011)}]{Rmp1}%
  \BibitemOpen
  \bibfield  {author} {\bibinfo {author} {\bibfnamefont {Xiao-Liang}\
  \bibnamefont {Qi}}\ and\ \bibinfo {author} {\bibfnamefont {Shou-Cheng}\
  \bibnamefont {Zhang}},\ }\bibfield  {title} {\enquote {\bibinfo {title}
  {Topological insulators and superconductors},}\ }\href {\doibase
  10.1103/RevModPhys.83.1057} {\bibfield  {journal} {\bibinfo  {journal} {Rev.
  Mod. Phys.}\ }\textbf {\bibinfo {volume} {83}},\ \bibinfo {pages}
  {1057--1110} (\bibinfo {year} {2011})}\BibitemShut {NoStop}%
\bibitem [{\citenamefont {Hasan}\ and\ \citenamefont {Kane}(2010)}]{Rmp2}%
  \BibitemOpen
  \bibfield  {author} {\bibinfo {author} {\bibfnamefont {M.~Z.}\ \bibnamefont
  {Hasan}}\ and\ \bibinfo {author} {\bibfnamefont {C.~L.}\ \bibnamefont
  {Kane}},\ }\bibfield  {title} {\enquote {\bibinfo {title} {Colloquium:
  {T}opological {I}nsulators},}\ }\href {\doibase 10.1103/RevModPhys.82.3045}
  {\bibfield  {journal} {\bibinfo  {journal} {Rev. Mod. Phys.}\ }\textbf
  {\bibinfo {volume} {82}},\ \bibinfo {pages} {3045--3067} (\bibinfo {year}
  {2010})}\BibitemShut {NoStop}%
\bibitem [{\citenamefont {Schnyder}\ \emph {et~al.}(2008)\citenamefont
  {Schnyder}, \citenamefont {Ryu}, \citenamefont {Furusaki},\ and\
  \citenamefont {Ludwig}}]{Schnyder08}%
  \BibitemOpen
  \bibfield  {author} {\bibinfo {author} {\bibfnamefont {Andreas~P.}\
  \bibnamefont {Schnyder}}, \bibinfo {author} {\bibfnamefont {Shinsei}\
  \bibnamefont {Ryu}}, \bibinfo {author} {\bibfnamefont {Akira}\ \bibnamefont
  {Furusaki}}, \ and\ \bibinfo {author} {\bibfnamefont {Andreas W.~W.}\
  \bibnamefont {Ludwig}},\ }\bibfield  {title} {\enquote {\bibinfo {title}
  {Classification of topological insulators and superconductors in three
  spatial dimensions},}\ }\href {\doibase 10.1103/PhysRevB.78.195125}
  {\bibfield  {journal} {\bibinfo  {journal} {Phys. Rev. B}\ }\textbf {\bibinfo
  {volume} {78}},\ \bibinfo {pages} {195125} (\bibinfo {year}
  {2008})}\BibitemShut {NoStop}%
\bibitem [{\citenamefont {Kitaev}(2009)}]{Kitaev}%
  \BibitemOpen
  \bibfield  {author} {\bibinfo {author} {\bibfnamefont {A.}~\bibnamefont
  {Kitaev}},\ }\bibfield  {title} {\enquote {\bibinfo {title} {{P}eriodic table
  for topological insulators and superconductors},}\ }\href@noop {} {\bibfield
  {journal} {\bibinfo  {journal} {AIP Conf. Proc.}\ }\textbf {\bibinfo {volume}
  {1134}},\ \bibinfo {pages} {22} (\bibinfo {year} {2009})}\BibitemShut
  {NoStop}%
\bibitem [{\citenamefont {Hughes}\ \emph {et~al.}(2011)\citenamefont {Hughes},
  \citenamefont {Prodan},\ and\ \citenamefont {Bernevig}}]{InvTIBernevig}%
  \BibitemOpen
  \bibfield  {author} {\bibinfo {author} {\bibfnamefont {Taylor~L.}\
  \bibnamefont {Hughes}}, \bibinfo {author} {\bibfnamefont {Emil}\ \bibnamefont
  {Prodan}}, \ and\ \bibinfo {author} {\bibfnamefont {B.~Andrei}\ \bibnamefont
  {Bernevig}},\ }\bibfield  {title} {\enquote {\bibinfo {title}
  {Inversion-symmetric topological insulators},}\ }\href {\doibase
  10.1103/PhysRevB.83.245132} {\bibfield  {journal} {\bibinfo  {journal} {Phys.
  Rev. B}\ }\textbf {\bibinfo {volume} {83}},\ \bibinfo {pages} {245132}
  (\bibinfo {year} {2011})}\BibitemShut {NoStop}%
\bibitem [{\citenamefont {Fu}(2011)}]{Clas1}%
  \BibitemOpen
  \bibfield  {author} {\bibinfo {author} {\bibfnamefont {Liang}\ \bibnamefont
  {Fu}},\ }\bibfield  {title} {\enquote {\bibinfo {title} {Topological
  {C}rystalline {I}nsulators},}\ }\href {\doibase
  10.1103/PhysRevLett.106.106802} {\bibfield  {journal} {\bibinfo  {journal}
  {Phys. Rev. Lett.}\ }\textbf {\bibinfo {volume} {106}},\ \bibinfo {pages}
  {106802} (\bibinfo {year} {2011})}\BibitemShut {NoStop}%
\bibitem [{\citenamefont {Turner}\ \emph {et~al.}(2012)\citenamefont {Turner},
  \citenamefont {Zhang}, \citenamefont {Mong},\ and\ \citenamefont
  {Vishwanath}}]{InvTIVish}%
  \BibitemOpen
  \bibfield  {author} {\bibinfo {author} {\bibfnamefont {Ari~M.}\ \bibnamefont
  {Turner}}, \bibinfo {author} {\bibfnamefont {Yi}~\bibnamefont {Zhang}},
  \bibinfo {author} {\bibfnamefont {Roger S.~K.}\ \bibnamefont {Mong}}, \ and\
  \bibinfo {author} {\bibfnamefont {Ashvin}\ \bibnamefont {Vishwanath}},\
  }\bibfield  {title} {\enquote {\bibinfo {title} {Quantized response and
  topology of magnetic insulators with inversion symmetry},}\ }\href {\doibase
  10.1103/PhysRevB.85.165120} {\bibfield  {journal} {\bibinfo  {journal} {Phys.
  Rev. B}\ }\textbf {\bibinfo {volume} {85}},\ \bibinfo {pages} {165120}
  (\bibinfo {year} {2012})}\BibitemShut {NoStop}%
\bibitem [{\citenamefont {Slager}\ \emph {et~al.}(2012)\citenamefont {Slager},
  \citenamefont {Mesaros}, \citenamefont {Juri{\v c}i{\'c}},\ and\
  \citenamefont {Zaanen}}]{Clas2}%
  \BibitemOpen
  \bibfield  {author} {\bibinfo {author} {\bibfnamefont {Robert-Jan}\
  \bibnamefont {Slager}}, \bibinfo {author} {\bibfnamefont {Andrej}\
  \bibnamefont {Mesaros}}, \bibinfo {author} {\bibfnamefont {Vladimir}\
  \bibnamefont {Juri{\v c}i{\'c}}}, \ and\ \bibinfo {author} {\bibfnamefont
  {Jan}\ \bibnamefont {Zaanen}},\ }\bibfield  {title} {\enquote {\bibinfo
  {title} {The space group classification of topological band-insulators},}\
  }\href {http://dx.doi.org/10.1038/nphys2513} {\bibfield  {journal} {\bibinfo
  {journal} {Nat. Phys.}\ }\textbf {\bibinfo {volume} {9}},\ \bibinfo {pages}
  {98} (\bibinfo {year} {2012})}\BibitemShut {NoStop}%
\bibitem [{\citenamefont {Juri\ifmmode \check{c}\else
  \v{c}\fi{}i\ifmmode~\acute{c}\else \'{c}\fi{}}\ \emph
  {et~al.}(2012)\citenamefont {Juri\ifmmode \check{c}\else
  \v{c}\fi{}i\ifmmode~\acute{c}\else \'{c}\fi{}}, \citenamefont {Mesaros},
  \citenamefont {Slager},\ and\ \citenamefont {Zaanen}}]{probes_2D}%
  \BibitemOpen
  \bibfield  {author} {\bibinfo {author} {\bibfnamefont {Vladimir}\
  \bibnamefont {Juri\ifmmode \check{c}\else \v{c}\fi{}i\ifmmode~\acute{c}\else
  \'{c}\fi{}}}, \bibinfo {author} {\bibfnamefont {Andrej}\ \bibnamefont
  {Mesaros}}, \bibinfo {author} {\bibfnamefont {Robert-Jan}\ \bibnamefont
  {Slager}}, \ and\ \bibinfo {author} {\bibfnamefont {Jan}\ \bibnamefont
  {Zaanen}},\ }\bibfield  {title} {\enquote {\bibinfo {title} {Universal
  {P}robes of {T}wo-{D}imensional {T}opological {I}nsulators: {D}islocation and
  $\ensuremath{\pi}$ {F}lux},}\ }\href {\doibase
  10.1103/PhysRevLett.108.106403} {\bibfield  {journal} {\bibinfo  {journal}
  {Phys. Rev. Lett.}\ }\textbf {\bibinfo {volume} {108}},\ \bibinfo {pages}
  {106403} (\bibinfo {year} {2012})}\BibitemShut {NoStop}%
\bibitem [{\citenamefont {Fang}\ \emph
  {et~al.}(2012{\natexlab{a}})\citenamefont {Fang}, \citenamefont {Gilbert},\
  and\ \citenamefont {Bernevig}}]{Chenprb2012}%
  \BibitemOpen
  \bibfield  {author} {\bibinfo {author} {\bibfnamefont {Chen}\ \bibnamefont
  {Fang}}, \bibinfo {author} {\bibfnamefont {Matthew~J.}\ \bibnamefont
  {Gilbert}}, \ and\ \bibinfo {author} {\bibfnamefont {B.~Andrei}\ \bibnamefont
  {Bernevig}},\ }\bibfield  {title} {\enquote {\bibinfo {title} {Bulk
  topological invariants in noninteracting point group symmetric insulators},}\
  }\href {\doibase 10.1103/PhysRevB.86.115112} {\bibfield  {journal} {\bibinfo
  {journal} {Phys. Rev. B}\ }\textbf {\bibinfo {volume} {86}},\ \bibinfo
  {pages} {115112} (\bibinfo {year} {2012}{\natexlab{a}})}\BibitemShut
  {NoStop}%
\bibitem [{\citenamefont {Morimoto}\ and\ \citenamefont
  {Furusaki}(2013)}]{Morimoto_2013}%
  \BibitemOpen
  \bibfield  {author} {\bibinfo {author} {\bibfnamefont {Takahiro}\
  \bibnamefont {Morimoto}}\ and\ \bibinfo {author} {\bibfnamefont {Akira}\
  \bibnamefont {Furusaki}},\ }\bibfield  {title} {\enquote {\bibinfo {title}
  {Topological classification with additional symmetries from {C}lifford
  algebras},}\ }\href {\doibase 10.1103/PhysRevB.88.125129} {\bibfield
  {journal} {\bibinfo  {journal} {Phys. Rev. B}\ }\textbf {\bibinfo {volume}
  {88}},\ \bibinfo {pages} {125129} (\bibinfo {year} {2013})}\BibitemShut
  {NoStop}%
\bibitem [{\citenamefont {Shiozaki}\ and\ \citenamefont
  {Sato}(2014)}]{Shiozaki14}%
  \BibitemOpen
  \bibfield  {author} {\bibinfo {author} {\bibfnamefont {Ken}\ \bibnamefont
  {Shiozaki}}\ and\ \bibinfo {author} {\bibfnamefont {Masatoshi}\ \bibnamefont
  {Sato}},\ }\bibfield  {title} {\enquote {\bibinfo {title} {Topology of
  crystalline insulators and superconductors},}\ }\href {\doibase
  10.1103/PhysRevB.90.165114} {\bibfield  {journal} {\bibinfo  {journal} {Phys.
  Rev. B}\ }\textbf {\bibinfo {volume} {90}},\ \bibinfo {pages} {165114}
  (\bibinfo {year} {2014})}\BibitemShut {NoStop}%
\bibitem [{\citenamefont {Chiu}\ \emph {et~al.}(2016)\citenamefont {Chiu},
  \citenamefont {Teo}, \citenamefont {Schnyder},\ and\ \citenamefont
  {Ryu}}]{SchnyderClass}%
  \BibitemOpen
  \bibfield  {author} {\bibinfo {author} {\bibfnamefont {Ching-Kai}\
  \bibnamefont {Chiu}}, \bibinfo {author} {\bibfnamefont {Jeffrey C.~Y.}\
  \bibnamefont {Teo}}, \bibinfo {author} {\bibfnamefont {Andreas~P.}\
  \bibnamefont {Schnyder}}, \ and\ \bibinfo {author} {\bibfnamefont {Shinsei}\
  \bibnamefont {Ryu}},\ }\bibfield  {title} {\enquote {\bibinfo {title}
  {Classification of topological quantum matter with symmetries},}\ }\href
  {\doibase 10.1103/RevModPhys.88.035005} {\bibfield  {journal} {\bibinfo
  {journal} {Rev. Mod. Phys.}\ }\textbf {\bibinfo {volume} {88}},\ \bibinfo
  {pages} {035005} (\bibinfo {year} {2016})}\BibitemShut {NoStop}%
\bibitem [{\citenamefont {Slager}\ \emph {et~al.}(2014)\citenamefont {Slager},
  \citenamefont {Mesaros}, \citenamefont {Juri\ifmmode \check{c}\else
  \v{c}\fi{}i\ifmmode~\acute{c}\else \'{c}\fi{}},\ and\ \citenamefont
  {Zaanen}}]{Mode2}%
  \BibitemOpen
  \bibfield  {author} {\bibinfo {author} {\bibfnamefont {Robert-Jan}\
  \bibnamefont {Slager}}, \bibinfo {author} {\bibfnamefont {Andrej}\
  \bibnamefont {Mesaros}}, \bibinfo {author} {\bibfnamefont {Vladimir}\
  \bibnamefont {Juri\ifmmode \check{c}\else \v{c}\fi{}i\ifmmode~\acute{c}\else
  \'{c}\fi{}}}, \ and\ \bibinfo {author} {\bibfnamefont {Jan}\ \bibnamefont
  {Zaanen}},\ }\bibfield  {title} {\enquote {\bibinfo {title} {Interplay
  between electronic topology and crystal symmetry: {D}islocation-line modes in
  topological band insulators},}\ }\href {\doibase 10.1103/PhysRevB.90.241403}
  {\bibfield  {journal} {\bibinfo  {journal} {Phys. Rev. B}\ }\textbf {\bibinfo
  {volume} {90}},\ \bibinfo {pages} {241403} (\bibinfo {year}
  {2014})}\BibitemShut {NoStop}%
\bibitem [{\citenamefont {Bouhon}\ and\ \citenamefont
  {Sigrist}(2014)}]{Bouhon_CurrentInversion}%
  \BibitemOpen
  \bibfield  {author} {\bibinfo {author} {\bibfnamefont {Adrien}\ \bibnamefont
  {Bouhon}}\ and\ \bibinfo {author} {\bibfnamefont {Manfred}\ \bibnamefont
  {Sigrist}},\ }\bibfield  {title} {\enquote {\bibinfo {title} {Current
  inversion at the edges of a chiral $p$-wave superconductor},}\ }\href
  {\doibase 10.1103/PhysRevB.90.220511} {\bibfield  {journal} {\bibinfo
  {journal} {Phys. Rev. B}\ }\textbf {\bibinfo {volume} {90}},\ \bibinfo
  {pages} {220511} (\bibinfo {year} {2014})}\BibitemShut {NoStop}%
\bibitem [{\citenamefont {Slager}\ \emph {et~al.}(2015)\citenamefont {Slager},
  \citenamefont {Rademaker}, \citenamefont {Zaanen},\ and\ \citenamefont
  {Balents}}]{Codefects1}%
  \BibitemOpen
  \bibfield  {author} {\bibinfo {author} {\bibfnamefont {Robert-Jan}\
  \bibnamefont {Slager}}, \bibinfo {author} {\bibfnamefont {Louk}\ \bibnamefont
  {Rademaker}}, \bibinfo {author} {\bibfnamefont {Jan}\ \bibnamefont {Zaanen}},
  \ and\ \bibinfo {author} {\bibfnamefont {Leon}\ \bibnamefont {Balents}},\
  }\bibfield  {title} {\enquote {\bibinfo {title} {Impurity-bound states and
  {G}reen's function zeros as local signatures of topology},}\ }\href {\doibase
  10.1103/PhysRevB.92.085126} {\bibfield  {journal} {\bibinfo  {journal} {Phys.
  Rev. B}\ }\textbf {\bibinfo {volume} {92}},\ \bibinfo {pages} {085126}
  (\bibinfo {year} {2015})}\BibitemShut {NoStop}%
\bibitem [{\citenamefont {Alexandradinata}\ \emph {et~al.}(2014)\citenamefont
  {Alexandradinata}, \citenamefont {Dai},\ and\ \citenamefont
  {Bernevig}}]{Wi1}%
  \BibitemOpen
  \bibfield  {author} {\bibinfo {author} {\bibfnamefont {A.}~\bibnamefont
  {Alexandradinata}}, \bibinfo {author} {\bibfnamefont {Xi}~\bibnamefont
  {Dai}}, \ and\ \bibinfo {author} {\bibfnamefont {B.~Andrei}\ \bibnamefont
  {Bernevig}},\ }\bibfield  {title} {\enquote {\bibinfo {title} {Wilson-loop
  characterization of inversion-symmetric topological insulators},}\ }\href
  {\doibase 10.1103/PhysRevB.89.155114} {\bibfield  {journal} {\bibinfo
  {journal} {Phys. Rev. B}\ }\textbf {\bibinfo {volume} {89}},\ \bibinfo
  {pages} {155114} (\bibinfo {year} {2014})}\BibitemShut {NoStop}%
\bibitem [{\citenamefont {Alexandradinata}\ \emph {et~al.}(2016)\citenamefont
  {Alexandradinata}, \citenamefont {Wang},\ and\ \citenamefont
  {Bernevig}}]{Wi3}%
  \BibitemOpen
  \bibfield  {author} {\bibinfo {author} {\bibfnamefont {A.}~\bibnamefont
  {Alexandradinata}}, \bibinfo {author} {\bibfnamefont {Zhijun}\ \bibnamefont
  {Wang}}, \ and\ \bibinfo {author} {\bibfnamefont {B.~Andrei}\ \bibnamefont
  {Bernevig}},\ }\bibfield  {title} {\enquote {\bibinfo {title} {Topological
  {I}nsulators from {G}roup {C}ohomology},}\ }\href {\doibase
  10.1103/PhysRevX.6.021008} {\bibfield  {journal} {\bibinfo  {journal} {Phys.
  Rev. X}\ }\textbf {\bibinfo {volume} {6}},\ \bibinfo {pages} {021008}
  (\bibinfo {year} {2016})}\BibitemShut {NoStop}%
\bibitem [{\citenamefont {Bzdu{\v s}ek}\ \emph {et~al.}(2016)\citenamefont
  {Bzdu{\v s}ek}, \citenamefont {Wu}, \citenamefont {R{\"u}egg}, \citenamefont
  {Sigrist},\ and\ \citenamefont {Soluyanov}}]{Nodal_chains}%
  \BibitemOpen
  \bibfield  {author} {\bibinfo {author} {\bibfnamefont {Tom{\'a}{\v s}}\
  \bibnamefont {Bzdu{\v s}ek}}, \bibinfo {author} {\bibfnamefont {QuanSheng}\
  \bibnamefont {Wu}}, \bibinfo {author} {\bibfnamefont {Andreas}\ \bibnamefont
  {R{\"u}egg}}, \bibinfo {author} {\bibfnamefont {Manfred}\ \bibnamefont
  {Sigrist}}, \ and\ \bibinfo {author} {\bibfnamefont {Alexey~A.}\ \bibnamefont
  {Soluyanov}},\ }\bibfield  {title} {\enquote {\bibinfo {title} {Nodal-chain
  metals},}\ }\href {http://dx.doi.org/10.1038/nature19099} {\bibfield
  {journal} {\bibinfo  {journal} {Nature}\ }\textbf {\bibinfo {volume} {538}},\
  \bibinfo {pages} {75 EP --} (\bibinfo {year} {2016})}\BibitemShut {NoStop}%
\bibitem [{\citenamefont {Slager}\ \emph {et~al.}(2016)\citenamefont {Slager},
  \citenamefont {Juri\ifmmode \check{c}\else \v{c}\fi{}i\ifmmode~\acute{c}\else
  \'{c}\fi{}}, \citenamefont {Lahtinen},\ and\ \citenamefont
  {Zaanen}}]{Modebbcst}%
  \BibitemOpen
  \bibfield  {author} {\bibinfo {author} {\bibfnamefont {Robert-Jan}\
  \bibnamefont {Slager}}, \bibinfo {author} {\bibfnamefont {Vladimir}\
  \bibnamefont {Juri\ifmmode \check{c}\else \v{c}\fi{}i\ifmmode~\acute{c}\else
  \'{c}\fi{}}}, \bibinfo {author} {\bibfnamefont {Ville}\ \bibnamefont
  {Lahtinen}}, \ and\ \bibinfo {author} {\bibfnamefont {Jan}\ \bibnamefont
  {Zaanen}},\ }\bibfield  {title} {\enquote {\bibinfo {title} {Self-organized
  pseudo-graphene on grain boundaries in topological band insulators},}\ }\href
  {\doibase 10.1103/PhysRevB.93.245406} {\bibfield  {journal} {\bibinfo
  {journal} {Phys. Rev. B}\ }\textbf {\bibinfo {volume} {93}},\ \bibinfo
  {pages} {245406} (\bibinfo {year} {2016})}\BibitemShut {NoStop}%
\bibitem [{\citenamefont {Kruthoff}\ \emph {et~al.}(2017)\citenamefont
  {Kruthoff}, \citenamefont {de~Boer}, \citenamefont {van Wezel}, \citenamefont
  {Kane},\ and\ \citenamefont {Slager}}]{Clas3}%
  \BibitemOpen
  \bibfield  {author} {\bibinfo {author} {\bibfnamefont {Jorrit}\ \bibnamefont
  {Kruthoff}}, \bibinfo {author} {\bibfnamefont {Jan}\ \bibnamefont {de~Boer}},
  \bibinfo {author} {\bibfnamefont {Jasper}\ \bibnamefont {van Wezel}},
  \bibinfo {author} {\bibfnamefont {Charles~L.}\ \bibnamefont {Kane}}, \ and\
  \bibinfo {author} {\bibfnamefont {Robert-Jan}\ \bibnamefont {Slager}},\
  }\bibfield  {title} {\enquote {\bibinfo {title} {Topological {C}lassification
  of {C}rystalline {I}nsulators through {B}and {S}tructure {C}ombinatorics},}\
  }\href {\doibase 10.1103/PhysRevX.7.041069} {\bibfield  {journal} {\bibinfo
  {journal} {Phys. Rev. X}\ }\textbf {\bibinfo {volume} {7}},\ \bibinfo {pages}
  {041069} (\bibinfo {year} {2017})}\BibitemShut {NoStop}%
\bibitem [{\citenamefont {Shiozaki}\ \emph {et~al.}(2017)\citenamefont
  {Shiozaki}, \citenamefont {Sato},\ and\ \citenamefont
  {Gomi}}]{ShiozakiSatoGomiK}%
  \BibitemOpen
  \bibfield  {author} {\bibinfo {author} {\bibfnamefont {Ken}\ \bibnamefont
  {Shiozaki}}, \bibinfo {author} {\bibfnamefont {Masatoshi}\ \bibnamefont
  {Sato}}, \ and\ \bibinfo {author} {\bibfnamefont {Kiyonori}\ \bibnamefont
  {Gomi}},\ }\bibfield  {title} {\enquote {\bibinfo {title} {Topological
  crystalline materials: {G}eneral formulation, module structure, and wallpaper
  groups},}\ }\href {\doibase 10.1103/PhysRevB.95.235425} {\bibfield  {journal}
  {\bibinfo  {journal} {Phys. Rev. B}\ }\textbf {\bibinfo {volume} {95}},\
  \bibinfo {pages} {235425} (\bibinfo {year} {2017})}\BibitemShut {NoStop}%
\bibitem [{\citenamefont {Bouhon}\ and\ \citenamefont
  {Black-Schaffer}(2017)}]{Wi2}%
  \BibitemOpen
  \bibfield  {author} {\bibinfo {author} {\bibfnamefont {Adrien}\ \bibnamefont
  {Bouhon}}\ and\ \bibinfo {author} {\bibfnamefont {Annica~M.}\ \bibnamefont
  {Black-Schaffer}},\ }\bibfield  {title} {\enquote {\bibinfo {title} {Global
  band topology of simple and double {D}irac-point semimetals},}\ }\href
  {\doibase 10.1103/PhysRevB.95.241101} {\bibfield  {journal} {\bibinfo
  {journal} {Phys. Rev. B}\ }\textbf {\bibinfo {volume} {95}},\ \bibinfo
  {pages} {241101} (\bibinfo {year} {2017})}\BibitemShut {NoStop}%
\bibitem [{\citenamefont {Geilhufe}\ \emph {et~al.}(2017)\citenamefont
  {Geilhufe}, \citenamefont {Bouhon}, \citenamefont {Borysov},\ and\
  \citenamefont {Balatsky}}]{NodalLines1}%
  \BibitemOpen
  \bibfield  {author} {\bibinfo {author} {\bibfnamefont {R.~Matthias}\
  \bibnamefont {Geilhufe}}, \bibinfo {author} {\bibfnamefont {Adrien}\
  \bibnamefont {Bouhon}}, \bibinfo {author} {\bibfnamefont {Stanislav~S.}\
  \bibnamefont {Borysov}}, \ and\ \bibinfo {author} {\bibfnamefont
  {Alexander~V.}\ \bibnamefont {Balatsky}},\ }\bibfield  {title} {\enquote
  {\bibinfo {title} {Three-dimensional organic {D}irac-line materials due to
  nonsymmorphic symmetry: {A} data mining approach},}\ }\href {\doibase
  10.1103/PhysRevB.95.041103} {\bibfield  {journal} {\bibinfo  {journal} {Phys.
  Rev. B}\ }\textbf {\bibinfo {volume} {95}},\ \bibinfo {pages} {041103}
  (\bibinfo {year} {2017})}\BibitemShut {NoStop}%
\bibitem [{\citenamefont {Po}\ \emph {et~al.}(2017)\citenamefont {Po},
  \citenamefont {Vishwanath},\ and\ \citenamefont {Watanabe}}]{Clas4}%
  \BibitemOpen
  \bibfield  {author} {\bibinfo {author} {\bibfnamefont {Hoi~Chun}\
  \bibnamefont {Po}}, \bibinfo {author} {\bibfnamefont {Ashvin}\ \bibnamefont
  {Vishwanath}}, \ and\ \bibinfo {author} {\bibfnamefont {Haruki}\ \bibnamefont
  {Watanabe}},\ }\bibfield  {title} {\enquote {\bibinfo {title} {Symmetry-based
  indicators of band topology in the 230 space groups},}\ }\href {\doibase
  10.1038/s41467-017-00133-2} {\bibfield  {journal} {\bibinfo  {journal} {Nat.
  Commun.}\ }\textbf {\bibinfo {volume} {8}},\ \bibinfo {pages} {50} (\bibinfo
  {year} {2017})}\BibitemShut {NoStop}%
\bibitem [{\citenamefont {Bradlyn}\ \emph {et~al.}(2017)\citenamefont
  {Bradlyn}, \citenamefont {Elcoro}, \citenamefont {Cano}, \citenamefont
  {Vergniory}, \citenamefont {Wang}, \citenamefont {Felser}, \citenamefont
  {Aroyo},\ and\ \citenamefont {Bernevig}}]{Clas5}%
  \BibitemOpen
  \bibfield  {author} {\bibinfo {author} {\bibfnamefont {Barry}\ \bibnamefont
  {Bradlyn}}, \bibinfo {author} {\bibfnamefont {L.}~\bibnamefont {Elcoro}},
  \bibinfo {author} {\bibfnamefont {Jennifer}\ \bibnamefont {Cano}}, \bibinfo
  {author} {\bibfnamefont {M.~G.}\ \bibnamefont {Vergniory}}, \bibinfo {author}
  {\bibfnamefont {Zhijun}\ \bibnamefont {Wang}}, \bibinfo {author}
  {\bibfnamefont {C.}~\bibnamefont {Felser}}, \bibinfo {author} {\bibfnamefont
  {M.~I.}\ \bibnamefont {Aroyo}}, \ and\ \bibinfo {author} {\bibfnamefont
  {B.~Andrei}\ \bibnamefont {Bernevig}},\ }\bibfield  {title} {\enquote
  {\bibinfo {title} {Topological quantum chemistry},}\ }\href
  {http://dx.doi.org/10.1038/nature23268} {\bibfield  {journal} {\bibinfo
  {journal} {Nature}\ }\textbf {\bibinfo {volume} {547}},\ \bibinfo {pages}
  {298} (\bibinfo {year} {2017})}\BibitemShut {NoStop}%
\bibitem [{\citenamefont {{Bzdu{\v s}ek}}\ and\ \citenamefont
  {Sigrist}(2017)}]{BzduSigristRobust}%
  \BibitemOpen
  \bibfield  {author} {\bibinfo {author} {\bibfnamefont {Tom\'{a}\v{s}}\
  \bibnamefont {{Bzdu{\v s}ek}}}\ and\ \bibinfo {author} {\bibfnamefont
  {Manfred}\ \bibnamefont {Sigrist}},\ }\bibfield  {title} {\enquote {\bibinfo
  {title} {Robust doubly charged nodal lines and nodal surfaces in
  centrosymmetric systems},}\ }\href {\doibase 10.1103/PhysRevB.96.155105}
  {\bibfield  {journal} {\bibinfo  {journal} {Phys. Rev. B}\ }\textbf {\bibinfo
  {volume} {96}},\ \bibinfo {pages} {155105} (\bibinfo {year}
  {2017})}\BibitemShut {NoStop}%
\bibitem [{\citenamefont {Slager}\ \emph {et~al.}(2017)\citenamefont {Slager},
  \citenamefont {Juri\ifmmode \check{c}\else \v{c}\fi{}i\ifmmode~\acute{c}\else
  \'{c}\fi{}},\ and\ \citenamefont {Roy}}]{BbcWeyl}%
  \BibitemOpen
  \bibfield  {author} {\bibinfo {author} {\bibfnamefont {Robert-Jan}\
  \bibnamefont {Slager}}, \bibinfo {author} {\bibfnamefont {Vladimir}\
  \bibnamefont {Juri\ifmmode \check{c}\else \v{c}\fi{}i\ifmmode~\acute{c}\else
  \'{c}\fi{}}}, \ and\ \bibinfo {author} {\bibfnamefont {Bitan}\ \bibnamefont
  {Roy}},\ }\bibfield  {title} {\enquote {\bibinfo {title} {Dissolution of
  topological {F}ermi arcs in a dirty {W}eyl semimetal},}\ }\href {\doibase
  10.1103/PhysRevB.96.201401} {\bibfield  {journal} {\bibinfo  {journal} {Phys.
  Rev. B}\ }\textbf {\bibinfo {volume} {96}},\ \bibinfo {pages} {201401}
  (\bibinfo {year} {2017})}\BibitemShut {NoStop}%
\bibitem [{\citenamefont {Bouhon}\ \emph {et~al.}(2018)\citenamefont {Bouhon},
  \citenamefont {Schmidt},\ and\ \citenamefont {Black-Schaffer}}]{Bouhon_HHL}%
  \BibitemOpen
  \bibfield  {author} {\bibinfo {author} {\bibfnamefont {Adrien}\ \bibnamefont
  {Bouhon}}, \bibinfo {author} {\bibfnamefont {Johann}\ \bibnamefont
  {Schmidt}}, \ and\ \bibinfo {author} {\bibfnamefont {Annica~M.}\ \bibnamefont
  {Black-Schaffer}},\ }\bibfield  {title} {\enquote {\bibinfo {title}
  {Topological nodal superconducting phases and topological phase transition in
  the hyperhoneycomb lattice},}\ }\href {\doibase 10.1103/PhysRevB.97.104508}
  {\bibfield  {journal} {\bibinfo  {journal} {Phys. Rev. B}\ }\textbf {\bibinfo
  {volume} {97}},\ \bibinfo {pages} {104508} (\bibinfo {year}
  {2018})}\BibitemShut {NoStop}%
\bibitem [{\citenamefont {Rhim}\ \emph {et~al.}(2018)\citenamefont {Rhim},
  \citenamefont {Bardarson},\ and\ \citenamefont {Slager}}]{UnifiedBBc}%
  \BibitemOpen
  \bibfield  {author} {\bibinfo {author} {\bibfnamefont {Jun-Won}\ \bibnamefont
  {Rhim}}, \bibinfo {author} {\bibfnamefont {Jens~H.}\ \bibnamefont
  {Bardarson}}, \ and\ \bibinfo {author} {\bibfnamefont {Robert-Jan}\
  \bibnamefont {Slager}},\ }\bibfield  {title} {\enquote {\bibinfo {title}
  {Unified bulk-boundary correspondence for band insulators},}\ }\href
  {\doibase 10.1103/PhysRevB.97.115143} {\bibfield  {journal} {\bibinfo
  {journal} {Phys. Rev. B}\ }\textbf {\bibinfo {volume} {97}},\ \bibinfo
  {pages} {115143} (\bibinfo {year} {2018})}\BibitemShut {NoStop}%
\bibitem [{\citenamefont {{H{\"o}ller}}\ and\ \citenamefont
  {Alexandradinata}(2018)}]{HolAlex_Bloch_Oscillations}%
  \BibitemOpen
  \bibfield  {author} {\bibinfo {author} {\bibfnamefont {J.}~\bibnamefont
  {{H{\"o}ller}}}\ and\ \bibinfo {author} {\bibfnamefont {A.}~\bibnamefont
  {Alexandradinata}},\ }\bibfield  {title} {\enquote {\bibinfo {title}
  {Topological {B}loch oscillations},}\ }\href {\doibase
  10.1103/PhysRevB.98.024310} {\bibfield  {journal} {\bibinfo  {journal} {Phys.
  Rev. B}\ }\textbf {\bibinfo {volume} {98}},\ \bibinfo {pages} {024310}
  (\bibinfo {year} {2018})}\BibitemShut {NoStop}%
\bibitem [{\citenamefont {Sun}\ \emph {et~al.}(2018{\natexlab{a}})\citenamefont
  {Sun}, \citenamefont {Zhang},\ and\ \citenamefont
  {Bzdu\ifmmode~\check{s}\else \v{s}\fi{}ek}}]{Bzdusek_conversion}%
  \BibitemOpen
  \bibfield  {author} {\bibinfo {author} {\bibfnamefont {Xiao-Qi}\ \bibnamefont
  {Sun}}, \bibinfo {author} {\bibfnamefont {Shou-Cheng}\ \bibnamefont {Zhang}},
  \ and\ \bibinfo {author} {\bibfnamefont {Tom\'{a}\v{s}}\ \bibnamefont
  {Bzdu\ifmmode~\check{s}\else \v{s}\fi{}ek}},\ }\bibfield  {title} {\enquote
  {\bibinfo {title} {Conversion rules for {W}eyl points and nodal lines in
  topological media},}\ }\href {\doibase 10.1103/PhysRevLett.121.106402}
  {\bibfield  {journal} {\bibinfo  {journal} {Phys. Rev. Lett.}\ }\textbf
  {\bibinfo {volume} {121}},\ \bibinfo {pages} {106402} (\bibinfo {year}
  {2018}{\natexlab{a}})}\BibitemShut {NoStop}%
\bibitem [{\citenamefont {Slager}(2019)}]{Codefects2}%
  \BibitemOpen
  \bibfield  {author} {\bibinfo {author} {\bibfnamefont {Robert-Jan}\
  \bibnamefont {Slager}},\ }\bibfield  {title} {\enquote {\bibinfo {title} {The
  translational side of topological band insulators},}\ }\href {\doibase
  https://doi.org/10.1016/j.jpcs.2018.01.023} {\bibfield  {journal} {\bibinfo
  {journal} {J. Phys. Chem. Solids}\ }\textbf {\bibinfo {volume} {128}},\
  \bibinfo {pages} {24 -- 38} (\bibinfo {year} {2019})},\ \bibinfo {note}
  {spin-Orbit Coupled Materials}\BibitemShut {NoStop}%
\bibitem [{\citenamefont {Kariyado}\ and\ \citenamefont
  {Slager}(2019)}]{semimetals}%
  \BibitemOpen
  \bibfield  {author} {\bibinfo {author} {\bibfnamefont {Toshikaze}\
  \bibnamefont {Kariyado}}\ and\ \bibinfo {author} {\bibfnamefont {Robert-Jan}\
  \bibnamefont {Slager}},\ }\bibfield  {title} {\enquote {\bibinfo {title}
  {$\ensuremath{\pi}$-fluxes, semimetals, and flat bands in artificial
  materials},}\ }\href {\doibase 10.1103/PhysRevResearch.1.032027} {\bibfield
  {journal} {\bibinfo  {journal} {Phys. Rev. Research}\ }\textbf {\bibinfo
  {volume} {1}},\ \bibinfo {pages} {032027} (\bibinfo {year}
  {2019})}\BibitemShut {NoStop}%
\bibitem [{\citenamefont {Po}\ \emph {et~al.}(2018)\citenamefont {Po},
  \citenamefont {Watanabe},\ and\ \citenamefont {Vishwanath}}]{Ft1}%
  \BibitemOpen
  \bibfield  {author} {\bibinfo {author} {\bibfnamefont {Hoi~Chun}\
  \bibnamefont {Po}}, \bibinfo {author} {\bibfnamefont {Haruki}\ \bibnamefont
  {Watanabe}}, \ and\ \bibinfo {author} {\bibfnamefont {Ashvin}\ \bibnamefont
  {Vishwanath}},\ }\bibfield  {title} {\enquote {\bibinfo {title} {Fragile
  {T}opology and {W}annier {O}bstructions},}\ }\href {\doibase
  10.1103/PhysRevLett.121.126402} {\bibfield  {journal} {\bibinfo  {journal}
  {Phys. Rev. Lett.}\ }\textbf {\bibinfo {volume} {121}},\ \bibinfo {pages}
  {126402} (\bibinfo {year} {2018})}\BibitemShut {NoStop}%
\bibitem [{\citenamefont {Bouhon}\ \emph
  {et~al.}(2019{\natexlab{a}})\citenamefont {Bouhon}, \citenamefont
  {Black-Schaffer},\ and\ \citenamefont {Slager}}]{bouhon2018wilson}%
  \BibitemOpen
  \bibfield  {author} {\bibinfo {author} {\bibfnamefont {Adrien}\ \bibnamefont
  {Bouhon}}, \bibinfo {author} {\bibfnamefont {Annica~M.}\ \bibnamefont
  {Black-Schaffer}}, \ and\ \bibinfo {author} {\bibfnamefont {Robert-Jan}\
  \bibnamefont {Slager}},\ }\bibfield  {title} {\enquote {\bibinfo {title}
  {Wilson loop approach to fragile topology of split elementary band
  representations and topological crystalline insulators with time-reversal
  symmetry},}\ }\href {\doibase 10.1103/PhysRevB.100.195135} {\bibfield
  {journal} {\bibinfo  {journal} {Phys. Rev. B}\ }\textbf {\bibinfo {volume}
  {100}},\ \bibinfo {pages} {195135} (\bibinfo {year}
  {2019}{\natexlab{a}})}\BibitemShut {NoStop}%
\bibitem [{\citenamefont {Alexandradinata}\ and\ \citenamefont
  {H\"oller}(2018)}]{AlexHol_nogo}%
  \BibitemOpen
  \bibfield  {author} {\bibinfo {author} {\bibfnamefont {A.}~\bibnamefont
  {Alexandradinata}}\ and\ \bibinfo {author} {\bibfnamefont {J.}~\bibnamefont
  {H\"oller}},\ }\bibfield  {title} {\enquote {\bibinfo {title} {No-go theorem
  for topological insulators and high-throughput identification of {C}hern
  insulators},}\ }\href {\doibase 10.1103/PhysRevB.98.184305} {\bibfield
  {journal} {\bibinfo  {journal} {Phys. Rev. B}\ }\textbf {\bibinfo {volume}
  {98}},\ \bibinfo {pages} {184305} (\bibinfo {year} {2018})}\BibitemShut
  {NoStop}%
\bibitem [{\citenamefont {Bradlyn}\ \emph {et~al.}(2019)\citenamefont
  {Bradlyn}, \citenamefont {Wang}, \citenamefont {Cano},\ and\ \citenamefont
  {Bernevig}}]{Bradlyn_fragile}%
  \BibitemOpen
  \bibfield  {author} {\bibinfo {author} {\bibfnamefont {Barry}\ \bibnamefont
  {Bradlyn}}, \bibinfo {author} {\bibfnamefont {Zhijun}\ \bibnamefont {Wang}},
  \bibinfo {author} {\bibfnamefont {Jennifer}\ \bibnamefont {Cano}}, \ and\
  \bibinfo {author} {\bibfnamefont {B.~Andrei}\ \bibnamefont {Bernevig}},\
  }\bibfield  {title} {\enquote {\bibinfo {title} {Disconnected elementary band
  representations, fragile topology, and wilson loops as topological indices:
  {A}n example on the triangular lattice},}\ }\href {\doibase
  10.1103/PhysRevB.99.045140} {\bibfield  {journal} {\bibinfo  {journal} {Phys.
  Rev. B}\ }\textbf {\bibinfo {volume} {99}},\ \bibinfo {pages} {045140}
  (\bibinfo {year} {2019})}\BibitemShut {NoStop}%
\bibitem [{\citenamefont {Song}\ \emph {et~al.}(2019)\citenamefont {Song},
  \citenamefont {Elcoro}, \citenamefont {Regnault},\ and\ \citenamefont
  {Bernevig}}]{song2019fragile}%
  \BibitemOpen
  \bibfield  {author} {\bibinfo {author} {\bibfnamefont {Zhida}\ \bibnamefont
  {Song}}, \bibinfo {author} {\bibfnamefont {L.}~\bibnamefont {Elcoro}},
  \bibinfo {author} {\bibfnamefont {Nicolas}\ \bibnamefont {Regnault}}, \ and\
  \bibinfo {author} {\bibfnamefont {B.~Andrei}\ \bibnamefont {Bernevig}},\
  }\href@noop {} {\enquote {\bibinfo {title} {Fragile {P}hases as {A}ffine
  {M}onoids: {F}ull {C}lassification and {M}aterial {E}xamples},}\ } (\bibinfo
  {year} {2019}),\ \Eprint {http://arxiv.org/abs/1905.03262} {arXiv:1905.03262
  [cond-mat.mes-hall]} \BibitemShut {NoStop}%
\bibitem [{\citenamefont {Hwang}\ \emph {et~al.}(2019)\citenamefont {Hwang},
  \citenamefont {Ahn},\ and\ \citenamefont {Yang}}]{Hwang_inversion_fragile}%
  \BibitemOpen
  \bibfield  {author} {\bibinfo {author} {\bibfnamefont {Yoonseok}\
  \bibnamefont {Hwang}}, \bibinfo {author} {\bibfnamefont {Junyeong}\
  \bibnamefont {Ahn}}, \ and\ \bibinfo {author} {\bibfnamefont {Bohm-Jung}\
  \bibnamefont {Yang}},\ }\bibfield  {title} {\enquote {\bibinfo {title}
  {Fragile topology protected by inversion symmetry: {D}iagnosis, bulk-boundary
  correspondence, and {W}ilson loop},}\ }\href {\doibase
  10.1103/PhysRevB.100.205126} {\bibfield  {journal} {\bibinfo  {journal}
  {Phys. Rev. B}\ }\textbf {\bibinfo {volume} {100}},\ \bibinfo {pages}
  {205126} (\bibinfo {year} {2019})}\BibitemShut {NoStop}%
\bibitem [{\citenamefont {Alexandradinata}\ \emph
  {et~al.}(2019{\natexlab{a}})\citenamefont {Alexandradinata}, \citenamefont
  {Holler}, \citenamefont {Wang}, \citenamefont {Cheng},\ and\ \citenamefont
  {Lu}}]{alex2019crystallographic}%
  \BibitemOpen
  \bibfield  {author} {\bibinfo {author} {\bibfnamefont {A.}~\bibnamefont
  {Alexandradinata}}, \bibinfo {author} {\bibfnamefont {J.}~\bibnamefont
  {Holler}}, \bibinfo {author} {\bibfnamefont {Chong}\ \bibnamefont {Wang}},
  \bibinfo {author} {\bibfnamefont {Hengbin}\ \bibnamefont {Cheng}}, \ and\
  \bibinfo {author} {\bibfnamefont {Ling}\ \bibnamefont {Lu}},\ }\href@noop {}
  {\enquote {\bibinfo {title} {Crystallographic splitting theorem for band
  representations and fragile topological photonic crystals},}\ } (\bibinfo
  {year} {2019}{\natexlab{a}}),\ \Eprint {http://arxiv.org/abs/1908.08541}
  {arXiv:1908.08541 [cond-mat.str-el]} \BibitemShut {NoStop}%
\bibitem [{\citenamefont {Elcoro}\ \emph {et~al.}(2020)\citenamefont {Elcoro},
  \citenamefont {Song},\ and\ \citenamefont {Bernevig}}]{Elcoro_SmithDecomp}%
  \BibitemOpen
  \bibfield  {author} {\bibinfo {author} {\bibfnamefont {Luis}\ \bibnamefont
  {Elcoro}}, \bibinfo {author} {\bibfnamefont {Zhida}\ \bibnamefont {Song}}, \
  and\ \bibinfo {author} {\bibfnamefont {B.~Andrei}\ \bibnamefont {Bernevig}},\
  }\href@noop {} {\enquote {\bibinfo {title} {Application of the induction
  procedure and the {S}mith decomposition in the calculation and topological
  classification of electronic band structures in the 230 space groups},}\ }
  (\bibinfo {year} {2020}),\ \Eprint {http://arxiv.org/abs/arXiv:2002.03836}
  {arXiv:2002.03836} \BibitemShut {NoStop}%
\bibitem [{\citenamefont {Song}\ \emph {et~al.}(2020)\citenamefont {Song},
  \citenamefont {Elcoro},\ and\ \citenamefont {Bernevig}}]{Song794}%
  \BibitemOpen
  \bibfield  {author} {\bibinfo {author} {\bibfnamefont {Zhi-Da}\ \bibnamefont
  {Song}}, \bibinfo {author} {\bibfnamefont {Luis}\ \bibnamefont {Elcoro}}, \
  and\ \bibinfo {author} {\bibfnamefont {B.~Andrei}\ \bibnamefont {Bernevig}},\
  }\bibfield  {title} {\enquote {\bibinfo {title} {Twisted bulk-boundary
  correspondence of fragile topology},}\ }\href {\doibase
  10.1126/science.aaz7650} {\bibfield  {journal} {\bibinfo  {journal}
  {Science}\ }\textbf {\bibinfo {volume} {367}},\ \bibinfo {pages} {794--797}
  (\bibinfo {year} {2020})}\BibitemShut {NoStop}%
\bibitem [{\citenamefont {Peri}\ \emph {et~al.}(2020)\citenamefont {Peri},
  \citenamefont {Song}, \citenamefont {Serra-Garcia}, \citenamefont {Engeler},
  \citenamefont {Queiroz}, \citenamefont {Huang}, \citenamefont {Deng},
  \citenamefont {Liu}, \citenamefont {Bernevig},\ and\ \citenamefont
  {Huber}}]{Peri797}%
  \BibitemOpen
  \bibfield  {author} {\bibinfo {author} {\bibfnamefont {Valerio}\ \bibnamefont
  {Peri}}, \bibinfo {author} {\bibfnamefont {Zhi-Da}\ \bibnamefont {Song}},
  \bibinfo {author} {\bibfnamefont {Marc}\ \bibnamefont {Serra-Garcia}},
  \bibinfo {author} {\bibfnamefont {Pascal}\ \bibnamefont {Engeler}}, \bibinfo
  {author} {\bibfnamefont {Raquel}\ \bibnamefont {Queiroz}}, \bibinfo {author}
  {\bibfnamefont {Xueqin}\ \bibnamefont {Huang}}, \bibinfo {author}
  {\bibfnamefont {Weiyin}\ \bibnamefont {Deng}}, \bibinfo {author}
  {\bibfnamefont {Zhengyou}\ \bibnamefont {Liu}}, \bibinfo {author}
  {\bibfnamefont {B.~Andrei}\ \bibnamefont {Bernevig}}, \ and\ \bibinfo
  {author} {\bibfnamefont {Sebastian~D.}\ \bibnamefont {Huber}},\ }\bibfield
  {title} {\enquote {\bibinfo {title} {Experimental characterization of fragile
  topology in an acoustic metamaterial},}\ }\href {\doibase
  10.1126/science.aaz7654} {\bibfield  {journal} {\bibinfo  {journal}
  {Science}\ }\textbf {\bibinfo {volume} {367}},\ \bibinfo {pages} {797--800}
  (\bibinfo {year} {2020})}\BibitemShut {NoStop}%
\bibitem [{\citenamefont {Kane}\ and\ \citenamefont
  {Mele}(2005)}]{KaneMele_Z2}%
  \BibitemOpen
  \bibfield  {author} {\bibinfo {author} {\bibfnamefont {C.~L.}\ \bibnamefont
  {Kane}}\ and\ \bibinfo {author} {\bibfnamefont {E.~J.}\ \bibnamefont
  {Mele}},\ }\bibfield  {title} {\enquote {\bibinfo {title} {${Z}_{2}$
  topological order and the quantum spin hall effect},}\ }\href {\doibase
  10.1103/PhysRevLett.95.146802} {\bibfield  {journal} {\bibinfo  {journal}
  {Phys. Rev. Lett.}\ }\textbf {\bibinfo {volume} {95}},\ \bibinfo {pages}
  {146802} (\bibinfo {year} {2005})}\BibitemShut {NoStop}%
\bibitem [{\citenamefont {Ahn}\ \emph {et~al.}(2019)\citenamefont {Ahn},
  \citenamefont {Park},\ and\ \citenamefont {Yang}}]{BJY_nielsen}%
  \BibitemOpen
  \bibfield  {author} {\bibinfo {author} {\bibfnamefont {Junyeong}\
  \bibnamefont {Ahn}}, \bibinfo {author} {\bibfnamefont {Sungjoon}\
  \bibnamefont {Park}}, \ and\ \bibinfo {author} {\bibfnamefont {Bohm-Jung}\
  \bibnamefont {Yang}},\ }\bibfield  {title} {\enquote {\bibinfo {title}
  {Failure of {N}ielsen-{N}inomiya {T}heorem and {F}ragile {T}opology in
  {T}wo-{D}imensional {S}ystems with {S}pace-{T}ime {I}nversion {S}ymmetry:
  {A}pplication to {T}wisted {B}ilayer {G}raphene at {M}agic {A}ngle},}\ }\href
  {\doibase 10.1103/PhysRevX.9.021013} {\bibfield  {journal} {\bibinfo
  {journal} {Phys. Rev. X}\ }\textbf {\bibinfo {volume} {9}},\ \bibinfo {pages}
  {021013} (\bibinfo {year} {2019})}\BibitemShut {NoStop}%
\bibitem [{\citenamefont {Wieder}\ and\ \citenamefont
  {Bernevig}(2018)}]{Wieder_axion}%
  \BibitemOpen
  \bibfield  {author} {\bibinfo {author} {\bibfnamefont {Benjamin~J.}\
  \bibnamefont {Wieder}}\ and\ \bibinfo {author} {\bibfnamefont {B.~Andrei}\
  \bibnamefont {Bernevig}},\ }\href@noop {} {\enquote {\bibinfo {title} {The
  axion insulator as a pump of fragile topology},}\ } (\bibinfo {year}
  {2018}),\ \Eprint {http://arxiv.org/abs/arXiv:1810.02373} {arXiv:1810.02373}
  \BibitemShut {NoStop}%
\bibitem [{\citenamefont {Kooi}\ \emph {et~al.}(2019)\citenamefont {Kooi},
  \citenamefont {van Miert},\ and\ \citenamefont {Ortix}}]{Kooi_nested}%
  \BibitemOpen
  \bibfield  {author} {\bibinfo {author} {\bibfnamefont {Sander~H.}\
  \bibnamefont {Kooi}}, \bibinfo {author} {\bibfnamefont {Guido}\ \bibnamefont
  {van Miert}}, \ and\ \bibinfo {author} {\bibfnamefont {Carmine}\ \bibnamefont
  {Ortix}},\ }\bibfield  {title} {\enquote {\bibinfo {title} {Classification of
  crystalline insulators without symmetry indicators: {A}tomic and fragile
  topological phases in twofold rotation symmetric systems},}\ }\href {\doibase
  10.1103/PhysRevB.100.115160} {\bibfield  {journal} {\bibinfo  {journal}
  {Phys. Rev. B}\ }\textbf {\bibinfo {volume} {100}},\ \bibinfo {pages}
  {115160} (\bibinfo {year} {2019})}\BibitemShut {NoStop}%
\bibitem [{\citenamefont {Ahn}\ \emph {et~al.}(2018)\citenamefont {Ahn},
  \citenamefont {Kim}, \citenamefont {Kim},\ and\ \citenamefont
  {Yang}}]{BJY_linking}%
  \BibitemOpen
  \bibfield  {author} {\bibinfo {author} {\bibfnamefont {Junyeong}\
  \bibnamefont {Ahn}}, \bibinfo {author} {\bibfnamefont {Dongwook}\
  \bibnamefont {Kim}}, \bibinfo {author} {\bibfnamefont {Youngkuk}\
  \bibnamefont {Kim}}, \ and\ \bibinfo {author} {\bibfnamefont {Bohm-Jung}\
  \bibnamefont {Yang}},\ }\bibfield  {title} {\enquote {\bibinfo {title} {Band
  topology and linking structure of nodal line semimetals with ${Z}_{2}$
  monopole charges},}\ }\href {\doibase 10.1103/PhysRevLett.121.106403}
  {\bibfield  {journal} {\bibinfo  {journal} {Phys. Rev. Lett.}\ }\textbf
  {\bibinfo {volume} {121}},\ \bibinfo {pages} {106403} (\bibinfo {year}
  {2018})}\BibitemShut {NoStop}%
\bibitem [{\citenamefont {Wu}\ \emph {et~al.}(2019)\citenamefont {Wu},
  \citenamefont {Soluyanov},\ and\ \citenamefont {Bzdu{\v s}ek}}]{Wu1273}%
  \BibitemOpen
  \bibfield  {author} {\bibinfo {author} {\bibfnamefont {QuanSheng}\
  \bibnamefont {Wu}}, \bibinfo {author} {\bibfnamefont {Alexey~A.}\
  \bibnamefont {Soluyanov}}, \ and\ \bibinfo {author} {\bibfnamefont
  {Tom{\'a}{\v s}}\ \bibnamefont {Bzdu{\v s}ek}},\ }\bibfield  {title}
  {\enquote {\bibinfo {title} {Non-{A}belian band topology in noninteracting
  metals},}\ }\href {\doibase 10.1126/science.aau8740} {\bibfield  {journal}
  {\bibinfo  {journal} {Science}\ }\textbf {\bibinfo {volume} {365}},\ \bibinfo
  {pages} {1273--1277} (\bibinfo {year} {2019})}\BibitemShut {NoStop}%
\bibitem [{\citenamefont {Tiwari}\ and\ \citenamefont
  {Bzdu\v{s}ek}(2020)}]{Tiwari:2019}%
  \BibitemOpen
  \bibfield  {author} {\bibinfo {author} {\bibfnamefont {Apoorv}\ \bibnamefont
  {Tiwari}}\ and\ \bibinfo {author} {\bibfnamefont {Tom\'{a}\v{s}}\
  \bibnamefont {Bzdu\v{s}ek}},\ }\bibfield  {title} {\enquote {\bibinfo {title}
  {Non-{A}belian topology of nodal-line rings in $\mathcal{PT}$-symmetric
  systems},}\ }\href {\doibase 10.1103/PhysRevB.101.195130} {\bibfield
  {journal} {\bibinfo  {journal} {Phys. Rev. B}\ }\textbf {\bibinfo {volume}
  {101}},\ \bibinfo {pages} {195130} (\bibinfo {year} {2020})}\BibitemShut
  {NoStop}%
\bibitem [{\citenamefont {Bouhon}\ \emph
  {et~al.}(2019{\natexlab{b}})\citenamefont {Bouhon}, \citenamefont {Wu},
  \citenamefont {Slager}, \citenamefont {Weng}, \citenamefont {Yazyev},\ and\
  \citenamefont {Bzdu\v{s}ek}}]{bouhon2019nonabelian}%
  \BibitemOpen
  \bibfield  {author} {\bibinfo {author} {\bibfnamefont {Adrien}\ \bibnamefont
  {Bouhon}}, \bibinfo {author} {\bibfnamefont {QuanSheng~Wu}\ \bibnamefont
  {Wu}}, \bibinfo {author} {\bibfnamefont {Robert-Jan}\ \bibnamefont {Slager}},
  \bibinfo {author} {\bibfnamefont {Hongming}\ \bibnamefont {Weng}}, \bibinfo
  {author} {\bibfnamefont {Oleg V.~Yazyev}\ \bibnamefont {Yazyev}}, \ and\
  \bibinfo {author} {\bibfnamefont {Tom\'{a}\v{s}}\ \bibnamefont
  {Bzdu\v{s}ek}},\ }\href@noop {} {\enquote {\bibinfo {title} {Non-{A}belian
  {R}eciprocal {B}raiding of {W}eyl {N}odes and its {M}anifestation in
  {Z}r{T}e},}\ } (\bibinfo {year} {2019}{\natexlab{b}}),\ \Eprint
  {http://arxiv.org/abs/1907.10611} {arXiv:1907.10611 [cond-mat.mes-hall]}
  \BibitemShut {NoStop}%
\bibitem [{\citenamefont {Po}\ \emph {et~al.}(2019)\citenamefont {Po},
  \citenamefont {Zou}, \citenamefont {Senthil},\ and\ \citenamefont
  {Vishwanath}}]{Potwisted}%
  \BibitemOpen
  \bibfield  {author} {\bibinfo {author} {\bibfnamefont {Hoi~Chun}\
  \bibnamefont {Po}}, \bibinfo {author} {\bibfnamefont {Liujun}\ \bibnamefont
  {Zou}}, \bibinfo {author} {\bibfnamefont {T.}~\bibnamefont {Senthil}}, \ and\
  \bibinfo {author} {\bibfnamefont {Ashvin}\ \bibnamefont {Vishwanath}},\
  }\bibfield  {title} {\enquote {\bibinfo {title} {Faithful tight-binding
  models and fragile topology of magic-angle bilayer graphene},}\ }\href
  {\doibase 10.1103/PhysRevB.99.195455} {\bibfield  {journal} {\bibinfo
  {journal} {Phys. Rev. B}\ }\textbf {\bibinfo {volume} {99}},\ \bibinfo
  {pages} {195455} (\bibinfo {year} {2019})}\BibitemShut {NoStop}%
\bibitem [{\citenamefont {Cao}\ \emph {et~al.}(2018)\citenamefont {Cao},
  \citenamefont {Fatemi}, \citenamefont {Fang}, \citenamefont {Watanabe},
  \citenamefont {Taniguchi}, \citenamefont {Kaxiras},\ and\ \citenamefont
  {Jarillo-Herrero}}]{Cao2018}%
  \BibitemOpen
  \bibfield  {author} {\bibinfo {author} {\bibfnamefont {Yuan}\ \bibnamefont
  {Cao}}, \bibinfo {author} {\bibfnamefont {Valla}\ \bibnamefont {Fatemi}},
  \bibinfo {author} {\bibfnamefont {Shiang}\ \bibnamefont {Fang}}, \bibinfo
  {author} {\bibfnamefont {Kenji}\ \bibnamefont {Watanabe}}, \bibinfo {author}
  {\bibfnamefont {Takashi}\ \bibnamefont {Taniguchi}}, \bibinfo {author}
  {\bibfnamefont {Efthimios}\ \bibnamefont {Kaxiras}}, \ and\ \bibinfo {author}
  {\bibfnamefont {Pablo}\ \bibnamefont {Jarillo-Herrero}},\ }\bibfield  {title}
  {\enquote {\bibinfo {title} {Unconventional superconductivity in magic-angle
  graphene superlattices},}\ }\href {\doibase 10.1038/nature26160} {\bibfield
  {journal} {\bibinfo  {journal} {Nature}\ }\textbf {\bibinfo {volume} {556}},\
  \bibinfo {pages} {43--50} (\bibinfo {year} {2018})}\BibitemShut {NoStop}%
\bibitem [{\citenamefont {Kang}\ and\ \citenamefont
  {Vafek}(2020)}]{Kang_Dirac_node_braiding}%
  \BibitemOpen
  \bibfield  {author} {\bibinfo {author} {\bibfnamefont {Jian}\ \bibnamefont
  {Kang}}\ and\ \bibinfo {author} {\bibfnamefont {Oskar}\ \bibnamefont
  {Vafek}},\ }\href@noop {} {\enquote {\bibinfo {title} {Non-{A}belian {D}irac
  node braiding and near-degeneracy of correlated phases at odd integer filling
  in magic angle twisted bilayer graphene},}\ } (\bibinfo {year} {2020}),\
  \Eprint {http://arxiv.org/abs/arXiv:2002.10360} {arXiv:2002.10360}
  \BibitemShut {NoStop}%
\bibitem [{\citenamefont {Ünal}\ \emph {et~al.}(2020)\citenamefont {Ünal},
  \citenamefont {Bouhon},\ and\ \citenamefont {Slager}}]{Unal_quenched_Euler}%
  \BibitemOpen
  \bibfield  {author} {\bibinfo {author} {\bibfnamefont {F.~Nur}\ \bibnamefont
  {\"Unal}}, \bibinfo {author} {\bibfnamefont {Adrien}\ \bibnamefont {Bouhon}},
  \ and\ \bibinfo {author} {\bibfnamefont {Robert-Jan}\ \bibnamefont
  {Slager}},\ }\href@noop {} {\enquote {\bibinfo {title} {Quench dynamics of
  topological euler class in optical lattices},}\ } (\bibinfo {year} {2020}),\
  \Eprint {http://arxiv.org/abs/arXiv:2005.03033} {arXiv:2005.03033}
  \BibitemShut {NoStop}%
\bibitem [{\citenamefont {Wang}\ \emph {et~al.}(2019)\citenamefont {Wang},
  \citenamefont {Wieder}, \citenamefont {Li}, \citenamefont {Yan},\ and\
  \citenamefont {Bernevig}}]{Wieder_HOTI}%
  \BibitemOpen
  \bibfield  {author} {\bibinfo {author} {\bibfnamefont {Zhijun}\ \bibnamefont
  {Wang}}, \bibinfo {author} {\bibfnamefont {Benjamin~J.}\ \bibnamefont
  {Wieder}}, \bibinfo {author} {\bibfnamefont {Jian}\ \bibnamefont {Li}},
  \bibinfo {author} {\bibfnamefont {Binghai}\ \bibnamefont {Yan}}, \ and\
  \bibinfo {author} {\bibfnamefont {B.~Andrei}\ \bibnamefont {Bernevig}},\
  }\bibfield  {title} {\enquote {\bibinfo {title} {Higher-order topology,
  monopole nodal lines, and the origin of large {F}ermi arcs in transition
  metal dichalcogenides $x{\mathrm{te}}_{2}$ ($x=\mathrm{Mo},\mathrm{W}$)},}\
  }\href {\doibase 10.1103/PhysRevLett.123.186401} {\bibfield  {journal}
  {\bibinfo  {journal} {Phys. Rev. Lett.}\ }\textbf {\bibinfo {volume} {123}},\
  \bibinfo {pages} {186401} (\bibinfo {year} {2019})}\BibitemShut {NoStop}%
\bibitem [{\citenamefont {Bouhon}(2020)}]{abouhon_EulerClassTightBinding}%
  \BibitemOpen
  \bibfield  {author} {\bibinfo {author} {\bibfnamefont {Adrien}\ \bibnamefont
  {Bouhon}},\ }\href {https://github.com/abouhon/EulerClassTightBinding}
  {\enquote {\bibinfo {title} {{3-band and 4-band real symmetric tight-binding
  models with arbitrary {E}uler class based on the {P}l\"{u}cker embedding}},}\
  }\bibinfo {howpublished} {GitHub} (\bibinfo {year} {2020}),\ \bibinfo {note}
  {publicly available Mathematica code,
  https://github.com/abouhon/EulerClassTightBinding}\BibitemShut {NoStop}%
\bibitem [{\citenamefont {Yu}\ \emph {et~al.}(2011)\citenamefont {Yu},
  \citenamefont {Qi}, \citenamefont {Bernevig}, \citenamefont {Fang},\ and\
  \citenamefont {Dai}}]{WindingKMZ2}%
  \BibitemOpen
  \bibfield  {author} {\bibinfo {author} {\bibfnamefont {Rui}\ \bibnamefont
  {Yu}}, \bibinfo {author} {\bibfnamefont {Xiao~Liang}\ \bibnamefont {Qi}},
  \bibinfo {author} {\bibfnamefont {Andrei}\ \bibnamefont {Bernevig}}, \bibinfo
  {author} {\bibfnamefont {Zhong}\ \bibnamefont {Fang}}, \ and\ \bibinfo
  {author} {\bibfnamefont {Xi}~\bibnamefont {Dai}},\ }\bibfield  {title}
  {\enquote {\bibinfo {title} {Equivalent expression of ${\mathbb{z}}_{2}$
  topological invariant for band insulators using the non-{A}belian {B}erry
  connection},}\ }\href {\doibase 10.1103/PhysRevB.84.075119} {\bibfield
  {journal} {\bibinfo  {journal} {Phys. Rev. B}\ }\textbf {\bibinfo {volume}
  {84}},\ \bibinfo {pages} {075119} (\bibinfo {year} {2011})}\BibitemShut
  {NoStop}%
\bibitem [{\citenamefont {Fang}\ \emph
  {et~al.}(2012{\natexlab{b}})\citenamefont {Fang}, \citenamefont {Gilbert},\
  and\ \citenamefont {Bernevig}}]{PointGroupsTI}%
  \BibitemOpen
  \bibfield  {author} {\bibinfo {author} {\bibfnamefont {Chen}\ \bibnamefont
  {Fang}}, \bibinfo {author} {\bibfnamefont {Matthew~J.}\ \bibnamefont
  {Gilbert}}, \ and\ \bibinfo {author} {\bibfnamefont {B.~Andrei}\ \bibnamefont
  {Bernevig}},\ }\bibfield  {title} {\enquote {\bibinfo {title} {Bulk
  topological invariants in noninteracting point group symmetric insulators},}\
  }\href {\doibase 10.1103/PhysRevB.86.115112} {\bibfield  {journal} {\bibinfo
  {journal} {Phys. Rev. B}\ }\textbf {\bibinfo {volume} {86}},\ \bibinfo
  {pages} {115112} (\bibinfo {year} {2012}{\natexlab{b}})}\BibitemShut
  {NoStop}%
\bibitem [{\citenamefont {Alexandradinata}\ and\ \citenamefont
  {Bernevig}(2016)}]{Alex_BerryPhase}%
  \BibitemOpen
  \bibfield  {author} {\bibinfo {author} {\bibfnamefont {A.}~\bibnamefont
  {Alexandradinata}}\ and\ \bibinfo {author} {\bibfnamefont {B.~Andrei}\
  \bibnamefont {Bernevig}},\ }\bibfield  {title} {\enquote {\bibinfo {title}
  {Berry-phase description of topological crystalline insulators},}\ }\href
  {\doibase 10.1103/PhysRevB.93.205104} {\bibfield  {journal} {\bibinfo
  {journal} {Phys. Rev. B}\ }\textbf {\bibinfo {volume} {93}},\ \bibinfo
  {pages} {205104} (\bibinfo {year} {2016})}\BibitemShut {NoStop}%
\bibitem [{\citenamefont {Muechler}\ \emph {et~al.}(2016)\citenamefont
  {Muechler}, \citenamefont {Alexandradinata}, \citenamefont {Neupert},\ and\
  \citenamefont {Car}}]{AlexAdiabaticSOC}%
  \BibitemOpen
  \bibfield  {author} {\bibinfo {author} {\bibfnamefont {Lukas}\ \bibnamefont
  {Muechler}}, \bibinfo {author} {\bibfnamefont {A.}~\bibnamefont
  {Alexandradinata}}, \bibinfo {author} {\bibfnamefont {Titus}\ \bibnamefont
  {Neupert}}, \ and\ \bibinfo {author} {\bibfnamefont {Roberto}\ \bibnamefont
  {Car}},\ }\bibfield  {title} {\enquote {\bibinfo {title} {Topological
  {N}onsymmorphic {M}etals from {B}and {I}nversion},}\ }\href {\doibase
  10.1103/PhysRevX.6.041069} {\bibfield  {journal} {\bibinfo  {journal} {Phys.
  Rev. X}\ }\textbf {\bibinfo {volume} {6}},\ \bibinfo {pages} {041069}
  (\bibinfo {year} {2016})}\BibitemShut {NoStop}%
\bibitem [{\citenamefont {{Bouhon}}\ and\ \citenamefont
  {{Black-Schaffer}}(2017)}]{BBS_nodal_lines}%
  \BibitemOpen
  \bibfield  {author} {\bibinfo {author} {\bibfnamefont {A.}~\bibnamefont
  {{Bouhon}}}\ and\ \bibinfo {author} {\bibfnamefont {A.~M.}\ \bibnamefont
  {{Black-Schaffer}}},\ }\bibfield  {title} {\enquote {\bibinfo {title} {{Bulk
  topology of line-nodal structures protected by space group symmetries in
  class {A}{I}}},}\ }\href@noop {} {\bibfield  {journal} {\bibinfo  {journal}
  {ArXiv e-prints}\ } (\bibinfo {year} {2017})},\ \Eprint
  {http://arxiv.org/abs/1710.04871} {arXiv:1710.04871 [cond-mat.mtrl-sci]}
  \BibitemShut {NoStop}%
\bibitem [{\citenamefont {Hatcher}(2003)}]{Hatcher_2}%
  \BibitemOpen
  \bibfield  {author} {\bibinfo {author} {\bibfnamefont {A.}~\bibnamefont
  {Hatcher}},\ }\href@noop {} {\emph {\bibinfo {title} {{V}ector {B}undles and
  {K}-{T}heory}}}\ (\bibinfo  {publisher} {Unpublished},\ \bibinfo {year}
  {2003})\BibitemShut {NoStop}%
\bibitem [{\citenamefont {Panati}(2007)}]{Panati_Chern}%
  \BibitemOpen
  \bibfield  {author} {\bibinfo {author} {\bibfnamefont {Gianluca}\
  \bibnamefont {Panati}},\ }\bibfield  {title} {\enquote {\bibinfo {title}
  {Triviality of {B}loch and {B}loch-{D}irac bundles},}\ }\href@noop {}
  {\bibfield  {journal} {\bibinfo  {journal} {Ann. Henri Poincar\'e}\ }\textbf
  {\bibinfo {volume} {8}},\ \bibinfo {pages} {995Ð1011} (\bibinfo {year}
  {2007})}\BibitemShut {NoStop}%
\bibitem [{\citenamefont {Budich}\ and\ \citenamefont
  {Diehl}(2015)}]{Budich_DensityMatrix}%
  \BibitemOpen
  \bibfield  {author} {\bibinfo {author} {\bibfnamefont {Jan~Carl}\
  \bibnamefont {Budich}}\ and\ \bibinfo {author} {\bibfnamefont {Sebastian}\
  \bibnamefont {Diehl}},\ }\bibfield  {title} {\enquote {\bibinfo {title}
  {Topology of density matrices},}\ }\href {\doibase
  10.1103/PhysRevB.91.165140} {\bibfield  {journal} {\bibinfo  {journal} {Phys.
  Rev. B}\ }\textbf {\bibinfo {volume} {91}},\ \bibinfo {pages} {165140}
  (\bibinfo {year} {2015})}\BibitemShut {NoStop}%
\bibitem [{\citenamefont {Milnor}\ and\ \citenamefont
  {Stasheff}(1974)}]{Milnor:1974}%
  \BibitemOpen
  \bibfield  {author} {\bibinfo {author} {\bibfnamefont {John~W.}\ \bibnamefont
  {Milnor}}\ and\ \bibinfo {author} {\bibfnamefont {James~D.}\ \bibnamefont
  {Stasheff}},\ }\href@noop {} {\emph {\bibinfo {title} {{C}haracteristic
  classes}}}\ (\bibinfo  {publisher} {Princeton University Press},\ \bibinfo
  {address} {Princeton, New Jersey},\ \bibinfo {year} {1974})\BibitemShut
  {NoStop}%
\bibitem [{\citenamefont {Thiang}(2016)}]{Thiang_KTheory}%
  \BibitemOpen
  \bibfield  {author} {\bibinfo {author} {\bibfnamefont {Guo~Chuan}\
  \bibnamefont {Thiang}},\ }\bibfield  {title} {\enquote {\bibinfo {title} {On
  the {K}-theoretic classification of topological phases of matter},}\ }\href
  {\doibase 10.1007/s00023-015-0418-9} {\bibfield  {journal} {\bibinfo
  {journal} {Annales Henri Poincar{\'e}}\ }\textbf {\bibinfo {volume} {17}},\
  \bibinfo {pages} {757--794} (\bibinfo {year} {2016})}\BibitemShut {NoStop}%
\bibitem [{\citenamefont {Montambaux}\ \emph {et~al.}(2018)\citenamefont
  {Montambaux}, \citenamefont {Lim}, \citenamefont {Fuchs},\ and\ \citenamefont
  {Pi\'echon}}]{Montambaux_2018}%
  \BibitemOpen
  \bibfield  {author} {\bibinfo {author} {\bibfnamefont {Gilles}\ \bibnamefont
  {Montambaux}}, \bibinfo {author} {\bibfnamefont {Lih-King}\ \bibnamefont
  {Lim}}, \bibinfo {author} {\bibfnamefont {Jean-No\"el}\ \bibnamefont
  {Fuchs}}, \ and\ \bibinfo {author} {\bibfnamefont {Fr\'ed\'eric}\
  \bibnamefont {Pi\'echon}},\ }\bibfield  {title} {\enquote {\bibinfo {title}
  {Winding {V}ector: {H}ow to {A}nnihilate {T}wo {D}irac {P}oints with the
  {S}ame {C}harge},}\ }\href {\doibase 10.1103/PhysRevLett.121.256402}
  {\bibfield  {journal} {\bibinfo  {journal} {Phys. Rev. Lett.}\ }\textbf
  {\bibinfo {volume} {121}},\ \bibinfo {pages} {256402} (\bibinfo {year}
  {2018})}\BibitemShut {NoStop}%
\bibitem [{\citenamefont {Gottlieb}(2003)}]{gottlieb_2003}%
  \BibitemOpen
  \bibfield  {author} {\bibinfo {author} {\bibfnamefont {Daniel~Henry}\
  \bibnamefont {Gottlieb}},\ }\bibfield  {title} {\enquote {\bibinfo {title}
  {{E}igenbundles, {Q}uaternions, and {B}erry's {P}hase},}\ }\href@noop {}
  {\bibfield  {journal} {\bibinfo  {journal} {ArXiv e-prints}\ } (\bibinfo
  {year} {2003})},\ \Eprint {http://arxiv.org/abs/math/0304281}
  {arXiv:math/0304281} \BibitemShut {NoStop}%
\bibitem [{\citenamefont {Avron}\ \emph {et~al.}(1983)\citenamefont {Avron},
  \citenamefont {Seiler},\ and\ \citenamefont {Simon}}]{Simon_homotopy}%
  \BibitemOpen
  \bibfield  {author} {\bibinfo {author} {\bibfnamefont {J.~E.}\ \bibnamefont
  {Avron}}, \bibinfo {author} {\bibfnamefont {R.}~\bibnamefont {Seiler}}, \
  and\ \bibinfo {author} {\bibfnamefont {B.}~\bibnamefont {Simon}},\ }\bibfield
   {title} {\enquote {\bibinfo {title} {Homotopy and {Q}uantization in
  {C}ondensed {M}atter {P}hysics},}\ }\href {\doibase
  10.1103/PhysRevLett.51.51} {\bibfield  {journal} {\bibinfo  {journal} {Phys.
  Rev. Lett.}\ }\textbf {\bibinfo {volume} {51}},\ \bibinfo {pages} {51--53}
  (\bibinfo {year} {1983})}\BibitemShut {NoStop}%
\bibitem [{\citenamefont {Kennedy}\ and\ \citenamefont
  {Guggenheim}(2015)}]{Kennedy_homotopy}%
  \BibitemOpen
  \bibfield  {author} {\bibinfo {author} {\bibfnamefont {Ricardo}\ \bibnamefont
  {Kennedy}}\ and\ \bibinfo {author} {\bibfnamefont {Charles}\ \bibnamefont
  {Guggenheim}},\ }\bibfield  {title} {\enquote {\bibinfo {title} {Homotopy
  theory of strong and weak topological insulators},}\ }\href {\doibase
  10.1103/PhysRevB.91.245148} {\bibfield  {journal} {\bibinfo  {journal} {Phys.
  Rev. B}\ }\textbf {\bibinfo {volume} {91}},\ \bibinfo {pages} {245148}
  (\bibinfo {year} {2015})}\BibitemShut {NoStop}%
\bibitem [{\citenamefont {Wojcik}\ \emph {et~al.}(2020)\citenamefont {Wojcik},
  \citenamefont {Sun}, \citenamefont {Bzdu\v{s}ek},\ and\ \citenamefont
  {Fan}}]{Wojcik:2019}%
  \BibitemOpen
  \bibfield  {author} {\bibinfo {author} {\bibfnamefont {Charles~C.}\
  \bibnamefont {Wojcik}}, \bibinfo {author} {\bibfnamefont {Xiao-Qi}\
  \bibnamefont {Sun}}, \bibinfo {author} {\bibfnamefont {Tom\'{a}\v{s}}\
  \bibnamefont {Bzdu\v{s}ek}}, \ and\ \bibinfo {author} {\bibfnamefont
  {Shanhui}\ \bibnamefont {Fan}},\ }\bibfield  {title} {\enquote {\bibinfo
  {title} {Homotopy characterization of non-{H}ermitian hamiltonians},}\ }\href
  {\doibase 10.1103/PhysRevB.101.205417} {\bibfield  {journal} {\bibinfo
  {journal} {Phys. Rev. B}\ }\textbf {\bibinfo {volume} {101}},\ \bibinfo
  {pages} {205417} (\bibinfo {year} {2020})}\BibitemShut {NoStop}%
\bibitem [{\citenamefont {Hatcher}(2001)}]{Hatcher_1}%
  \BibitemOpen
  \bibfield  {author} {\bibinfo {author} {\bibfnamefont {A.}~\bibnamefont
  {Hatcher}},\ }\href@noop {} {\emph {\bibinfo {title} {{A}lgebraic
  {T}opology}}}\ (\bibinfo  {publisher} {Cambridge University Press},\ \bibinfo
  {year} {2001})\BibitemShut {NoStop}%
\bibitem [{\citenamefont {Sun}\ \emph {et~al.}(2020)\citenamefont {Sun},
  \citenamefont {Wojcik}, \citenamefont {Fan},\ and\ \citenamefont
  {Bzdu\v{s}ek}}]{Sun:2019}%
  \BibitemOpen
  \bibfield  {author} {\bibinfo {author} {\bibfnamefont {Xiao-Qi}\ \bibnamefont
  {Sun}}, \bibinfo {author} {\bibfnamefont {Charles~C.}\ \bibnamefont
  {Wojcik}}, \bibinfo {author} {\bibfnamefont {Shanhui}\ \bibnamefont {Fan}}, \
  and\ \bibinfo {author} {\bibfnamefont {Tom\'{a}\v{s}}\ \bibnamefont
  {Bzdu\v{s}ek}},\ }\bibfield  {title} {\enquote {\bibinfo {title} {Alice
  strings in non-{H}ermitian systems},}\ }\href {\doibase
  10.1103/PhysRevResearch.2.023226} {\bibfield  {journal} {\bibinfo  {journal}
  {Phys. Rev. Research}\ }\textbf {\bibinfo {volume} {2}},\ \bibinfo {pages}
  {023226} (\bibinfo {year} {2020})}\BibitemShut {NoStop}%
\bibitem [{\citenamefont {Zhao}\ and\ \citenamefont {Lu}(2017)}]{Zhao_PT}%
  \BibitemOpen
  \bibfield  {author} {\bibinfo {author} {\bibfnamefont {Y.~X.}\ \bibnamefont
  {Zhao}}\ and\ \bibinfo {author} {\bibfnamefont {Y.}~\bibnamefont {Lu}},\
  }\bibfield  {title} {\enquote {\bibinfo {title} {${P}{T}$-{S}ymmetric real
  {D}irac {F}ermions and {S}emimetals},}\ }\href {\doibase
  10.1103/PhysRevLett.118.056401} {\bibfield  {journal} {\bibinfo  {journal}
  {Phys. Rev. Lett.}\ }\textbf {\bibinfo {volume} {118}},\ \bibinfo {pages}
  {056401} (\bibinfo {year} {2017})}\BibitemShut {NoStop}%
\bibitem [{\citenamefont {Ahn}\ and\ \citenamefont {Yang}(2019)}]{Ahn2019}%
  \BibitemOpen
  \bibfield  {author} {\bibinfo {author} {\bibfnamefont {Junyeong}\
  \bibnamefont {Ahn}}\ and\ \bibinfo {author} {\bibfnamefont {Bohm-Jung}\
  \bibnamefont {Yang}},\ }\bibfield  {title} {\enquote {\bibinfo {title}
  {Symmetry representation approach to topological invariants in
  ${C}_{2z}{T}$-symmetric systems},}\ }\href {\doibase
  10.1103/PhysRevB.99.235125} {\bibfield  {journal} {\bibinfo  {journal} {Phys.
  Rev. B}\ }\textbf {\bibinfo {volume} {99}},\ \bibinfo {pages} {235125}
  (\bibinfo {year} {2019})}\BibitemShut {NoStop}%
\bibitem [{\citenamefont {Moore}\ \emph {et~al.}(2008)\citenamefont {Moore},
  \citenamefont {Ran},\ and\ \citenamefont {Wen}}]{Hopf_1}%
  \BibitemOpen
  \bibfield  {author} {\bibinfo {author} {\bibfnamefont {Joel~E.}\ \bibnamefont
  {Moore}}, \bibinfo {author} {\bibfnamefont {Ying}\ \bibnamefont {Ran}}, \
  and\ \bibinfo {author} {\bibfnamefont {Xiao-Gang}\ \bibnamefont {Wen}},\
  }\bibfield  {title} {\enquote {\bibinfo {title} {Topological {S}urface states
  in {T}hree-{D}imensional {M}agnetic {I}nsulators},}\ }\href {\doibase
  10.1103/PhysRevLett.101.186805} {\bibfield  {journal} {\bibinfo  {journal}
  {Phys. Rev. Lett.}\ }\textbf {\bibinfo {volume} {101}},\ \bibinfo {pages}
  {186805} (\bibinfo {year} {2008})}\BibitemShut {NoStop}%
\bibitem [{\citenamefont {Deng}\ \emph {et~al.}(2013)\citenamefont {Deng},
  \citenamefont {Wang}, \citenamefont {Shen},\ and\ \citenamefont
  {Duan}}]{Hopf_2}%
  \BibitemOpen
  \bibfield  {author} {\bibinfo {author} {\bibfnamefont {D.-L.}\ \bibnamefont
  {Deng}}, \bibinfo {author} {\bibfnamefont {S.-T.}\ \bibnamefont {Wang}},
  \bibinfo {author} {\bibfnamefont {C.}~\bibnamefont {Shen}}, \ and\ \bibinfo
  {author} {\bibfnamefont {L.-M.}\ \bibnamefont {Duan}},\ }\bibfield  {title}
  {\enquote {\bibinfo {title} {Hopf insulators and their topologically
  protected surface states},}\ }\href {\doibase 10.1103/PhysRevB.88.201105}
  {\bibfield  {journal} {\bibinfo  {journal} {Phys. Rev. B}\ }\textbf {\bibinfo
  {volume} {88}},\ \bibinfo {pages} {201105} (\bibinfo {year}
  {2013})}\BibitemShut {NoStop}%
\bibitem [{\citenamefont {\"Unal}\ \emph {et~al.}(2019)\citenamefont {\"Unal},
  \citenamefont {Eckardt},\ and\ \citenamefont {Slager}}]{Unal2019}%
  \BibitemOpen
  \bibfield  {author} {\bibinfo {author} {\bibfnamefont {F.~Nur}\ \bibnamefont
  {\"Unal}}, \bibinfo {author} {\bibfnamefont {Andr\'e}\ \bibnamefont
  {Eckardt}}, \ and\ \bibinfo {author} {\bibfnamefont {Robert-Jan}\
  \bibnamefont {Slager}},\ }\bibfield  {title} {\enquote {\bibinfo {title}
  {Hopf characterization of two-dimensional {F}loquet topological
  insulators},}\ }\href {\doibase 10.1103/PhysRevResearch.1.022003} {\bibfield
  {journal} {\bibinfo  {journal} {Phys. Rev. Research}\ }\textbf {\bibinfo
  {volume} {1}},\ \bibinfo {pages} {022003} (\bibinfo {year}
  {2019})}\BibitemShut {NoStop}%
\bibitem [{\citenamefont {Alexandradinata}\ \emph
  {et~al.}(2019{\natexlab{b}})\citenamefont {Alexandradinata}, \citenamefont
  {Nelson},\ and\ \citenamefont {Soluyanov}}]{Alex:2019}%
  \BibitemOpen
  \bibfield  {author} {\bibinfo {author} {\bibfnamefont {A.}~\bibnamefont
  {Alexandradinata}}, \bibinfo {author} {\bibfnamefont {Aleksandra}\
  \bibnamefont {Nelson}}, \ and\ \bibinfo {author} {\bibfnamefont {Alexey~A.}\
  \bibnamefont {Soluyanov}},\ }\bibfield  {title} {\enquote {\bibinfo {title}
  {The actually robust surface signature of a {H}opf insulator:
  {B}ulk-to-boundary flow of {B}erry curvature beyond the anomaly inflow
  paradigm},}\ }\href {https://arxiv.org/abs/1910.10717} {\bibfield  {journal}
  {\bibinfo  {journal} {arXiv:1910.10717}\ } (\bibinfo {year}
  {2019}{\natexlab{b}})}\BibitemShut {NoStop}%
\bibitem [{\citenamefont {Kozlov}(2000)}]{Kozlov_Gr}%
  \BibitemOpen
  \bibfield  {author} {\bibinfo {author} {\bibfnamefont {S.~E.}\ \bibnamefont
  {Kozlov}},\ }\bibfield  {title} {\enquote {\bibinfo {title} {{G}eometry of
  real grassmann manifolds. {P}arts {I}, {II}.}}\ }\href
  {https://doi.org/10.1007/s10958-000-0008-2} {\bibfield  {journal} {\bibinfo
  {journal} {J. Math. Sci.}\ }\textbf {\bibinfo {volume} {math/0304281}},\
  \bibinfo {pages} {2239} (\bibinfo {year} {2000})}\BibitemShut {NoStop}%
\bibitem [{\citenamefont {Volovik}\ and\ \citenamefont
  {Mineev}(2018)}]{volovik2018investigation}%
  \BibitemOpen
  \bibfield  {author} {\bibinfo {author} {\bibfnamefont {G.~E.}\ \bibnamefont
  {Volovik}}\ and\ \bibinfo {author} {\bibfnamefont {V.~P.}\ \bibnamefont
  {Mineev}},\ }\bibfield  {title} {\enquote {\bibinfo {title} {Investigation of
  singularities in superfluid {H}e$^3$ in liquid crystals by the homotopic
  topology methods},}\ }in\ \href@noop {} {\emph {\bibinfo {booktitle} {Basic
  Notions Of Condensed Matter Physics}}}\ (\bibinfo  {publisher} {CRC Press},\
  \bibinfo {year} {2018})\ pp.\ \bibinfo {pages} {392--401}\BibitemShut
  {NoStop}%
\bibitem [{\citenamefont {Beekman}\ \emph {et~al.}(2017)\citenamefont
  {Beekman}, \citenamefont {Nissinen}, \citenamefont {Wu}, \citenamefont {Liu},
  \citenamefont {Slager}, \citenamefont {Nussinov}, \citenamefont {Cvetkovic},\
  and\ \citenamefont {Zaanen}}]{Beekman20171}%
  \BibitemOpen
  \bibfield  {author} {\bibinfo {author} {\bibfnamefont {Aron~J.}\ \bibnamefont
  {Beekman}}, \bibinfo {author} {\bibfnamefont {Jaakko}\ \bibnamefont
  {Nissinen}}, \bibinfo {author} {\bibfnamefont {Kai}\ \bibnamefont {Wu}},
  \bibinfo {author} {\bibfnamefont {Ke}~\bibnamefont {Liu}}, \bibinfo {author}
  {\bibfnamefont {Robert-Jan}\ \bibnamefont {Slager}}, \bibinfo {author}
  {\bibfnamefont {Zohar}\ \bibnamefont {Nussinov}}, \bibinfo {author}
  {\bibfnamefont {Vladimir}\ \bibnamefont {Cvetkovic}}, \ and\ \bibinfo
  {author} {\bibfnamefont {Jan}\ \bibnamefont {Zaanen}},\ }\bibfield  {title}
  {\enquote {\bibinfo {title} {Dual gauge field theory of quantum liquid
  crystals in two dimensions},}\ }\href {\doibase
  https://doi.org/10.1016/j.physrep.2017.03.004} {\bibfield  {journal}
  {\bibinfo  {journal} {Phys. Rep.}\ }\textbf {\bibinfo {volume} {683}},\
  \bibinfo {pages} {1 -- 110} (\bibinfo {year} {2017})},\ \bibinfo {note} {dual
  gauge field theory of quantum liquid crystals in two dimensions}\BibitemShut
  {NoStop}%
\bibitem [{\citenamefont {Alexander}\ \emph {et~al.}(2012)\citenamefont
  {Alexander}, \citenamefont {Chen}, \citenamefont {Matsumoto},\ and\
  \citenamefont {Kamien}}]{Kamienrmp}%
  \BibitemOpen
  \bibfield  {author} {\bibinfo {author} {\bibfnamefont {Gareth~P.}\
  \bibnamefont {Alexander}}, \bibinfo {author} {\bibfnamefont {Bryan Gin-ge}\
  \bibnamefont {Chen}}, \bibinfo {author} {\bibfnamefont {Elisabetta~A.}\
  \bibnamefont {Matsumoto}}, \ and\ \bibinfo {author} {\bibfnamefont
  {Randall~D.}\ \bibnamefont {Kamien}},\ }\bibfield  {title} {\enquote
  {\bibinfo {title} {{C}olloquium: {D}isclination loops, point defects, and all
  that in nematic liquid crystals},}\ }\href {\doibase
  10.1103/RevModPhys.84.497} {\bibfield  {journal} {\bibinfo  {journal} {Rev.
  Mod. Phys.}\ }\textbf {\bibinfo {volume} {84}},\ \bibinfo {pages} {497--514}
  (\bibinfo {year} {2012})}\BibitemShut {NoStop}%
\bibitem [{\citenamefont {Liu}\ \emph {et~al.}(2016)\citenamefont {Liu},
  \citenamefont {Nissinen}, \citenamefont {Slager}, \citenamefont {Wu},\ and\
  \citenamefont {Zaanen}}]{Genqcs2016}%
  \BibitemOpen
  \bibfield  {author} {\bibinfo {author} {\bibfnamefont {Ke}~\bibnamefont
  {Liu}}, \bibinfo {author} {\bibfnamefont {Jaakko}\ \bibnamefont {Nissinen}},
  \bibinfo {author} {\bibfnamefont {Robert-Jan}\ \bibnamefont {Slager}},
  \bibinfo {author} {\bibfnamefont {Kai}\ \bibnamefont {Wu}}, \ and\ \bibinfo
  {author} {\bibfnamefont {Jan}\ \bibnamefont {Zaanen}},\ }\bibfield  {title}
  {\enquote {\bibinfo {title} {Generalized {L}iquid {C}rystals: {G}iant
  {F}luctuations and the {V}estigial {C}hiral {O}rder of ${I}$, ${O}$, and
  ${T}$ {M}atter},}\ }\href {\doibase 10.1103/PhysRevX.6.041025} {\bibfield
  {journal} {\bibinfo  {journal} {Phys. Rev. X}\ }\textbf {\bibinfo {volume}
  {6}},\ \bibinfo {pages} {041025} (\bibinfo {year} {2016})}\BibitemShut
  {NoStop}%
\bibitem [{\citenamefont {Machon}\ and\ \citenamefont
  {Alexander}(2016)}]{Machon2016}%
  \BibitemOpen
  \bibfield  {author} {\bibinfo {author} {\bibfnamefont {Thomas}\ \bibnamefont
  {Machon}}\ and\ \bibinfo {author} {\bibfnamefont {Gareth~P.}\ \bibnamefont
  {Alexander}},\ }\bibfield  {title} {\enquote {\bibinfo {title} {Global defect
  topology in nematic liquid crystals},}\ }\href {\doibase
  10.1098/rspa.2016.0265} {\bibfield  {journal} {\bibinfo  {journal} {Proc. R.
  Soc. A}\ }\textbf {\bibinfo {volume} {472}},\ \bibinfo {pages} {20160265}
  (\bibinfo {year} {2016})}\BibitemShut {NoStop}%
\bibitem [{\citenamefont {Frankel}(2011)}]{Frankel}%
  \BibitemOpen
  \bibfield  {author} {\bibinfo {author} {\bibfnamefont {Theodore}\
  \bibnamefont {Frankel}},\ }\href@noop {} {\emph {\bibinfo {title} {{G}eometry
  of {P}hysics}}}\ (\bibinfo  {publisher} {Cambridge University Press},\
  \bibinfo {year} {2011})\BibitemShut {NoStop}%
\bibitem [{\citenamefont {Roy}\ and\ \citenamefont {Harper}(2017)}]{Floquet1}%
  \BibitemOpen
  \bibfield  {author} {\bibinfo {author} {\bibfnamefont {Rahul}\ \bibnamefont
  {Roy}}\ and\ \bibinfo {author} {\bibfnamefont {Fenner}\ \bibnamefont
  {Harper}},\ }\bibfield  {title} {\enquote {\bibinfo {title} {Periodic table
  for {F}loquet topological insulators},}\ }\href {\doibase
  10.1103/PhysRevB.96.155118} {\bibfield  {journal} {\bibinfo  {journal} {Phys.
  Rev. B}\ }\textbf {\bibinfo {volume} {96}},\ \bibinfo {pages} {155118}
  (\bibinfo {year} {2017})}\BibitemShut {NoStop}%
\bibitem [{\citenamefont {Nakagawa}\ \emph {et~al.}(2020)\citenamefont
  {Nakagawa}, \citenamefont {Slager}, \citenamefont {Higashikawa},\ and\
  \citenamefont {Oka}}]{Floquet2}%
  \BibitemOpen
  \bibfield  {author} {\bibinfo {author} {\bibfnamefont {Masaya}\ \bibnamefont
  {Nakagawa}}, \bibinfo {author} {\bibfnamefont {Robert-Jan}\ \bibnamefont
  {Slager}}, \bibinfo {author} {\bibfnamefont {Sho}\ \bibnamefont
  {Higashikawa}}, \ and\ \bibinfo {author} {\bibfnamefont {Takashi}\
  \bibnamefont {Oka}},\ }\bibfield  {title} {\enquote {\bibinfo {title}
  {Wannier representation of {F}loquet topological states},}\ }\href {\doibase
  10.1103/PhysRevB.101.075108} {\bibfield  {journal} {\bibinfo  {journal}
  {Phys. Rev. B}\ }\textbf {\bibinfo {volume} {101}},\ \bibinfo {pages}
  {075108} (\bibinfo {year} {2020})}\BibitemShut {NoStop}%
\bibitem [{\citenamefont {Harper}\ \emph {et~al.}(2020)\citenamefont {Harper},
  \citenamefont {Roy}, \citenamefont {Rudner},\ and\ \citenamefont
  {Sondhi}}]{Floquet3}%
  \BibitemOpen
  \bibfield  {author} {\bibinfo {author} {\bibfnamefont {Fenner}\ \bibnamefont
  {Harper}}, \bibinfo {author} {\bibfnamefont {Rahul}\ \bibnamefont {Roy}},
  \bibinfo {author} {\bibfnamefont {Mark~S.}\ \bibnamefont {Rudner}}, \ and\
  \bibinfo {author} {\bibfnamefont {S.L.}\ \bibnamefont {Sondhi}},\ }\bibfield
  {title} {\enquote {\bibinfo {title} {Topology and {B}roken {S}ymmetry in
  {F}loquet systems},}\ }\href {\doibase
  10.1146/annurev-conmatphys-031218-013721} {\bibfield  {journal} {\bibinfo
  {journal} {Annu. Rev. Condens. Matter Phys.}\ }\textbf {\bibinfo {volume}
  {11}},\ \bibinfo {pages} {345--368} (\bibinfo {year} {2020})}\BibitemShut
  {NoStop}%
\bibitem [{\citenamefont {Sun}\ \emph {et~al.}(2018{\natexlab{b}})\citenamefont
  {Sun}, \citenamefont {Xiao}, \citenamefont {Bzdu\v{s}ek}, \citenamefont
  {Zhang},\ and\ \citenamefont {Fan}}]{Sun:2018b}%
  \BibitemOpen
  \bibfield  {author} {\bibinfo {author} {\bibfnamefont {Xiao-Qi}\ \bibnamefont
  {Sun}}, \bibinfo {author} {\bibfnamefont {Meng}\ \bibnamefont {Xiao}},
  \bibinfo {author} {\bibfnamefont {Tom\'a\v{s}}\ \bibnamefont {Bzdu\v{s}ek}},
  \bibinfo {author} {\bibfnamefont {Shou-Cheng}\ \bibnamefont {Zhang}}, \ and\
  \bibinfo {author} {\bibfnamefont {Shanhui}\ \bibnamefont {Fan}},\ }\bibfield
  {title} {\enquote {\bibinfo {title} {Three-{D}imensional {C}hiral {L}attice
  {F}ermion in {F}loquet {S}ystems},}\ }\href {\doibase
  10.1103/PhysRevLett.121.196401} {\bibfield  {journal} {\bibinfo  {journal}
  {Phys. Rev. Lett.}\ }\textbf {\bibinfo {volume} {121}},\ \bibinfo {pages}
  {196401} (\bibinfo {year} {2018}{\natexlab{b}})}\BibitemShut {NoStop}%
\bibitem [{\citenamefont {Borgnia}\ \emph {et~al.}(2020)\citenamefont
  {Borgnia}, \citenamefont {Kruchkov},\ and\ \citenamefont {Slager}}]{modes}%
  \BibitemOpen
  \bibfield  {author} {\bibinfo {author} {\bibfnamefont {Dan~S.}\ \bibnamefont
  {Borgnia}}, \bibinfo {author} {\bibfnamefont {Alex~Jura}\ \bibnamefont
  {Kruchkov}}, \ and\ \bibinfo {author} {\bibfnamefont {Robert-Jan}\
  \bibnamefont {Slager}},\ }\bibfield  {title} {\enquote {\bibinfo {title}
  {Non-{H}ermitian {B}oundary {M}odes and {T}opology},}\ }\href {\doibase
  10.1103/PhysRevLett.124.056802} {\bibfield  {journal} {\bibinfo  {journal}
  {Phys. Rev. Lett.}\ }\textbf {\bibinfo {volume} {124}},\ \bibinfo {pages}
  {056802} (\bibinfo {year} {2020})}\BibitemShut {NoStop}%
\bibitem [{\citenamefont {Kawabata}\ \emph {et~al.}(2019)\citenamefont
  {Kawabata}, \citenamefont {Shiozaki}, \citenamefont {Ueda},\ and\
  \citenamefont {Sato}}]{kawabatanonhermitian}%
  \BibitemOpen
  \bibfield  {author} {\bibinfo {author} {\bibfnamefont {Kohei}\ \bibnamefont
  {Kawabata}}, \bibinfo {author} {\bibfnamefont {Ken}\ \bibnamefont
  {Shiozaki}}, \bibinfo {author} {\bibfnamefont {Masahito}\ \bibnamefont
  {Ueda}}, \ and\ \bibinfo {author} {\bibfnamefont {Masatoshi}\ \bibnamefont
  {Sato}},\ }\bibfield  {title} {\enquote {\bibinfo {title} {Symmetry and
  {T}opology in {N}on-{H}ermitian {P}hysics},}\ }\href {\doibase
  10.1103/PhysRevX.9.041015} {\bibfield  {journal} {\bibinfo  {journal} {Phys.
  Rev. X}\ }\textbf {\bibinfo {volume} {9}},\ \bibinfo {pages} {041015}
  (\bibinfo {year} {2019})}\BibitemShut {NoStop}%
\bibitem [{\citenamefont {Zhou}\ and\ \citenamefont
  {Lee}(2019)}]{zhounonhermitian}%
  \BibitemOpen
  \bibfield  {author} {\bibinfo {author} {\bibfnamefont {Hengyun}\ \bibnamefont
  {Zhou}}\ and\ \bibinfo {author} {\bibfnamefont {Jong~Yeon}\ \bibnamefont
  {Lee}},\ }\bibfield  {title} {\enquote {\bibinfo {title} {Periodic table for
  topological bands with non-hermitian symmetries},}\ }\href {\doibase
  10.1103/PhysRevB.99.235112} {\bibfield  {journal} {\bibinfo  {journal} {Phys.
  Rev. B}\ }\textbf {\bibinfo {volume} {99}},\ \bibinfo {pages} {235112}
  (\bibinfo {year} {2019})}\BibitemShut {NoStop}%
\bibitem [{\citenamefont {Li}\ and\ \citenamefont {Mong}(2019)}]{li:2019}%
  \BibitemOpen
  \bibfield  {author} {\bibinfo {author} {\bibfnamefont {Zhi}\ \bibnamefont
  {Li}}\ and\ \bibinfo {author} {\bibfnamefont {Roger S.~K.}\ \bibnamefont
  {Mong}},\ }\bibfield  {title} {\enquote {\bibinfo {title} {Homotopical
  classification of non-{H}ermitian band structures},}\ }\href
  {https://arxiv.org/abs/1911.02697} {\bibfield  {journal} {\bibinfo  {journal}
  {arXiv:1911.02697}\ } (\bibinfo {year} {2019})}\BibitemShut {NoStop}%
\bibitem [{\citenamefont {Zhong}\ \emph {et~al.}(2018)\citenamefont {Zhong},
  \citenamefont {Khajavikhan}, \citenamefont {Christodoulides},\ and\
  \citenamefont {El-Ganainy}}]{ZhongNOnHerm}%
  \BibitemOpen
  \bibfield  {author} {\bibinfo {author} {\bibfnamefont {Qi}~\bibnamefont
  {Zhong}}, \bibinfo {author} {\bibfnamefont {Mercedeh}\ \bibnamefont
  {Khajavikhan}}, \bibinfo {author} {\bibfnamefont {Demetrios~N.}\ \bibnamefont
  {Christodoulides}}, \ and\ \bibinfo {author} {\bibfnamefont {Ramy}\
  \bibnamefont {El-Ganainy}},\ }\bibfield  {title} {\enquote {\bibinfo {title}
  {Winding around non-{H}ermitian singularities},}\ }\href {\doibase
  10.1038/s41467-018-07105-0} {\bibfield  {journal} {\bibinfo  {journal} {Nat.
  Commun.}\ }\textbf {\bibinfo {volume} {9}},\ \bibinfo {pages} {4808}
  (\bibinfo {year} {2018})}\BibitemShut {NoStop}%
\bibitem [{\citenamefont {Lee}(2011)}]{Lee_SM}%
  \BibitemOpen
  \bibfield  {author} {\bibinfo {author} {\bibfnamefont {John~M.}\ \bibnamefont
  {Lee}},\ }\href@noop {} {\emph {\bibinfo {title} {{I}ntroduction to
  {T}opological {M}anifolds}}},\ \bibinfo {edition} {2nd}\ ed.\ (\bibinfo
  {publisher} {Springer},\ \bibinfo {year} {2011})\BibitemShut {NoStop}%
\bibitem [{\citenamefont {Haber}()}]{Haber_antisym}%
  \BibitemOpen
  \bibfield  {author} {\bibinfo {author} {\bibfnamefont {Howard~E.}\
  \bibnamefont {Haber}},\ }\href@noop {} {\enquote {\bibinfo {title}
  {{P}arameterization of real orthogonal antisymmetric matrices},}\ }\bibinfo
  {howpublished} {\url{http://scipp.ucsc.edu/~haber/webpage/antiortho.pdf}},\
  \bibinfo {note} {accessed: 2020-01-07}\BibitemShut {NoStop}%
\bibitem [{\citenamefont {arctic tern}(2017)}]{mathSE:2215495orig}%
  \BibitemOpen
  \bibfield  {author} {\bibinfo {author} {\bibnamefont {arctic tern}},\
  }\href@noop {} {\enquote {\bibinfo {title} {Second homotopy group of real
  {G}rassmannians $\textrm{Gr}(n,m)$.}}\ }\bibinfo {howpublished} {Mathematics
  Stack Exchange} (\bibinfo {year} {2017}),\ \bibinfo {note} {{URL}:
  http://math.stackexchange.com/q/2215495 (version: 2017-04-03)}\BibitemShut
  {NoStop}%
\end{thebibliography}
\end{document}